\documentclass[epj]{svjour}
\usepackage{xcolor}
\usepackage{graphics}
\usepackage[dvips]{graphicx}
\usepackage{amssymb}
\usepackage{amsmath}
\usepackage[square, sort, numbers]{natbib}

\def\scr{\rm\scriptscriptstyle }
\def\scs{\scriptscriptstyle }

\begin{document}

\setcounter{secnumdepth}{2}
\setcounter{tocdepth}{2}

\title{The total reaction cross section of heavy-ion reactions induced by stable and unstable exotic beams: The low-energy regime}
\author{L. F. Canto\inst{1}, V. Guimar\~aes\inst{2}, J. Lubian\inst{3}, and M. S. Hussein\inst{4,5}\thanks{deceased}}
\institute{$^1$ Instituto de F\'{\i}sica, Universidade Federal do Rio de Janeiro, CP 68528,
Rio de Janeiro, Brazil, \email{canto@if.ufrj.br}\\
$^2$ Universidade de S\~ao Paulo, C.P. 66318, 05314-970, 
S\~ao Paulo, SP, Brazil, \\
$^3$ Instituto de F\'{\i}sica, Universidade Federal Fluminense, Av. Litoranea 
s/n, Gragoat\'{a}, Niter\'{o}i, R.J., 24210-340, Brazil, \\
$^4$ Instituto de Estudos Avan\c{c}ados, Universidade de S\~{a}o Paulo C. P.
72012, 05508-970 S\~{a}o Paulo-SP, Brazil, and Instituto de F\'{\i}sica,
Universidade de S\~{a}o Paulo, C. P. 66318, 05314-970 S\~{a}o Paulo,-SP,
Brazil, \\
$^5$ Departamento de F\'{i}sica, Instituto Tecnol\'{o}gico de Aeron\'{a}utica, CTA, S\~{a}o Jos\'{e} dos Campos, 
S\~ao Paulo, SP, Brazil \\
}

\date{\today}

\abstract{In this review paper we present  a detailed account of the extraction and the calculation of the total reaction cross section of  
strongly bound and weakly bound, stable and unstable, exotic, nuclei. We discuss the optical model and the 
more general coupled channels model of direct reactions, and how from fits to the data on elastic scattering supplies  the elastic element 
of (partial wave) S-matrix and correspondingly the differential cross section and the total reaction cross section. The effect of long-range 
absorption due to the coupling to excited states in the target and to the breakup  continuum in the projectile is also discussed. The 
semiclassical method is then analyzed and the Hill-Wheeler expression of the tunneling probability and the Wong formula for the fusion 
and the total reaction cross sections are discussed in details. The generalized optical theorem for charged particle scattering and the 
resulting sum-of differences method is then discussed. Also, the strong absorption model in its sharp cutoff form and  its generalization, 
the smooth cutoff, is discussed. The so-called "quarter-point recipe" is discussed next, and the quarter-point angle is introduced as 
a simple and rapid mean to obtain the total reaction cross section.  The last topic discussed is the reduction of the total reaction cross 
section that would allow a large body of data to sit on a single universal function. Such a universal function exists in the case of the fusion 
data, and the aim of this last topic of the review is to extend the fusion case to the total reaction, by adding the direct reaction contribution. 
Also discussed is 
the inclusive breakup cross section and how it can be used to extract the total reaction cross section of the interacting fragment with the 
target. This method is also known as the Surrogate method and represents a case of hybrid reactions. The sum of the integrated inclusive 
breakup cross section with the complete fusion cross section supplies the total fusion cross section. The main experimental methods to 
determine the total reaction cross section are also discussed, with emphasis in recent techniques developed to deal with reactions 
induced by unstable beams.
} 

\PACS {{24.10Eq}  {25.70.Bc} {25.60Gc}}

\authorrunning{L.F. Canto \textit{et al.}}

\titlerunning{Theory of the total reaction cross section ...}

\maketitle

\tableofcontents

\section{Introduction}

The field of nuclear reactions has evolved greatly over the last 6 or so decades. In most applications to low energy direct reactions the reliance has 
been on the use of the optical model in its single channel version and its coupled channels version. The more complex compound nuclear reactions 
are treated within the Statistical Theory and gives for the average cross section for a transition (a,b) that proceeds through the compound nucleus 
(CN) the Hauser-Feshbach form whose calculation involves precisely the  theory of direct reactions. This  closed theory has been recently pushed to 
the limits when confronted with reactions induced by weakly bound normal and exotic nuclei. These latter  nuclei, such as the Borromean two-neutron
halo ones $^6$He ($^4{\rm He} + 2n,\, S_{2n} = 0.973$ MeV), $^{11}$Li ($^9{\rm Li} + 2n,\, S_{2n} = 0.369$ MeV), $^{14}$Be ($^{12}{\rm Be} + 2n,\, 
S_{2n} = 1.27$ MeV ), $^{22}$C ($^{20}{\rm C} + 2n,\, S_{2n} = 0.42 \pm 0.94$ MeV), the one-proton halo isotope $^8$B ($^7{\rm Be} + p, \, S_{p} 
= 0.14$ MeV), and the one-neutron halo isotopes $^{11}$Be ($^{10}{\rm Be} + n,\, S_{n} = 0.503$ MeV), $^{15}$C ($^{14}{\rm C} + n,\, S_{n} = 
1.218$ MeV),  $^{19}$C ($^{18}{\rm C} + n,\, S_{n} = 0.242 \pm 0.095$ MeV) and $^{23}$O ($^{22}{\rm O} + n),\, S_{n} = 2.74$ MeV), are produced 
in several laboratories scattered in the world and used as secondary beams to react with target nuclei such as $^{12}$C, $^{58}$Ni and $^{208}$Pb. 
An important consequence of the small binding energy of the valence particles in these weakly-bound nuclei is the increase of the radii as compared 
to normal nuclei with the same mass number. Several review articles have been written on different reactions involving weakly bound 
nuclei \cite{CGD06,KRA07,KAK09,CGD15,KGA16,MoS17,Bon18,JPK20}.

What distinguishes the reactions induced by these very short lived nuclei as well as by stable weakly bound nuclei, such as $^6$Li ($^4{\rm He} +\, ^2{\rm H}
,\, S = 1.47$ MeV),  $^7$Li ($^4{\rm He} +\, ^3{\rm H},\, S = 2.47$ MeV) and $^9$Be ($^4{\rm He} +\, ^4{\rm He} + n,\, S = 1.67$ MeV), from the usual reactions 
involving stable strongly bound projectiles is the presence in the former of two important features: existence of appreciable dipole strength at low 
excitation energy (the Pygmy resonance), and strong couplings with the breakup channel even at energies close to the Coulomb barrier. The latter feature 
is shared by the well studied deuteron scattering, with breakup threshold of 2.2 MeV, which is to be compared to 0.369 MeV for $^{11}$Li and 0.973 MeV for 
$^6$He. This implies that the continuum must always be taken into account in any serious attempt to analyze the data with the direct reaction theory. This was not the 
case in the past where ordinary strongly bound projectiles such as $^{16}$O were used to induce the nuclear reaction. At low energies, the single channel optical 
model was found to be adequate for spherical targets, whereas its coupled channels version is required for deformed targets such as $^{152}$Sm or $^{238}$U, 
where, in these cases, the target's  rotational states must be taken into account.\\

One of the most important source of information about the size, radius, and geometry of nuclei comes from elastic scattering. A byproduct of the analysis of 
this reaction is the total reaction cross section, which accounts for all nonelastic processes. Whether the analysis of the elastic scattering data is done 
within the single channel or the coupled channel version of the optical model theory of direct reactions, the total reaction cross section is defined 
in terms of the modulus of the partial-wave projected elastic $S$-matrix.
 This cross section is the sum of the compound nucleus formation cross section (fusion) plus the sum of all direct nonelastic cross sections. In the case of 
 non-exotic and strongly bound projectiles, the fusion cross section is that which accounts for the capture of the whole projectile, while the direct processes 
 are inelastic excitations of the target and projectile and possibly transfer processes. In the case of exotic nuclei, the direct processes should include the elastic 
 and nonelastic breakup of the projectile as well. The fusion cross section should also be defined  differently here. The complete fusion (CF) is the compound
 nucleus formation
cross section. There are other processes that come from the direct part, which involve the capture of one of the fragments of the projectile after the breakup
process. This type of hybrid reaction is a new feature in the reaction and requires a new addition to the theory. This process is called incomplete fusion (ICF). 
The sum of CF and ICF is the total fusion TF. The cross section for these processes are denoted $\sigma_{\scr CF}$, $\sigma_{\scr ICF}$, and 
$\sigma_{\scr TF}$, with the sum rule, $\sigma_{\scr TF} = \sigma_{\scr CF} + \sigma_{\scr ICF}$. Experimentally, one usually measures $\sigma_{\scr TF}$.
However, separate measurements of  $\sigma_{\scr CF}$ and $\sigma_{\scr ICF}$ are available for a few particular projectile-target combinations (see 
Ref.~\cite{CGD15} and references therein). The analysis of the data, is usually done within the coupled channel version of the optical model theory of direct 
reactions, extended to include the breakup coupling. The resulting theory which reduced a three-body or four-body scattering system into an equivalent 
two-body system through a diligent discretization of the breakup continuum, is called the Continuum Discretized Coupled Channels (CDCC) 
theory~\cite{SYK86,AIK87} is widely used~\cite{HVD00,DiT02,DTB03}. Again from the analysis of the elastic scattering data one extracts the total reaction 
cross section which is of paramount importance in supplying a unitarity constraint on models used to calculate the different pieces of the direct reaction part 
and the CF part. For its importance in both strongly bound and weakly bound projectiles induced nuclear reactions, we felt the need to give an account of the 
different methods used for its extraction from the data. Back in 1991, Hussein, Rego and Bertulani~\cite{HRB91} wrote a comprehensive review of the theory 
of the total reaction cross section. At that time, very few data existed for the elastic scattering of exotic and other weakly bound nuclear projectiles. The
same happened for total reaction data. Thus, we
think the time has come to supply, not a review, but an overall account with new and original material concerning this cross section, in view of the existence 
of a reasonably large body of new elastic scattering data of these exotic nuclear system.\\

This paper is organized as follows. In section 2 we introduce the potential scattering approach to heavy-ion collisions, where the influence of intrinsic degrees of 
freedom is simulated by a complex optical potential. We then consider some frequently used semiclassical approximations to the scattering amplitude and to the 
fusion cross section, and discuss the optical theorem in the presence of absorption and Coulomb forces. The effects of intrinsic degrees of freedom on the 
collision dynamics are explicitly included in section 3. We introduce the coupled channel approach, and its generalization to deal with continuum states, the so-called 
{\it continuum discretized coupled channel} method. The latter is a basic tool to describe collisions of weakly bound projectiles, which may break up 
into fragments as it interacts with the target. In section 4 we discuss the surrogate method, which is very useful to determine inclusive cross sections 
in collisions of weakly bound projectiles. In section 5, we discuss experimental methods to determine total reaction cross sections in heavy-ion collisions. 
Special attention is devoted to recently developed techniques to handle reactions induced by low intensity radioactive beams. We discuss also the available
procedures to reduce fusion and total reaction data, for the purpose of comparing results for different systems. Finally, in section 6
we present a summary of the topics discussed in this review paper.


\section{Potential scattering and the optical Model}\label{PotScat}


The description of heavy-ion collisions by potential scattering relies on the optical model. In this approach, the intrinsic properties of the collision partners are not
explicitly taken into account. The attenuation of the incident wave owing to transitions to excited states of the system are then mocked up by a negative imaginary 
part in the interaction, which gives rise to a sub-unitary S-matrix.
In typical situations, the potential is complex, spherically symmetric, depending only on the modulus of the collision vector, r, and it can be written as:
\begin{equation}
U(r) = V(r) +\,i\,W(r).
\label{V-potscat}
\end{equation}
The real potential
\begin{equation}
V(r) = V_{\scr C}(r) +\, V_{\scr N}(r),
\label{UN-UCt}
\end{equation}
is the sum of the Coulomb and the nuclear interactions.
An immediate consequence of this simplified treatment is that it cannot give cross sections for a particular nonelastic channel. It can only predict the elastic cross 
section and the total reaction cross section. The latter, given by the absorption cross section, represents the inclusive cross section for all nonelastic events (fusion 
plus direct reactions).\\

The details of $W(r)$ depend on the physical processes responsible for the absorption of the incident flux. The formation of the CN, that is, fusion, always contribute 
to the absorption. Therefore, $W(r)$ must be very strong at small  values of $r$, i.e., at the inner region of the Coulomb barrier.  If fusion is the 
only process responsible for the absorption, the imaginary potential must be given by a function with large strength and short range. It can be represented, for instance, 
by the Woods-Saxon function,
\begin{equation}
W(r) = -\,\frac{W_0}{1 + \exp\left[ (r-R_{\rm 0i})/a_{\rm 0i}) \right]},
\label{WS-i}
\end{equation}
where $R_{\rm 0i}= r_{\rm 0i}\,\left[A_{\scr P}^{\scr 1/3}+A_{\scr T}^{\scr 1/3}\right]$, with $A_{\scr P}$ and $A_{\scr T}$ standing for the mass numbers of the 
projectile and the target, respectively. In this case, $W_0$ is of the order of tens of MeV and the radius and diffusivity parameters, $r_{\rm 0i}$ and 
$a_{\rm 0i}$, are chosen as to make $W(r)$ be of short range. Typical values are $W_0 = 50$ MeV, $r_{\rm 0i} \sim 1.0$ fm and $a_{\rm 0i} \sim 0.2$ fm. \\

However, in typical heavy-ion collisions, the elastic and the reaction cross sections are strongly affected by direct reactions. Thus, the imaginary 
potentials should account also for the influence of these processes. Since they take place in grazing collisions, the range of the imaginary potential 
should be longer, reaching the barrier region.  A common practice is to use the same radial dependence for the real and imaginary parts of the potential. 
Using this procedure with the double-folding S\~ao Paulo potential~\cite{CPH97,CCG02}, Gasques {\it et al.}~\cite{GCG06} successfully described 
elastic scattering and total reaction cross sections for systems in different mass regions. 

\smallskip

\begin{figure}
\begin{center}
\includegraphics*[width=7cm]{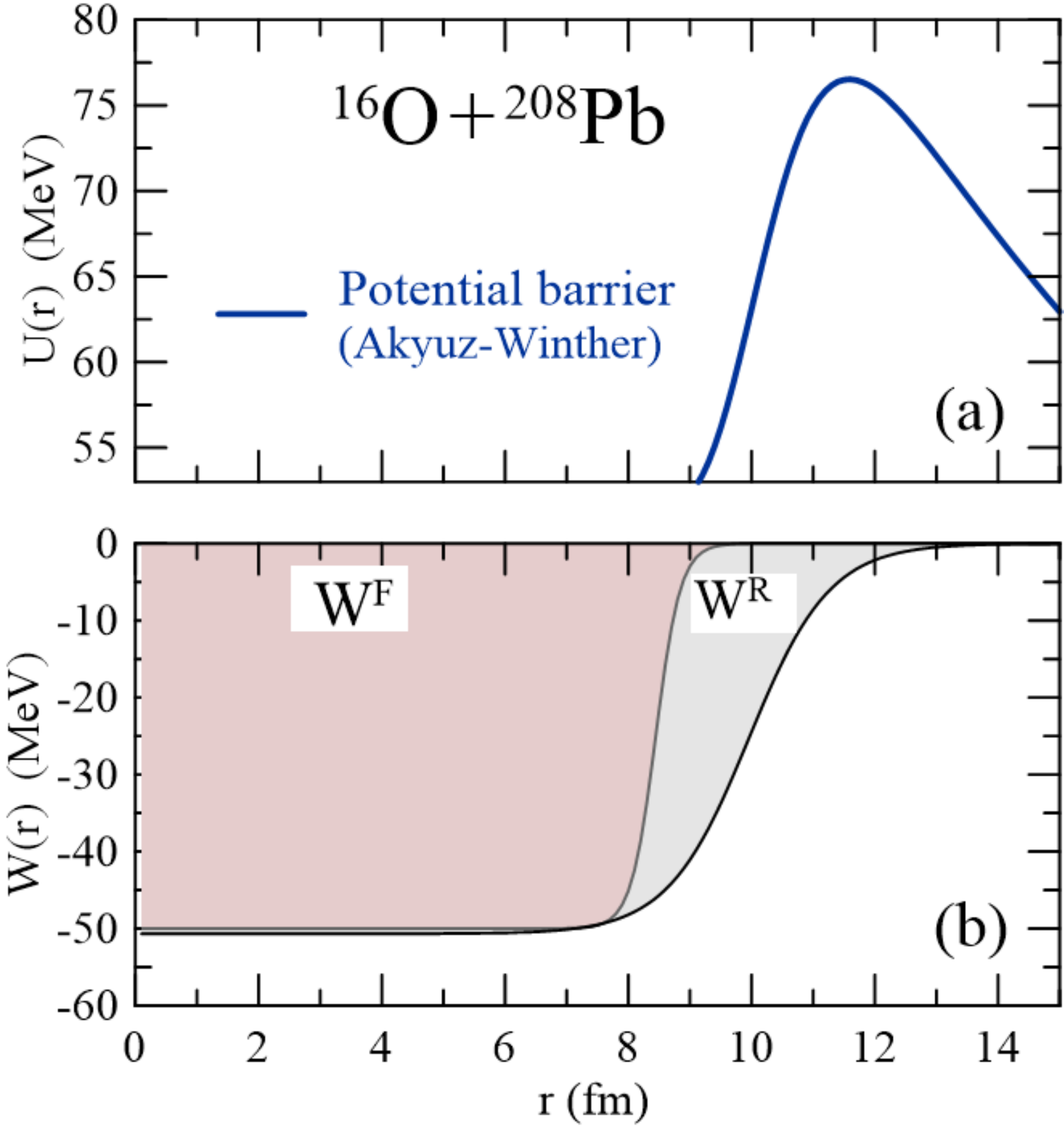}
\end{center}
\caption{(Color on line) Panel (a): The Coulomb barrier for the S\~ao Paulo potential~\cite{CPH97,CCG02} in $^{16}{\rm O} \,-\,^{208}{\rm Pb}$ scattering; Panel (b):
Imaginary potentials simulating pure fusion absorption ($W^{\scr F}$) and absorption arising from fusion and from direct reactions ($W^{\scr R}$). For detail, see the text.}
\label{WV-pots}
\end{figure}
An illustration of the imaginary potentials in the cases of pure fusion absorption ($W^{\scr F}$) and fusion plus direct reaction absorption ($W^{\scr R}$) for the 
$^{16}{\rm O} \,-\,^{208}{\rm Pb}$ scattering is presented in Fig.~\ref{WV-pots}. In this example, the real part of the interaction is given by the Aky\"uz-Winther 
potential~\cite{BrW04,AkW81}. Panel (a) shows the Coulomb barrier for this potential, whereas panel (b) shows the imaginary potentials $W^{\scr F}$ and
$W^{\scr R}$. The potential $W^{\scr F}$ was evaluated by Eq.~(\ref{WS-i}), with $W_0 = 50$ MeV, 
 $r_{\rm 0i} = 1.0$ fm and $a_{\rm 0i} = 0.2$ fm, and $W^{\scr R}$ was obtained by multiplying the real part of the potential by the factor 0.78. \\


\subsection{Fusion and total reaction cross sections}

In potential scattering the absorption cross section is given by the partial-wave series\footnote{For simplicity, we are neglecting spins of the collision partners.}
\begin{equation}
\sigma_{\rm abs}(E) = \frac{\pi}{k^2} \sum_{l = 0}^{\infty}\, (2l + 1)\, P_{\rm abs}\left(l,E\right) ,
\label{sigma-abs}
\end{equation}
where
\begin{equation}
P_{\rm abs}\left(l,E\right)= 1 - \left| S_{ l}(E) \right|^2
\label{modS}
\end{equation}
is the absorption probability at the $l^{\scr th}$-partial wave, which is given by the deviation of the partial-wave component of the S-matrix,  $S_{l}(E)$, 
from the unitary behaviour. \\

The absorption cross section can also be given in terms of the expectation value of the imaginary potential, through the expression
\begin{equation}
\sigma_{\rm abs}(E) = -\ \frac{1}{|A|^2}\ \frac{k}{E}\ \left\langle \psi^{\scr (+)}\left| \, W(r)\, \right| \, \psi^{\scr(+)} \right\rangle ,
\label{sigmapsi}
\end{equation}
where $A$ is the normalization constant of the scattering wave function, $\psi^{\scr (+)}({\bf r})$. Eq.~(\ref{sigmapsi}), can be easily derived from the continuity 
equation for the Schr\"odinger equation with the complex potential~\cite{CaH13}.
Carrying out the partial-wave expansion of Eq.~(\ref{sigmapsi}), the reaction cross section takes the form of Eq.~(\ref{sigma-abs}), with the absorption probabilities of 
Eq.~(\ref{modS}) given by the radial integral
\begin{equation}
P_{\rm abs}\left(l,E\right) = -\frac{4k}{E}\int_{0}^{\infty} dr\ W(r)\, |u_{l}(k, r)|^2 ,
\label{PW-l}
\end{equation}
with $u_{l}(k, r)$ standing for the radial wave function of the $l^{\rm th}$ partial wave, in a collision with wave number $k =\sqrt{2\mu E}/\hbar$, where $\mu$ stands 
for the reduced mass.\\

The relation between $\sigma_{\rm abs}(E)$ and the observable cross sections depends on the nature of the imaginary potential. When it simulates the influence of 
fusion plus direct reactions ($W(r)=W^{\scr R}(r)$), $\sigma_{\rm abs}(E)$ corresponds to the total reaction cross section. On the other hand, if one is interested in 
the fusion cross section, one should use a strong imaginary potential with a short-range. However, this does not guarantee fusion. For CN formation, the projectile and
the target must remain in close proximity for a long time, long enough for the full thermalization of the excitation energy. This happens when the system is caught inside 
a pocked of the effective potential,
\begin{equation}
 V_l(r) =  V(r) \,+\,  \frac{\hbar^2}{2\mu\,r^2}\ l(l+1),
\label {Vl(r)}
\end{equation}
which appears in the radial equation. Thus, the fusion cross section should be written,
\begin{equation}
\sigma_{\scr F}(E) = \frac{\pi}{k^2} \sum_{l = 0}^{\infty}\, (2l + 1)\, P_{\scr F}\left(l,E\right) ,
\label{sigma-F}
\end{equation}
with
\begin{equation}
P_{\scr F}\left(l,E\right)= P_{\rm abs}\left(l,E\right) \times P_{\scr CN}\left(l,E\right) .
\label{PF}
\end{equation}
Above, $P_{\scr CN}\left(l,E\right)$ is the probability of CN formation, after the system enters the strong absorption region. For low partial waves and near-barrier energies, 
this probability is very close to one. The situation is different for partial waves above the critical angular momentum,  $l_{\rm cr}$. This angular momentum is defined as the
highest $l$-value for which the potential $V_l(r)$ exhibits a pocket. Above $l_{\rm cr}$, $V_l(r)$ is strongly repulsive (dominated by the centrifugal term), decreasing 
monotonically with $r$. In this way, the system may enter the strong absorption region but it stays there for a short time, orders of magnitude shorter than that required 
for thermalization and CN formation. In this situation, the absorption corresponds to inelastic scattering, transfer or pre-equilibrium reactions, but definitely not to fusion. 
Therefore, partial-wave above $l_{\rm cr}$ should not be included in the calculation of fusion. This is is achieved by setting,\\
\begin{eqnarray}
\mathcal{P}_{\scr CN}(l ,E) &=& 1,\ \ \ {\rm for}\ l <l_{\rm cr} \nonumber \\
                                           &=& 0,\ \ \ {\rm for}\ \ge l_{\rm cr} .
\label{abs-PF}
\end{eqnarray}

There is an alternative to the use of a complex potential in the calculation of fusion cross sections. One can keep the potential real and adopt ingoing 
wave boundary conditions (IWBC)~\cite{FPW47,Raw64,Raw66,CGD06} for the radial wave functions at the bottom of the pocked of $V_l(r)$, $R_{\rm in}$.  
The wave functions and their derivatives at $r=R_{\rm in}$ are then evaluated within the WKB approximation,  and the radial equations are numerically 
integrated, from $R_{\rm in}$ to the matching radius. In an optical model calculation with a strong and short range imaginary potential, the incident waves reaching 
the inner region of the barrier are completely absorbed, so that there is no reflected wave coming out. In this way, radial wave functions obtained with 
$W^{\scr F}(r)$ are expected to be equivalent to those of a real potential with IWBC, and the same happens with the corresponding components of the S-matrix, 
$S_l(E)$.


\subsubsection{The WKB approximation and the Hill-Wheeler transmission coefficient}


The absorption probability in Eq.~(\ref{modS}), $P_{\rm abs}(l,E)$, acquires a simple form in the WKB (Wentzel, Kramers, and Brillouin) approximation, where the radial wave 
functions are written in terms of the local wave numbers, $k_{l}(r)$, defined as,

\begin{equation}
\kappa_{l}(r) = \frac{1}{\hbar}\ \sqrt{V_l(r) + i\,W(r) - E} .
\label{local-k}
\end{equation}
%
%
However, the explicit presence of the imaginary potential in $k_{l}(r)$ leads to extra difficulties in the calculation, as one has to deal with complex turning points (this will become
clear in Eqs.~(\ref{TK}) and (\ref{Phi})). 
Although, in principle, such calculation can be performed by resorting to the method of complex angular momenta (through the requirement that the imaginary part of the 
turning point is identically zero), practical usage of the results is limited. \\

The situation is better in calculations of fusion cross sections, where the imaginary potential is very strong but has a short-range. In this case, it absorbs completely the current 
that reaches the inner region of the barrier, but it is negligible elsewhere.  It is still simpler in IWBC calculations, which are based on real potentials. In this case, the
absorption probability at the partial-wave $l\,(\le l_{\rm cr})$ is equal to the transmission coefficient of the incident wave through the barrier of $V_l(r)$,
$T(l,E)$,  namely
\begin{equation}
P_{\rm abs}(l,E) = P_{\scr F}\left(l,E\right) \simeq T(l,E) .
\label{Pl-Tl}
\end{equation}

Within Kemble's improved version~\cite{Kem35} of the WKB approximation, the transmission coefficient is given by the expression
\begin{equation}
T(l,E) \simeq T^{\scr WKB}(l,E) = \frac{1}{1 + \exp[2\,\Phi_l^{\scr WKB}(E)]} ,
\label{TK}
\end{equation}
where $\Phi_l^{\scr WKB}(E)$ is the integral of the local wave number evaluated along the classically forbidden region,
\begin{equation}
\Phi^{\scr WKB}_l(E) = \int_{r_i}^{r_e} \, \kappa_l(r)\ dr ,
\label{Phi}
\end{equation}
with $r_i$ and $r_e$ standing respectively for the internal and external classical turning points for the potential $V_l(r)$. These turning points are real at sub-barrier 
energies but they become complex above the barrier. Kemble argued that this problem could be solved through an analytical continuation of the radial variable to the 
complex plane. He pointed out that in the case of a parabolic barrier, discussed in the next sub-section, the integral of Eq.~(\ref{Phi}) is given by an analytical expression and this expression is valid for any collision energy, below or above the barrier. Recently, the analytical continuation of the radial variable for typical heavy-ion potentials was 
discussed, and the applicability of the Wong formula (see section \ref{Poisson-Wong}) was extended to above-barrier energies~\cite{TCH17,TCH17a}.\\

Hill and Wheeler studied the transmission through the parabolic barrier
\begin{equation}
V(r) = V_{\scr B} -  \frac{1}{2}\, \mu\omega^2\ \left( r-R_{\scr B} \right)^2.
\end{equation}
In this case, the transmission coefficient can be evaluated exactly. The result, known as the Hill-Wheeler transmission factor, is
\begin{equation}
T(E) = \frac{1}{
1+\exp\left[ 
2\pi\,\left( V_{\scr B}-E \right)/\hbar\omega\right]
}.
\label{THW-1}
\end{equation}
Approximate expression for heavy-ion fusion cross sections can be obtained by taking the Hill-Wheeler transmission function to obtain the fusion probabilities.
For this purpose, the effective $l$-dependent potentials are approximated by a parabola, as 
\begin{equation}
V_l(r) = B_l -  \frac{1}{2}\, \mu\omega_l^2\ \left( r-R_l \right)^2,
\label{l-parabolae}
\end{equation}
where $B_l $, $R_l$ and $\hbar\omega_l$ are respectively the height, radius and curvature parameter of the barrier for the $l^{\rm th}$-partial wave.
The corresponding transmission coefficients are, then, given by the expression,
\begin{equation}
T^{\scr HW}(l,E) = \frac{1}{1+\exp\left[ 2\pi\,\left( B_l-E \right)/\hbar\omega_l \right]}.
\label{THW}
\end{equation}
If the imaginary potential has a short range, as the case of $W^{\scr F}(r)$ in Fig.~\ref{WV-pots}, absorption is equivalent to fusion and the fusion probabilities
should be very close to the corresponding transmission coefficients. 
However, this assumption is meaningless for partial-waves above $l_{\rm cr}$, where the potential $V_l(r)$ decreases monotonically as $r$ increases. To deal with this
problem, one truncates the partial-wave series for $\sigma_{\scr F}$ at $l = l_{\rm cr}$. In this way, the
WKB approximation for $\sigma_{\scr F}$ contains the implicit assumption that the factor $P_{\scr CN}(l,E)$ of Eq.~(\ref{PF}) is equal to one below $l_{\rm cr}$ and zero
otherwise. The fusion cross section in the WKB approximation can be written, \\
\begin{equation}
\sigma_{\scr F}(E) = \frac{\pi}{k^2} \sum_{l = 0}^{l_{\rm cr}} (2l + 1)\ T^{\scr HW}(l,E) .
\label{sigma-HW}
\end{equation}

The quality of the parabolic approximation for the potential barriers depends on the collision energy and on the mass of the system. It is reasonable at near-barrier
energies but becomes very inaccurate at energies well below the Coulomb barrier. It works fairly well for heavy systems even at collision energies several MeV below
the Coulomb barrier. However, this approximation is quite poor for light heavy ions~\cite{CGD06}. A common practice that leads to a very accurate fusion cross section
is to use Kemble's transmission coefficients (Eq.~(\ref{TK})) below the Coulomb barrier and the Hill-Wheeler transmission factors (Eq.~(\ref{THW})) above.


\subsubsection{Poisson series and the Wong formula}\label{Poisson-Wong}


The sum of partial waves giving the fusion cross section can be transformed into a rapidly converging series of integrals. According to the Poisson formula, we can
write
\begin{multline}
\sigma_{\scr F}(E) = \frac{\pi}{k^2}\ \sum_{l=0}^\infty (2l+1)\, T(l,E) \\
= \frac{2\pi}{k^2} \sum_{m = 0, \pm1, ...}(-)^{m}\int \lambda\  T\left( \lambda,E \right) 
e^{2i\pi m \, \lambda}\ d\lambda ,
\label{Poisson-1}
\end{multline}
where
\[
\lambda = l+1/2\ \ \ {\rm and}\ \ \ T_l(E) = T\left( \lambda,E \right) .
\]
The sum over $m$ converges very rapidly, so that, in most cases, it is enough to consider the leading term, with $m=0$.\\

Wong~\cite{Won73} obtained a very useful expression for the fusion cross section by considering  only the $m=0$ term of the Poisson series
and making additional approximations on the $l$-dependent effective potential. First, he adopted the parabolic approximations
of Eq.~(\ref{l-parabolae}). Next, he neglected the $l$-dependences of the $R_l$ and $\hbar\omega_l$ and made the approximations,
\begin{eqnarray}
R_l                    &=& R_{l=0}\equiv R_{\scr B}\nonumber\\
 \hbar\omega_l &=& \hbar\omega_{l=0} \equiv \hbar\omega, \nonumber\\  
B_{\scr \lambda} & \simeq &  V_{\scr B} +\frac{\hbar^2 \lambda^2}{2\mu\, R_{\scr B}^2} .
\label{Wong-approx}
\end{eqnarray}
With these approximations the integral over $\lambda$ can be carried out analytically and one gets the so-called Wong cross section~\cite{Won73} \\
\begin{equation}
\sigma^{\scr W}(E) = \frac{\hbar\omega\, R_{\scr B}^2}{2E}\ 
 \ln{\left[1 + \exp{\left(\frac{2\pi}{\hbar\omega}(E - V_{\scr B})\right)}\right]}.
\label{sigWong}
\end{equation}

At low enough sub-barrier energies ($E \ll  V_{\scr B} -\hbar\omega$), Eq.~(\ref{sigWong}) may be approximated by the simpler expression,
\begin{equation}
\sigma_{\scr F}(E) \simeq  \frac{\hbar\omega\, R_{\scr B}^2}{2E} \ \exp\left[ -2\pi\,\frac{\left| E-V_{\scr B}\right| }{\hbar\omega}  \right] .
\label{sigF-low}
\end{equation}
A simpler approximation for the Wong formula can also be derived at energies few MeV above the Coulomb barrier $(E \gtrsim V_{\scr B} + \hbar\omega $). 
In this energy region one can neglect the unity within the square bracket of Eq.~(\ref{sigWong}), and it becomes
\begin{equation}
\sigma_{\scr F}(E) \simeq \sigma_{\rm geo} \ \left[ 1 - \frac{V_{\scr B}}{E}   \right],
\label{sigF-high}
\end{equation}
where
\begin{equation}
\sigma_{\rm geo} =  \pi\ R_{\scr B}^2
\label{siggeo}
\end{equation}
is the geometric cross section. \\

Wong's formula is quite accurate in collisions  of heavy systems at near-barrier energies (say $V_{\scr B} + 5\,{\rm MeV} \ge E \ge V_{\scr B} - 5\, {\rm MeV}$). 
However, it is a poor approximation in collisions of light systems, both below and above the Coulomb barrier. 
The problem at sub-barrier energies is that the barrier for light systems is highly asymmetric, whereas the parabolic approximation is symmetric. The tail
of $V(r)$ falls off slowly, as the Coulomb potential ($\sim 1/r$), while the parabola goes to zero much faster. Thus, the external turning points for the two 
potentials become very different as the energy decreases, with $r_{\rm e}$ for the parabola being progressively smaller. In this way, the function 
$\Phi^{\scr WKB}_l(E)$ for the parabola is too small, and the transmission coefficient of Eq.~(\ref{TK}) is badly overestimated. For 
example, the fusion cross section for the $^6{\rm Li}\,+\,^{12}{\rm C}$ system predicted by the Wong formula at $E = 1$ MeV ($V_{\scr B}\simeq 3$ MeV)
is more than two orders of magnitude larger than the value obtained by a quantum mechanical calculations (see Fig. 26 of Ref.~\cite{CGD06}).
The inaccuracy of the Wong formula above the Coulomb barrier has another origin. Contrary to what was assumed in the derivation of the Wong formula, the 
parameters $R_l$ and $\omega_l$ for light systems have a strong $l$-dependence.\\

\begin{figure}
\begin{center}
\includegraphics*[width=7cm]{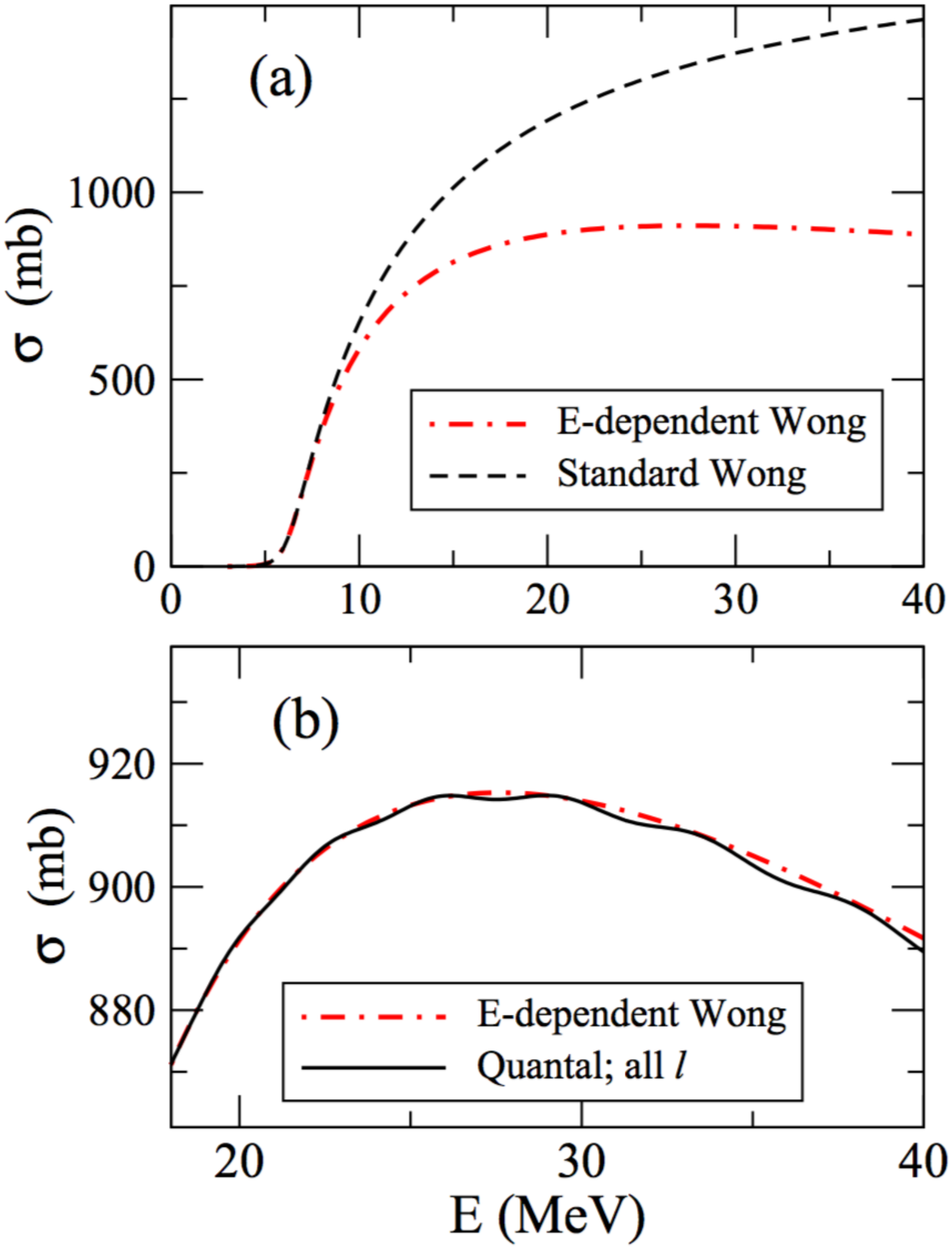}
\end{center}
\caption{(Color on line) Fusion cross sections for the $^{12}$C + $^{12}$C system. Panel (a): Comparison of the energy-dependent Wong formula (red
dot-dashed curve) with the standard Wong formula (dashed curve); Panel (b): Comparison of the energy-dependent Wong formula with the exact cross section,  
calculated by full quantum mechanics (solid line). The figure was taken from Ref.~\cite{RoH15}.}
\label{Wong-WongED}
\end{figure}

Rowley and Hagino~\cite{RoH15} have shown that the accuracy of the Wong formula for light systems at above-barrier energies is significantly improved if 
one replaces the s-wave barrier parameters in Eq.~(\ref{sigWong}) by the barrier parameters associated with the grazing angular momentum, $R_{\scs E}$,  
$\hbar\omega_{\scs E}$ and $V_{\scs E}$. This is illustrated in Fig.~\ref{Wong-WongED}. Panel (a) shows a comparison of the standard Wong cross section 
with the one obtained using the parameters associated with the grazing angular momentum at the corresponding collision energy, $E$. One observes that the 
two cross sections become progressively different as the energy increases. At 40 MeV, the standard Wong cross section is about 3/2 of the one obtained with 
the energy dependent parameters. 

Panel (b) shows a comparison of the energy-dependent Wong formula with exact results of full quantum mechanics. One concludes that the improved Wong 
formula of Ref.~\cite{RoH15} is a very good approximation to the quantum mechanical cross section.

Usually, higher order terms of the Poisson series give negligible contributions to the fusion cross section. The situation is slightly different in the case of 
identical nuclei. These contributions are responsible for the weak oscillations in the exact fusion cross section of the $^{12}$C + $^{12}$C system (solid line 
on panel (b) of Fig.~\ref{Wong-WongED}). Adding these contributions to the energy-dependent Wong formula, Rowley and Hagino were able to reproduce 
the quantum mechanical cross section of Fig.~\ref{Wong-WongED} with great accuracy~\cite{RoH15}. Recently, the study of Rowley and Hagino has been 
extended to collisions of identical nuclei with arbitrary spin~\cite{TCD17}.\\

\centerline {\it The Wong formula at $E > E_{\rm cr}$}

\bigskip

Eq.~(\ref{sigF-high}) indicates that the cross section tends to a constant value as the energy goes to infinity,  This is not consistent with the prediction of quantum
mechanics in potential scattering, where the cross section goes to zero in the $E\rightarrow \infty$ limit. The origin of this discrepancy is that Wong assumes that 
the radii and the shapes of the barriers of the effective potentials are independent of $l$. In this way the partial-wave series always gets important contributions 
from waves around the grazing angular momentum, given by the condition $V_{l_{\rm g}}(R_l) = E$. The situation is quite different for the actual potential. In the WKB
calculation, the fusion probability vanishes above $l_{\rm cr}$, since $V_l(r)$ has no barrier. On the other hand, in the quantum mechanical calculation, the partial-wave
series of Eq.~(\ref{sigma-F}) is limited by the factor $P_{\scr CN}(l,E)$, that vanishes above $l_{\rm cr}$. Owing to the repulsive nature of the potential, the collision
time is not long enough to allow the formation of an equilibrated CN. Thus, both in quantum mechanical calculation and in the WKB approximation with the actual
effective potential, the partial-wave series is truncated at $l = l_{\rm cr}$.  At high enough energies, all non-vanishing fusion probabilities are equal to one and the
partial-wave series can be summed analytically. This establishes a new energy regime, where the fusion cross section decreases monotonically with $E$, according 
to the expression,
\begin{equation}
\sigma_{\scr F} \left( E  \right) \simeq  \sigma_0\times  \frac{E_{\rm cr}}E,
\label{E>Ecr}
\end{equation}
with
\begin{equation}
\sigma_0 = \frac{\pi\hbar^2 \left( l_{\rm cr}+1 \right)^2}{2\mu E_{\scr cr}}.
\label{E>Ecr}
\end{equation}

\bigskip

\centerline {\it Wong formula vs. $\sigma_{\scr R}$}

\medskip

The derivation of the Wong formula is based on the assumption that the absorption probability is equal to the transmission coefficient through the barrier of  the real 
potential of Eq.~(\ref{Vl(r)}). This assumption is consistent with the IWBC or  calculations with a very strong imaginary potential acting exclusively in the inner region 
of the barrier. Thus, Wong's formula is an approximation for the fusion cross section. It is expected to be a poor approximation for the total reaction cross section, since it
may get important contributions from direct reactions, which correspond to absorption in the barrier region and beyond. Nevertheless, in Wong's original paper~\cite{Won73}
and in other publications it has been taken as an approximation for $\sigma_{\scr R}$. In such cases, $R_{\scr B}$, $V_{\scr B}$ and $\hbar\omega$ should be interpreted 
as effective quantities, rather than the parameters extracted from the parabolic fit of the real potential.\\


\subsection{The semiclassical scattering amplitude}


The partial-wave expansion of the nuclear part of the scattering amplitude is given by~\cite{CaH13}
\begin{equation}
f_{\scr N}(\theta) = \frac{1}{2ik}\, \sum_{l=0}^\infty (2l+1)\, P_l(\cos\theta)\ e^{2i\sigma_l} 
\Big[ \left| S_{\scr N}(l,E) \right| \,e^{2i\delta_l} -1\Big],
\label{fn-theta}
\end{equation}
where $\sigma_l$ and $\delta_l$ are respectively the Coulomb and the nuclear phase-shifts at the $l^{\rm th}$ partial-wave, $ \left| S_{\scr N}(l,E) \right|$ is the
modulus of the nuclear S-matrix and $P_l(\cos\theta)$ is the Legendre polynomial.
The semiclassical scattering amplitude is obtained through the following approximations.
\begin{enumerate}
\item use the Poisson series to evaluate the partial-wave sum; 

\item evaluate nuclear phase-shifts within the WKB approximation;

\item use Legendre polynomials of continuous order ($\lambda$) and adopt the large $\lambda$ approximation:
\begin{multline}
P_l(\cos \theta)\ \ \longrightarrow\ \  P(\lambda, \cos{\theta})\simeq 
\left(\frac{1}{2\pi\lambda\sin{\theta}}\right)^{1/2}\\
\left[ e^{i\left( \lambda\theta - \pi/4\right)} + e^{-i\left(\lambda\theta -\pi/4\right)}\right] ;
\label{LP-largel-2}
\end{multline}

\item evaluate integrals using the stationary phase approximation.

\end{enumerate}

With the above approximations, one can infer several characteristic features of heavy-ion scattering. The two terms within square
brackets in 
Eq.~(\ref{LP-largel-2}) give rise to a {\it near} and a {\it far} component of the scattering amplitude and to near-far and
rainbow oscillations of the cross section in heavy-ion elastic scattering.\\

\subsubsection{The Fresnel diffraction formula and the quarter point recipe} 

Heavy-ion scattering is dominated by Coulomb repulsion and strong absorption. This leads to the sharp cut-off model 
(black disk approximation) for charged particle scattering at low energies. In this model, nuclear phase shifts are neglected
and the $l$-projected components of the S-matrix are given by,
\begin{equation}
S(l) \equiv \left| S_{\scr N}(l,E)\right|\, e^{2i(\sigma_l + \delta_l)} \rightarrow S(\lambda) \simeq {\mathcal H}
\left( \lambda - \bar{\Lambda} \right)\ e^{2i\sigma(\lambda)} .   
\label{S-sc}
\end{equation}
Above, $\bar{\Lambda}$\, is the grazing angular momentum and 
\begin{eqnarray}
{\mathcal H}\left( \lambda - \bar{\Lambda} \right) & = & 1,\ {\rm for }\ \lambda \ge \bar{\Lambda} \\
                                                                              & = & 0,\ {\rm for }\ \lambda <  \bar{\Lambda}
\end{eqnarray}
is the Heaviside step function.\\

Within the sharp cut-off model of the semiclassical scattering amplitude, the elastic cross section is dominated by the interference between 
a refractive Coulomb wave at $\lambda > \bar{\Lambda}$ and a diffractive wave for smaller values of $\lambda$.  Frahn~\cite{Fra66,Fra85}  
demonstrated that the ratio of the elastic scattering cross section with respect to the corresponding Rutherford cross section, which will be denoted 
by $\sigma(\theta) / \sigma_{\scr C}(\theta)$, is given by the Fresnel diffraction formula
\begin{equation}
\frac{\sigma(\theta)}{\sigma_{\scr C}(\theta)} = \frac{1}{2}\left[\left(\frac{1}{2} - C({\scriptstyle W})\right)^2 + \left(\frac{1}{2} - S({\scriptstyle W})\right)^2\right] ,
\label{sig-sigR}
\end{equation}
where $C({\scriptstyle W})$ and $S({\scriptstyle W})$ are the Fresnel integrals given by, 
\begin{equation}
C({\scriptstyle W}) = \int_0^{\scriptstyle W} d\omega \cos{\left(\frac{\pi}{2}\omega^2\right)}
\label{fres-cos}
\end{equation}
and
\begin{equation}
S({\scriptstyle W}) = \int_0^{\scriptstyle W} d\omega \sin{\left(\frac{\pi}{2}\omega^2\right)} .
\label{fres-sin}
\end{equation}
The argument of the Fresnel integrals is
\begin{equation}
{\scriptstyle W} = \sqrt{\frac{2\eta}{\pi}}\ \left[\frac{\sin\left(\frac{\theta-\bar{\Theta}}{2}\right)}{\sin{\frac{\bar{\Theta}}{2}}}\right],
\label{W-arg}
\end{equation}
where $\eta$ is the Sommerfeld parameter, $\theta$ is the scattering angle and $\bar{\Theta}$ is the grazing angle.\\

An important property of the Fresnel integrals is that they vanish at ${\scriptstyle W}  = 0$ (this can be immediately checked in
Eqs.~(\ref{fres-cos}) and (\ref{fres-sin})). This value of ${\scriptstyle W}$ is reached at the scattering angle  $\theta = \bar{\Theta}$. 
Using these results in Eq.~(\ref{sig-sigR}), the ratio of the cross sections at the grazing angle becomes
\begin{equation}
\frac{\sigma(\bar{\Theta})}{\sigma_{\scr C}(\bar{\Theta})} = \frac{1}{4}.
\label{sig-theta-1/4}
\end{equation}
To emphasize this property, the grazing angle is denoted \\
\[
\bar{\Theta} = \theta_{\scr 1/4}.
\]

\bigskip

\begin{figure}
\begin{center}
\includegraphics*[width=8cm]{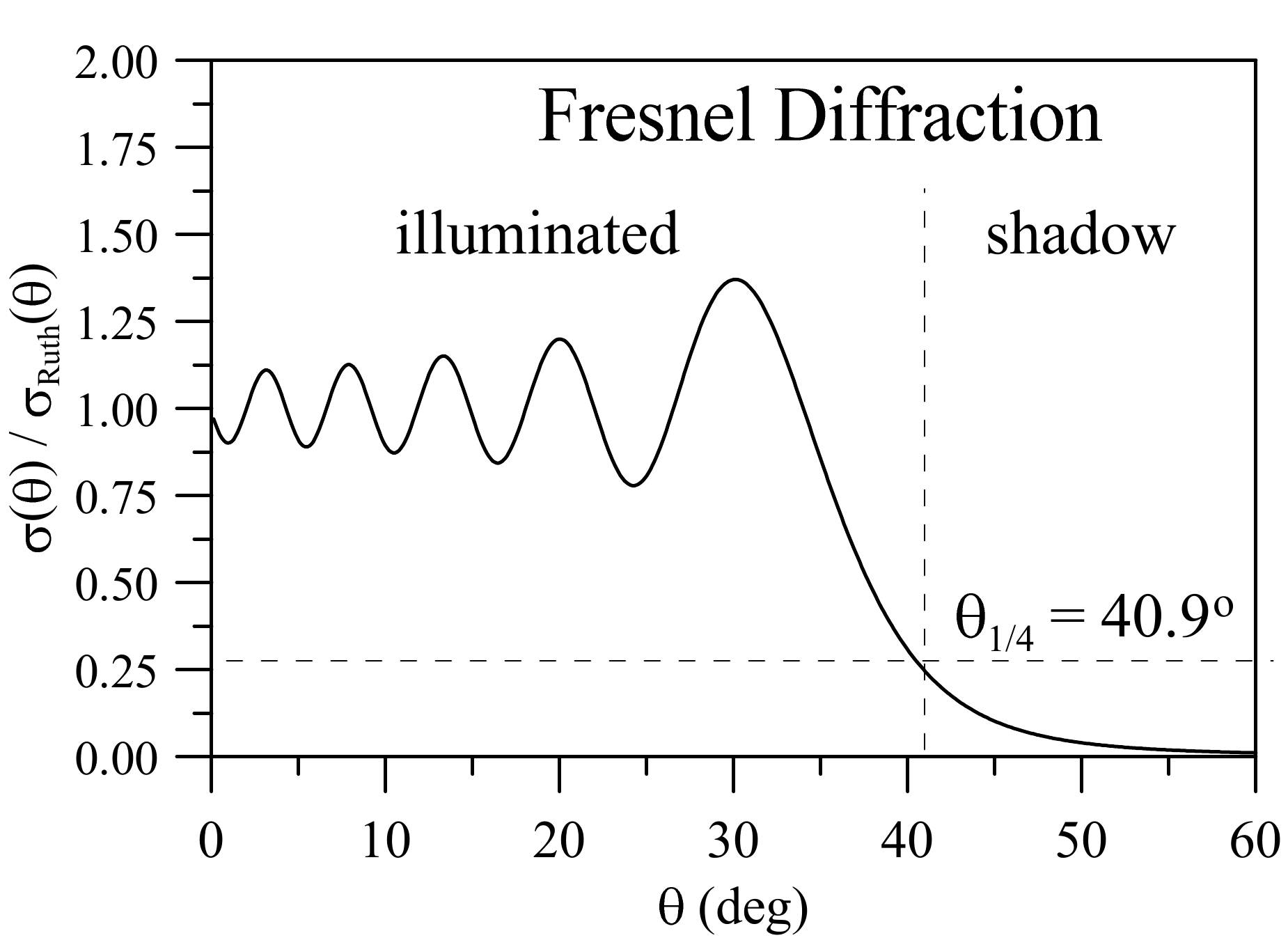}
\end{center}
\caption{Fresnel diffraction within the sharp cut-off approximation. Figure from Ref.~\cite{CaH13}.}
\label{Fres-dif}
\end{figure}
An important consequence of Eq.~(\ref{sig-theta-1/4}) is that the grazing angle can be determined directly from the scattering data:  $\bar{\Theta}$ is the angle where 
$\sigma(\theta)/\sigma{\scr C}(\theta)$ reaches the value 1/4. Having $\bar{\Theta}$, one can determine the argument ${\scr W}$ for each value of $\theta$, and 
evaluate the Fresnel integrals. Then, inserting them into Eq.~(\ref{sig-sigR}), one obtains $\sigma(\theta) / \sigma_{\scr C}(\theta)$, within the sharp cut-off 
approximation for the S-matrix. This result is known as Frahn's Fresnel diffraction formula for the angular distribution.\\

 An illustration of this procedure is given in Fig. \ref{Fres-dif}, for the $^{16}$O + $^{208}$Pb collision.
This example was discussed in detail by Frahn~\cite{Fra66}. In this case, the grazing angle is $\theta_{\scr 1/4} = 40.9^{\rm o}$, 
${\bar \Lambda}=86$ and the Sommerfeld parameter is $\eta = 32.05$.
Frahn~\cite{Fra66} has shown that the predictions of the sharp cut-off model are in qualitative agreement with the data. It predicts oscillations at low 
scattering angles (illuminated region) with increasing amplitudes, which ends in a pronounced maximum, followed by a rapid decrease of
$\sigma(\theta) / \sigma_{\scr C}(\theta)$ as $\theta$ increases (shadow region).  However, the Fresnel diffraction formula completely ignores
the effects of nuclear refraction. These effects are responsible for other kind of oscillations, like near-far interference and the rainbow. One
should stress that rainbow scattering with a unitary S-matrix leads also to the typical pattern of heavy-ion scattering, exhibited in 
Fig.~\ref{Fres-dif}. Quantum mechanical potential scattering calculations with complex potentials, which include both refractive and 
diffractive effects of the nuclear potential, lead to more quantitative prediction of the elastic angular distributions.\\

The sharp cut-off model can also be used to make qualitative prediction of total reaction cross sections. From the experimental 
$\sigma(\theta)/\sigma_{\scr C}(\theta)$ ratio at a given collision energy, $E$, one determines the grazing angle. Then, one finds the
grazing angular momentum from the Rutherford trajectory by the equation,
\begin{equation}
\bar{\Lambda}(E) = \eta\  \cot\left( \frac{\theta_{\scr 1/4}}{2} \right) .
\label{def1}
\end{equation}
Next, we evaluate the total reaction cross section taking Eq.~(\ref{sigma-abs}) and replacing the partial-wave sum by an integral
over $\lambda$. This corresponds to taking only the $m=0$ term in the Poisson series (see e.g. Ref.~\cite{CaH13}). One gets
\begin{equation}
\sigma_{\scr R}(E) = \frac{2\pi}{k^2} \int d\lambda \, \lambda \ P(\lambda,E) .
\label{sigma-TR}
\end{equation}
Using the sharp cut-off model, the absorption (reaction) probability of Eq.~(\ref{modS}) becomes
\begin{equation}
P_{\rm abs}(\lambda,E) = 1 - {\mathcal H}(\lambda - \bar{\Lambda}{\scr (E)}) = {\mathcal H}(\bar{\Lambda}{\scr (E)} - \lambda).
\label{P_cut-off}
\end{equation}
Then, the integral of Eq.~(\ref{sigma-TR}) can be immediately evaluated and one gets
\begin{equation}
\sigma_{\scr R} = \frac{\pi}{k^2}\ \bar{\Lambda}^2(E) .
\label{sigR}
\end{equation}

\bigskip

It is worth mentioning that at higher energies, the Coulomb effect becomes small and the angular distribution corresponding to the black disk or sharp cutoff model approximates the Fraunhoffer diffraction. The grazing angular momentum can still be obtained from the angular period of oscillations, $\Delta\theta = \pi/\bar{\Lambda}$.\\
%

\subsection{The Generalized Optical Theorem and the Sum of Differences Method}\label{SOD}


The Optical Theorem is an important result in scattering theory and it expresses unitarity in a useful mathematical form. For uncharged particle scattering, the theorem 
states that the total (angle integrated) elastic scattering cross section is proportional to the imaginary part of the elastic amplitude evaluated at $\theta = 0$. In potential
scattering from a real potential,  one has
\begin{equation}
\int \ \frac{d\sigma_{\rm el}(\theta)}{d\Omega} \ d\Omega= \frac{4 \pi}{k}\ {\rm Im} \left\{ f_{\rm el}(\theta = 0) \right\}.
\label{OT1}
\end{equation}
When absorption is present, the above becomes the Generalized Optical Theorem (GOT), 
\begin{equation}
\int \ \frac{d\sigma_{\rm el}(\theta)}{d\Omega}\ d\Omega + \sigma_{\scr R} = \frac{4 \pi}{k} {\rm Im}\left\{ f_{\rm el}(\theta = 0)\right\}.
\label{OT2}
\end{equation}

\bigskip

For charged particles, the GOT needs to be modified to cope with the point Coulomb singularity. The integral in Eqs.~(\ref{OT1}) and (\ref{OT2}) is divergent. 
First, the scattering amplitude is written as
\begin{equation}
f_{\rm el}(\theta) =  f_{\scr C}(\theta)\,+\, f_{\scr N}(\theta),
\label{fc-fn}
\end{equation}
where $f_{\scr C}(\theta)$ is the Coulomb scattering amplitude for two point charges, and $f_{\scr N}(\theta)$ is a correction arising from the short-range 
nuclear potential. The former is given by the analytical expression,
\begin{eqnarray}
f_{\scr C}(\theta) & = & \frac{\eta}{2k \sin^2{(\theta/2})}\  e^{i\, \left[ 2\sigma_0 + \pi - 2\eta\, \ln\left( \sin \theta/2 \right) \right] } \nonumber\\
                           & = & -\frac{\eta}{k}\ 2^{i\eta}\ \frac{e^{2i\sigma_0}}{(1 - \cos{\theta})^{i\eta + 1}} \ ,
\label{fc1}
\end{eqnarray}
and the latter is given by the partial-wave expansion\footnote{To simplify the notation, we omit the energy dependence of the S-matrix.},
\begin{equation}
f_{\scr N}(\theta)= \frac{1}{2ik}\ \sum_{l = 0}^{\infty} (2l + 1)\ e^{2i\sigma_{l}}\ \big[S_{\scr N}(l) - 1 \big]\  P_{l}(\cos{\theta}) .
\label{fn1}
\end{equation}

\medskip

To get rid of the singularity, the GOT for charged particles is expressed in terms of the difference
\begin{equation}
\sigma_{\scr SOD}\left( \theta_0 \right) = \ 2\pi\ \int_{\theta_0}^\pi\ \left[ \frac{d\sigma_{\scr Ruth}(\theta)}{d\Omega} \ 
- \  \frac{d\sigma_{\scr el}(\theta)}{d\Omega} \right] \, \sin\theta\, d\theta,
\label{Dif1}
\end{equation}
where $\theta_0$ is a very small angle. The cross section difference, $\sigma_{\scr SOD}\left( \theta_0 \right)$, is called the sum-of-difference (SOD) 
cross section. The replacement $\theta = 0 \rightarrow \theta = \theta_0$ in the lower limit of the integration is justified by the fact
that, in a realistic collision, the Coulomb potential is screened. \\

Using the explicit forms of the elastic and the Coulomb cross sections, namely
\begin{equation}
\frac{d\sigma_{\rm el}(\theta)} {d\Omega} = \big| f_{\scr C}(\theta)\,+\, f_{\scr N}(\theta) \big|^2;\ \ \ 
\frac{d\sigma_{\scr C}(\theta)} {d\Omega} = \big| f_{\scr C}(\theta) \big|^2,
\label{Dif2}
\end{equation}
Eq.~(\ref{Dif1}) becomes
\begin{multline}
\sigma_{\scr SOD}\left( \theta_0 \right) = -\ 2 \pi\,\int_{\theta_0}^\pi \left|f_{\scr N}(\theta) \right|^2\,\sin\theta\ d\theta\\
-\,4 \pi\,\int_{\theta_0}^\pi {\rm Re} \left\{f_{\scr C}^*(\theta)\cdot f_{\scr N}(\theta) \right\} \,\sin\theta\ d\theta.
\label{Dif3}
\end{multline}
The above cross section has been discussed by several author~\cite{HoT65,HoT65a,Mar83,UPH99,BaA85a,BaA85b,OTV91}. Marty evaluated the integrals of 
Eq.~(\ref{Dif3}) and obtained the SOD cross section,
\begin{multline}
 \sigma_{\scr SOD}(\theta_0)\, =\, \sigma_{\scr R} - \frac{4\pi}{k}\ \left| f_{\scr N}(0)\right| \, \\
\times \sin{ \Big[\arg \left\{ f_{\scr N}(0) \right\}\, -\, 2\sigma_0 +   \eta \ln\left\{\sin^2{\theta_{0}/2}\right\}\Big]} \ +\ O\left(\theta_{\scr 0}^{\scr 2}\right).
\label{SOD1}
\end{multline}
Above, $\eta$ is the Sommerfeld parameter, $\sigma_0$ is the s-wave Coulomb phase shift, and $O\left(\theta_0^2\right)$ is a correction\footnote{For a detailed discussion
of this correction, we refer to Ref.~\cite{BaA85b}.}, that becomes
negligible when $\theta_0$ is a very forward angle.\\

\begin{figure}
\begin{center}
\includegraphics*[width=9cm]{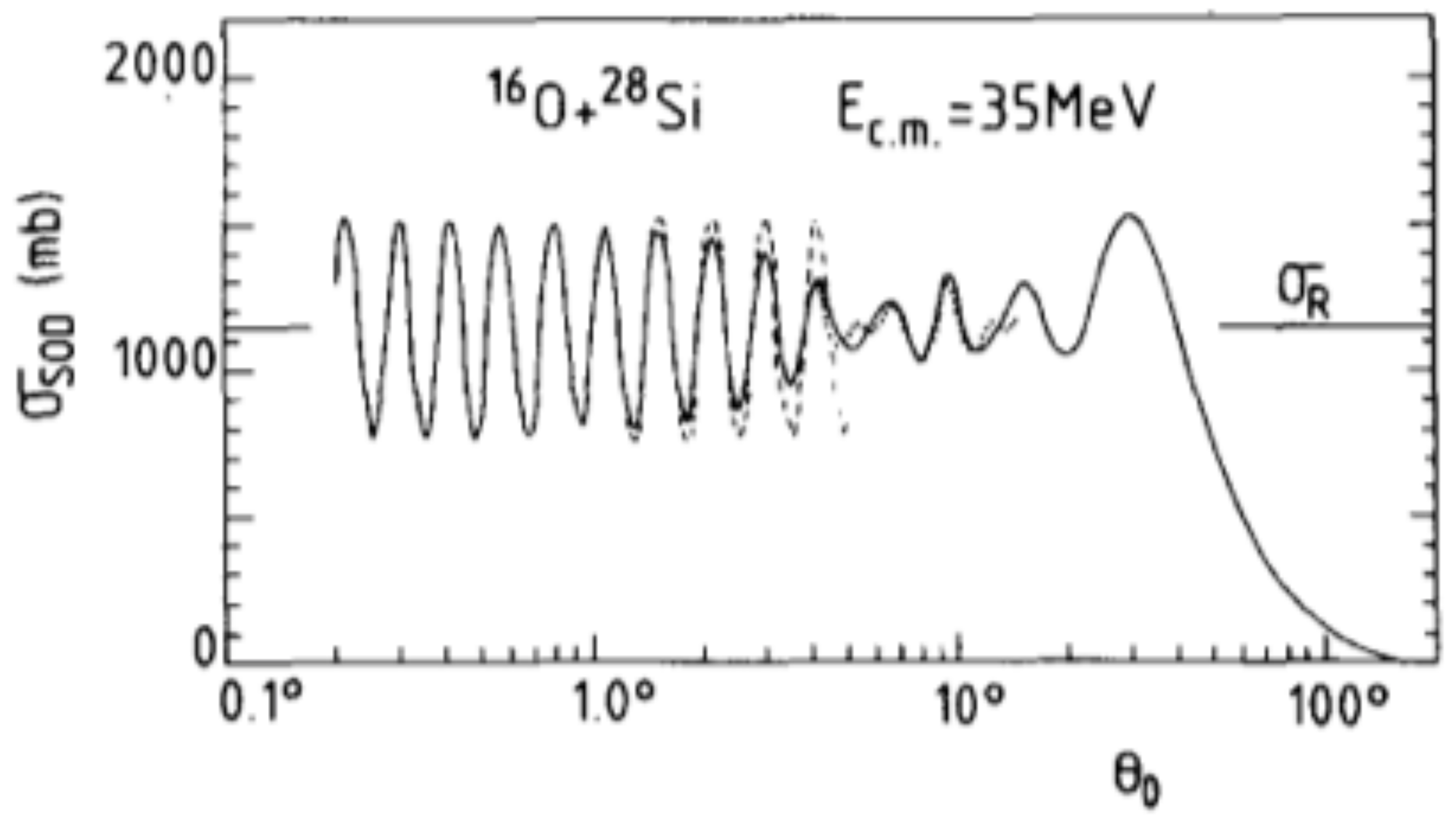}
\end{center}
\caption{The SOD cross section for the $^{16}{\rm O}\,+\,^{28}{\rm Si}$ system at $E_{\rm c.m.} = 35$ MeV. The solid line and the dashed lines correspond
respectively to the exact result of Eq.~(\ref{Dif1}), and the approximation of Eq.~(\ref{SOD1}). The calculations adopted the optical potential 
of Shkolnik {\it et al.}~\cite{SDK78} (figure taken from Ref.~\cite{BaA85a}).}
\label{approx-SOD}
\end{figure}
 The accuracy of Eq.~(\ref{SOD1}) is illustrated in Fig.~\ref{approx-SOD}, where the approximate cross section of  Eq.~(\ref{SOD1}) and the exact result of
 Eq.~(\ref{Dif1}) are compared. In this example, taken from Ref.~\cite{BaA85a}, $\sigma_{\scr el}(\theta)$ and 
 $f_{\scr N}(0)$ were calculated with the optical potential 
 of Ref.~\cite{SDK78}. Clearly, the approximation is quite accurate for small values of $\theta_0$. The oscillatory behaviour of $\sigma_{\scr SOD}(\theta_0)$ at 
 forward angles provide two important pieces of information.  First, the total reaction cross section is given by the SOD cross section averaged over a period of 
 oscillation in the low $\theta_0$ region. Thus, the total reaction cross section can be determined from accurate measurements of the elastic cross section at 
forward angles. The use of this technique is discussed in sect. \ref{SOD tech}.\\
 
 The second important consequence of Eq.~(\ref{SOD1}) is that the modulus of $ f_{\scr N}(0)$ can be extracted from the period of oscillation. This equation
has been fully analysed in the context of heavy-ions collisions~\cite{BaA85a,BaA85b,OTV92,UPH98,UPH99}. In most of the applications of
Eq.(\ref{SOD1}), the aim was to study the oscillations in the nuclear amplitude at forward angles, which can be traced to forward glory effects. The existence of a 
non-zero value of the impact parameter, $b_{\rm gl}$, at which the classical deflection angle defined through the relation, 
\[
\Theta(l) = 2\, \frac{\delta(\sigma_l + \delta_{l})}{\delta l}.
\]
satisfies the condition $\Theta(b_{\rm gl}) = 0$, corresponds to forward glory. Under this condition, the scattering amplitude at forward angles is enhanced and with 
conspicuous oscillations. In fact, this nuclear amplitude, $f_{\scr N}(\theta)$, in the vicinity of the forward glory angle ($\theta = 0^{o}$) is given by the product 
$A(\theta)J_{0}(\lambda_{\rm gl}\theta)$, where $J_{0}$ is the Bessel function of order zero and $A(\theta)$ is a smooth function of 
$\theta$. Thus the SOD method, in the presence of forward glory, would supply a 
reaction cross section with greatly enhanced energy oscillations~\cite{BaA85a, BaA85b, UPH99}. 
Besides this, the forward glory effect can be used to learn more about the nuclear interaction at the surface region, complementary to the information obtained from the study of nuclear rainbow scattering, as emphasised in Ref.~\cite{UPH99}. 


\subsection{Comparison of Quantum mechanical cross sections with the Wong and the quarter-point approximations}


In potential scattering, fusion and direct reactions are taken into account through the inclusion of an imaginary part in the nucleus-nucleus potential. This procedure
leads to reasonable predictions for $\sigma_{\scr F}$ and $\sigma_{\scr R}$. These cross sections are calculated by Eq.~(\ref{sigma-abs}), with absorption 
probabilities expressed in terms of the unitarity defect of the S-matrix (Eq.~(\ref{modS})), or with the radial integrals involving the imaginary potential (Eq.~(\ref{PW-l})). 
However, calculations of fusion and of total reaction cross sections must use different imaginary potentials, as discussed in section \ref{PotScat}. 
In the previous sections we discussed also the Wong formula and the quarter-point recipe, where the cross sections are approximated by simple 
analytical expressions. Now we discuss the validity of these approximations.\\

\begin{figure}
\begin{center}
\includegraphics*[width=8cm]{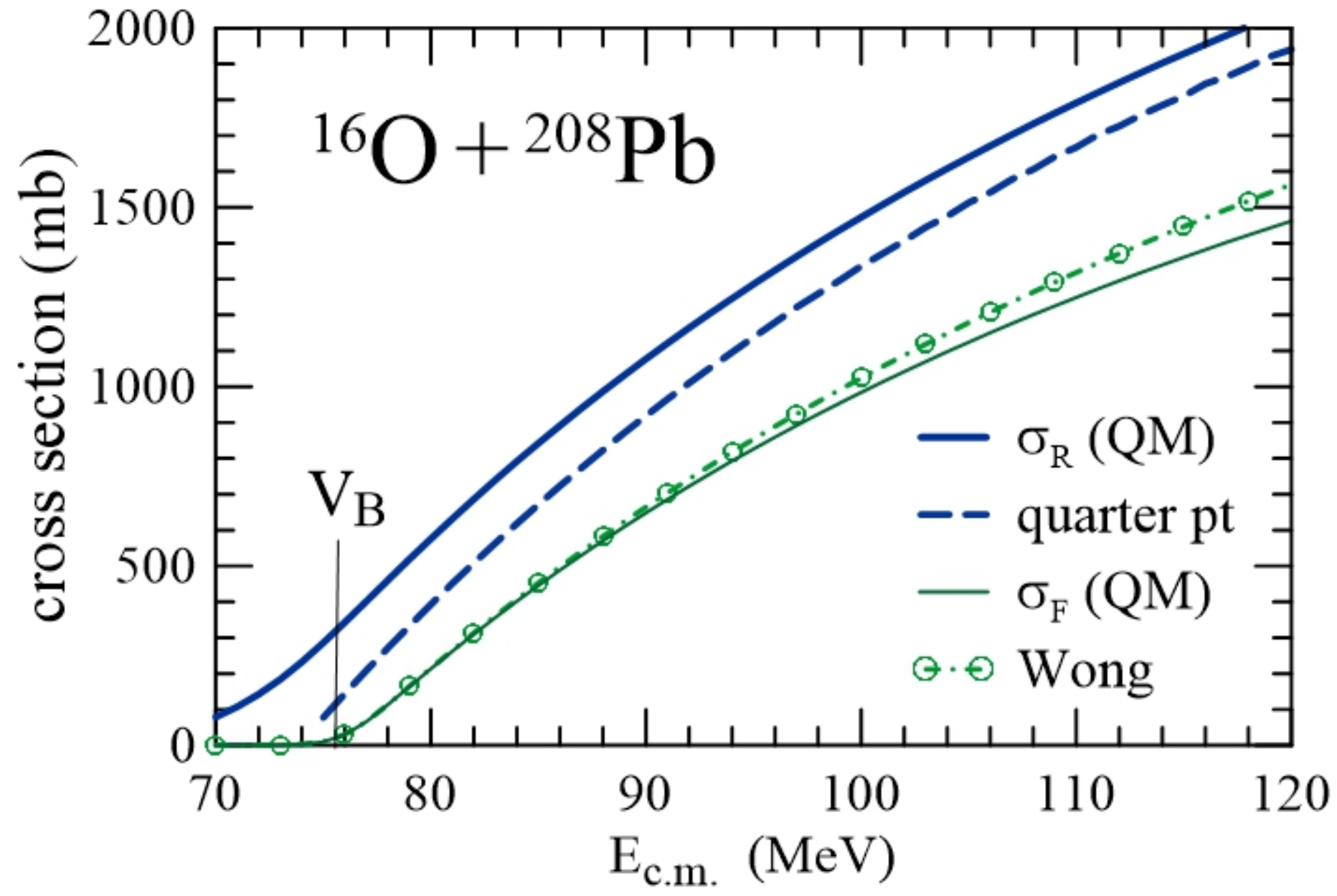}
\end{center}
\caption{(Color on line) Quantum mechanical total reaction (thick blue solid line) and fusion (thin solid green line) cross sections, in comparison with
the Wong cross section of Eq.~(\ref{Wong-approx}) (green dot-dashed line with open circles) and the quarter point recipe cross section of Eq.~(\ref{sigR}) (blue dashed line).}
\label{F4}
\end{figure}
As an example, let us consider the potential model for the $^{16}{\rm O}\, -\,^{208}{\rm Pb}$  collision, adopting the S\~ao Paulo Potential~\cite{CPH97,CCG02} for the real part 
of the nucleus-nucleus potential. 
In calculations of $\sigma_{\scr F}$, the imaginary potential is given by a short range WS function with the parameters: $W_0 = 50$ MeV, $r_{0i} = 1.0$ fm 
and $a_i = 0.2$ fm. In calculations of $\sigma_{\scr R}$, the imaginary part of the potential is proportional to its real part, $W(r) = 0.78\ V(r)$. 
This prescription has been successfully used to describe the average behaviour of total reaction cross sections of many systems ~\cite{GCG06}. In Fig~\ref{F4}, the two
quantum mechanical cross sections are shown in comparison with the ones obtained using the Wong formula and the quarter point recipe.  
The quarter point cross section (blue dashed line) was obtained by Eqs.~(\ref{sigR}) and (\ref{def1}), using in the latter the quarter point angle extracted from angular 
distributions of the quantum mechanical calculations, using $W^{\scr R}(r)$. 
Therefore, it is should be compared to $\sigma_{\scr R}$. We find that it underestimates the quantum mechanical cross section systematically. The difference between
the two cross section is roughly constant, except at energies just above the barrier, where this approximation cannot be applied. In this region, $\theta_{\scr 1/4}$ is not
defined, since the ratio between the elastic cross section and its Coulomb counterpart is above 1/4 for any scattering angle. 

\medskip

Now let us consider the Wong cross section. We note that it is very close to the fusion cross section, except at the higher energies ($E > 100$ MeV). Although this 
approximation is known to become progressively poorer as the energy falls well below the Coulomb barrier~\cite{CGD06}, this shortcoming cannot be observed in 
Fig.~\ref{F4}. However, it would be clear in a logarithmic scale plot. We should remark that it is not a surprise that the Wong cross section falls well below $\sigma_{\scr R}$.
In the derivation of his formula, Wong approximates the absorption probability at each partial-wave by the transmission coefficient through the barrier of the 
$l$-dependent effective potential. This approximation implies that there is total absorption of the current that reaches the inner region of the barrier, but no absorption 
in the barrier region. This procedure is justified in the case of fusion absorption but it does not account for absorption arising from direct reactions, which gives an
important contribution to $\sigma_{\scr R}$.\\

\begin{figure}
\begin{center}
\includegraphics*[width=8cm]{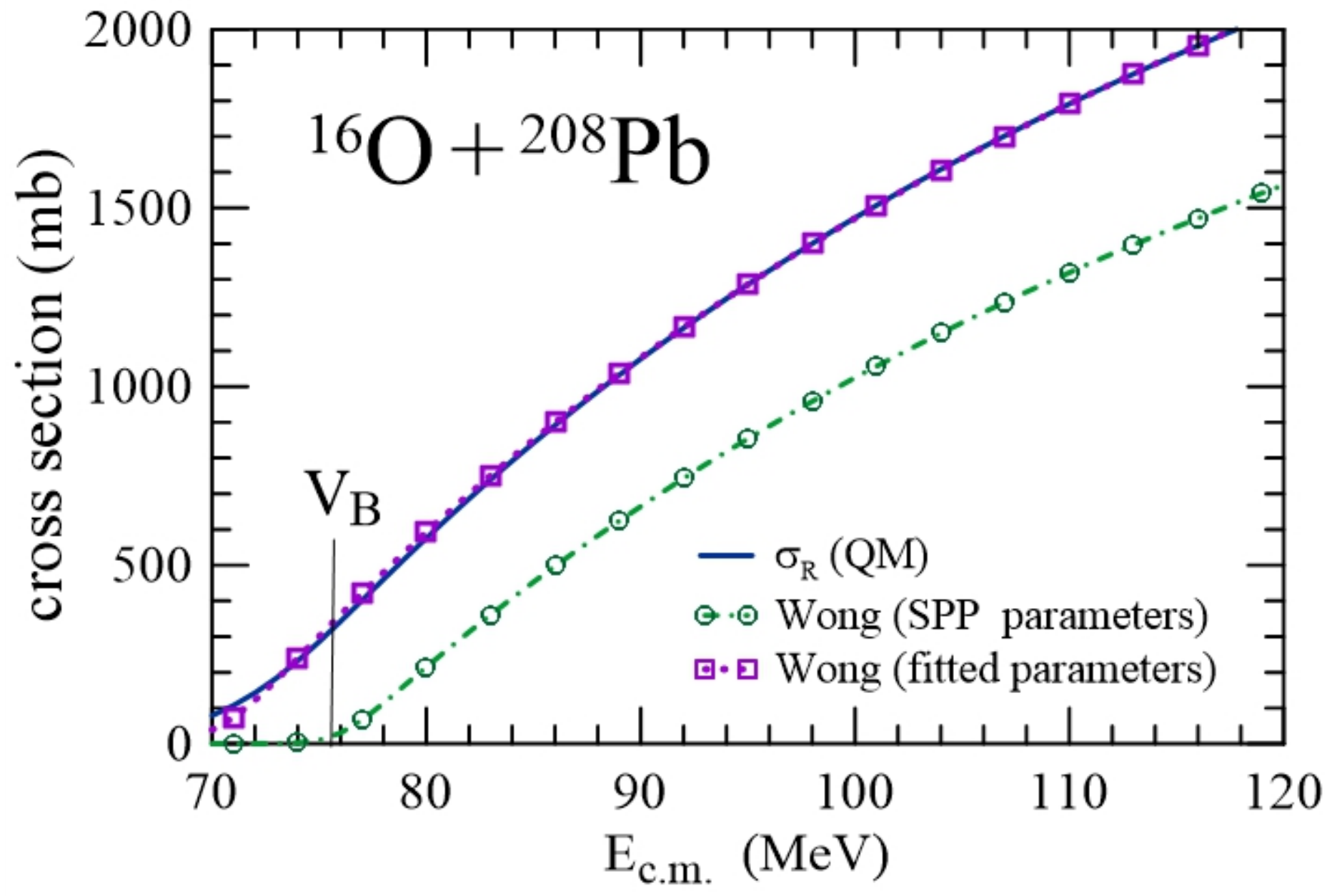}
\end{center}
\caption{(Color on line) The quantum mechanical total reaction cross section (thick blue solid line) and Wong's cross sections obtained with the barrier parameters 
of the s\~ao Paulo potential, and with barrier parameters fitted as to reproduce the quantum mechanical cross section. The notation is indicated in the legend of the figure.}
\label{F5}
\end{figure}

To finish this sub-section, we point out that, in the original paper of Wong, his formula was presented as an approximation for the total reaction cross section. This may
be reasonable if the reaction cross section is dominated by the fusion process. However, contributions of direct reactions can hardly be neglected. This becomes clear
when one tries to compare reduced reaction cross section for different systems. Although the reduction method of Ref.~\cite{CGL09a,CGL09b}, which is based
on the Wong formula works fine for fusion data, it fails when applied to total reaction data~\cite{CMG15}. Nevertheless, the Wong formula can be a very good parameterization
of the total reaction cross section, if $V_{\scr B}, R_{\scr B}$ and $\hbar\omega$ are treated as adjustable parameters, fitted to the total reaction data.
This is illustrated in Fig.~\ref{F5}, where the quantum mechanical cross section, $\sigma_{\scr R}$ (blue solid line), is compared with results of the 
Wong formula (green dot-dashed line with open circles) with the barrier parameters of the S\~ao Paulo potential, $R_{\scr B} = 11.7$ fm, $V_{\scr B} = 76.0$ MeV 
and $\hbar\omega = 4.8$ MeV, and with the Wong formula with barrier parameters fitted to reproduce the quantum mechanical cross section (purple dotted line and
open squares). The total reaction cross section given by the Wong formula with the parameters of the S\~ao Paulo potential falls well below its quantum mechanical
counterpart. On the other hand, the cross section given by the Wong formula with the fitted parameters can hardly be distinguished from the quantum mechanical result. 
However, we stress that using barrier parameters which do not correspond to the actual potential has no physical meaning. In this case, the Wong formula is just a 
smart parameterization for the total reaction cross section.

\section{Many-body scattering theory}


Potential scattering is a very limited theory for heavy-ion collisions. In typical situations, the dynamics is strongly influenced by the nuclear structure
of the collision partners, and then the coupled channel (CC) theory is a more suitable approach. In this treatment, the intrinsic degrees of freedom 
of the projectile and/or the target, denoted by $\xi$, are explicitly taken into account. The scattering wave function, ${\rm \Psi}^{\scr (+)}({\bf R},\xi)$, where ${\bf R}$ 
in the vector joining the centers of the collision partners{\footnote{From now on, the collision collision vector will be denoted by ${\bf R}$, instead of
${\bf r}$. This change will prove convenient when we discuss collisions of a two-cluster projectile.}, is the solution of the Schr\"odinger 
equation\footnote{Blackboard bold fonts are used to indicate operators acting on both the ${\bf R}$ and the $\xi$ spaces.},
\begin{equation}
\big[ E - \mathbb{H} \big] \, {\rm \Psi}^{\scr (+)}({\bf R},\xi)= 0,
\label{CC-Sch}
\end{equation}
with scattering boundary conditions. Above,  
\begin{equation}
 \mathbb{H} = h(\xi) +\hat{K}+ \mathbb{U}({\bf R},\xi)
 \label{total H}
\end{equation}
is the total Hamiltonian of the system, $\hat{K}$ is the kinetic energy operator associated with the projectile-target relative motion, $h(\xi)$ is the
intrinsic Hamiltonian, and $\mathbb{U}({\bf R},\xi)$ is the complex coupling interaction. In the CC method, the scattering wave function is expanded on a set of 
eigenstates of the intrinsic Hamiltonian (channels), $\varphi_\alpha(\xi)$, given by the eigenvalue equation,
\begin{equation} 
h\, \varphi_\alpha(\xi) = \varepsilon_\alpha\  \varphi_\alpha(\xi) ,
 \label{B-int}
\end{equation}
where $\alpha$ stands for the set of quantum numbers required to specify the intrinsic state (usually the energy and the appropriate angular momentum
quantum numbers, $l s j$ and $m$).  That is
\begin{equation}
{\rm \Psi}^{\scr (+)}({\bf R},\xi)= \sum_{\alpha=0}^{N} \varphi_\alpha(\xi)\,\psi_\alpha^{\scr (+)}({\bf R}).
\label{CC-exp}
\end{equation}
Although the channel expansion involves an infinite number of intrinsic states, there is a finite number of nonelastic states, denoted by $N$, that are relevant to the reaction dynamics. Thus, the series is truncated after $N+1$ terms (the elastic, labelled by $\alpha=0$, and $N$ nonelastic channels). 


\subsection{The CC equations}\label{CC eq}


Inserting the channel expansion into Eq.~(\ref{CC-Sch}), taking scalar products with each of the intrinsic states, and using their orthonormality properties, one gets the set of  
CC equations 
\begin{equation}
\big[E - H_{\alpha}({\bf R})  \big] \,\psi_\alpha^{\scr (+)}({\bf R}) = \sum_{\alpha^\prime} H_{\alpha,\alpha^\prime}({\bf R})\,\psi_{\alpha^\prime}^{\scr (+)}({\bf R}),
\label{CC-eq1}
\end{equation}
where $\alpha$ and $\alpha^\prime$ run from 0 to $N$. Above, $H_{\alpha,\alpha^\prime}({\bf R})$ is the matrix-element of the Hamiltonian in the basis of intrinsic states,
\begin{equation}
H_{\alpha,\alpha^\prime}({\bf R}) = \int d\xi \ \varphi^*_\alpha(\xi)\ \mathbb{H}({\bf R},\xi)\  \varphi_{\alpha^\prime}(\xi) .
\label{Hmat-el}
\end{equation}
Note that we used the short-hand notation for the diagonal matrix-elements of $\mathbb{H}$:\, $H_{\alpha,\alpha^\prime}({\bf R}) \equiv H_{\alpha}({\bf R})$.
The expansion of Eq.~(\ref{CC-exp}) is restricted to excited or transfer states with simple structure, reached through a small number of steps. They correspond
to the so called direct reactions. On the other hand, equilibrated CN states are too complicated to be included in the expansion. Nevertheless, they cannot
be totally neglected. This situation can be remedied by the generalized optical potential,
\begin{equation}
\mathbb{U}({\bf R},\xi) = \mathbb{V}({\bf R},\xi) +\,i\,\mathbb{W} ({\bf R},\xi).
\label{OPT-gen}
\end{equation}
The imaginary part of this potential accounts for the loss of flux going to CN formation.
Alternatively, the effects of the CN can be simulated by keeping the potential real but assuming ingoing IWBC for all radial wave 
functions at some radial distance inside the potential barrier.\\

If all relevant direct channels are included in the CC expansion, absorption is associated exclusively with the fusion process. On the other hand, the total reaction cross 
section results from absorption by the imaginary potential, and also from the population of the direct reaction channels. Thus, it can be expressed 
by the deviation of the modulus of the 
elastic S-matrix from unity. In the simpler situation of spin zero, where the angular momentum projected components
of the S-matrix are denoted by $S_{\alpha, l}(E)$, the total reaction cross section is given by 
\begin{equation}
\sigma_{\scr R}(E) = \frac{\pi}{k^2}\  \sum_{l = 0}^{\infty}\ (2l + 1)\ P^{\scr R}_{l}(E) ,
\label{sigma-abs-CC}
\end{equation}
with the reaction probability at the $l^{\rm th}$ partial-wave given by
\begin{equation}
P^{\scr R}_{l}(E)= 1 - \left| S_{0, l}(E) \right|^2,
\label{modS-CC}
\end{equation}
where $S_{0, l}(E)$ is the elastic $S$-matrix at the $l^{\rm th}$ partial wave.

\bigskip

The fusion cross section is then given by the difference between the reaction cross section and the summed cross sections for the direct channels involved in the
 CC calculation, $\sigma_\alpha (E)$. That is,
\begin{equation}
\sigma_{\scr F}(E) = \sigma_{\scr R}(E) - \sum_{\alpha =1}^N \ \sigma_\alpha(E).
\label{sigma-abs-CC1}
\end{equation}

\medskip

Since fusion corresponds to absorption by the short-range imaginary potential, the cross section can be extracted directly from the violated continuity equation. 
A straightforward generalization of  Eq.~(\ref{sigmapsi}) to the  CC space leads to the expression~\cite{CaH13},
\begin{eqnarray}
\sigma_{\scr F}(E) &=& -\ \frac{1}{|A|^2}\ \frac{k}{E}\ \left\langle {\rm \Psi}^{\scr (+)}\left| \, \mathbb{W}\, \right| \, {\rm \Psi}^{\scr(+)} \right\rangle \nonumber\\
                              &=& -\ \frac{1}{|A|^2}\ \frac{k}{E}\ \sum_{\alpha,\alpha^\prime }\ \left\langle \psi^{\scr (+)}_\alpha \left| \, W_{\alpha \alpha^\prime}\, 
\right| \, \psi^{\scr(+)}_{\alpha^\prime} \right\rangle .
\label{sigF-CC}
\end{eqnarray}
This equation takes a simpler form when the imaginary potential is diagonal in channel space and it is
channel-independent ($W_{\alpha \alpha^\prime} = W(r)\ \delta_{\alpha \alpha^\prime}$).  One gets
\begin{equation}
\sigma_{\scr F}= \sum_\alpha \sigma_{\scr F}^{\scr (\alpha )},
\label{sigF-CC1}
\end{equation}
with
\begin{equation}
\sigma_{\scr F}^{\scr (\alpha )}=  -\ \frac{1}{|A|^2}\ \frac{k}{E}\  \left\langle \psi^{\scr (+)}_\alpha \left| \, W\, 
\right| \, \psi^{\scr(+)}_{\alpha} \right\rangle .
\label{sigF-CC2}
\end{equation}

\bigskip

In potential scattering calculations with long range imaginary potentials, one might be tempted to associate fusion and direct reactions with absorption in 
the inner region of the barrier and absorption on the surface region, respectively.  This could be done as follows. First, one splits the long range imaginary
potential as the sum of two terms, namely
\[
W^{\scr R}(R) = W^{\scr F}(R) + W^{\scr S}(R),
\]
where $W^{\scr F}(R)$ is a short-range term and $W^{\scr S}(R)$ is a surface term. Then, the fusion and the direct reaction cross sections would be
evaluated by Eq.~(\ref{sigmapsi}), with the replacements $W(R) \rightarrow W^{\scr F}(R)$ and $W(R) \rightarrow W^{\scr R}(S)$, respectively. 

However, this procedure is misleading. This can be seen clearly in a comparison with the more reliable CC approach. An {\it ideal} potential scattering calculation
would use an exact polarization potential, which leads to the same wave function as the  elastic wave function obtained by the CC method, with the imaginary potential
$W^{\scr F}(R)$. The fusion cross section of the CC method would then be given by Eq.~(\ref{sigF-CC1}), which contains contributions from both the elastic 
and nonelastic channels. The fusion cross section of the {\it ideal} potential scattering calculation would give the exact contribution from the
elastic channel, but it would miss the contributions from nonelastic ones. This could be a very serious flaw. In typical CC calculations, the contributions from the 
main nonelastic channels may be comparable to that from the elastic one. Thus, the fusion cross section evaluated in this way may be
greatly underestimated. Although this procedure may lead to the correct total reaction cross section, it does not take into account the fact that the incident flux lost
to excited direct channels may, eventually, lead to fusion (the contribution from nonelastic channels to Eq.~(\ref{sigF-CC1})).

\subsection{Coupled channels in the continuum - The CDCC method}\label{CDCC}


In typical collisions of tightly bound nuclei, the channel expansion of Eq.~(\ref{CC-exp}) involves only bound intrinsic states. This is justified by the fact that the breakup 
threshold of these nuclei are typically of several MeV, which makes the couplings to unbound channels negligible, at near-barrier collision energies. A different situation 
is found in collisions of weakly bound nuclei. This is the case of the stable light nuclei $^{6}$Li, $^{9}$Be and $^{7}$Li, which have breakup thresholds of 1.47, 1.67 and 
2.48 MeV,  respectively, and radioactive nuclei like $^6$He, $^8$B, $^{11}$Li and $^{11}$Be, with breakup thresholds below 1 MeV. In such cases, couplings with channels 
in the continuum (the breakup channel) have strong influence on the reaction dynamics, and it is necessary to include the continuum in the channel expansion.  
Eq.~(\ref{CC-exp}) then becomes,

\begin{equation}
{\rm \Psi}^{\scr (+)}({\bf R},\xi)= {\rm \Psi}^{\scr (+)}_{\scr B}({\bf R},\xi)\ +\ {\rm \Psi}^{\scr (+)}_{\scr C}({\bf R},\xi),
\label{CC-BC}
\end{equation}
with
\begin{equation}
{\rm \Psi}^{\scr (+)}_{\scr B}({\bf R},\xi) =  \sum_{\alpha=0}^{N_{\scr B}} \varphi_\alpha(\xi)\,\psi_\alpha^{\scr (+)}({\bf R}) ,
\label{CC-B}
\end{equation}
where $N_{\scr B}$ is the number of bound states of the projectile, with $\alpha$ standing for the quantum numbers required to 
specify them (usually $\varepsilon_\alpha,l_\alpha, j_\alpha$ and $\nu$), and
\begin{equation}
{\rm \Psi}^{\scr (+)}_{\scr C}({\bf R},\xi) =  \sum_{\beta} \int d\varepsilon \ \varphi_{\varepsilon\beta}(\xi)\,\psi_{\varepsilon \beta}^{\scr (+)}({\bf R}).
\label{CC-C}
\end{equation}
Above, $\varepsilon$ is the intrinsic energy in the continuum, which runs from the breakup threshold to infinity, and $\beta$ stands for the remaining
quantum numbers of the states representing the scattering of the projectile's fragments. In principle, this label runs from 1 to infinity, independently of 
$\varepsilon$.

\bigskip

Coupled equations could be derived as described in section \ref{CC eq}. That is, taking scalar product of Eq.~(\ref{CC-Sch}) with each of
the intrinsic states ($\varphi_\alpha(\xi)$ and $\varphi_{\varepsilon\beta}(\xi)$), and using their orthonormality properties. However, this procedure
would lead to an infinite number of coupled equations, even truncating the intrinsic energy  and keeping only a few values of $\beta$. This problem 
can be traced back to the fact that the expansion involves the continuous quantum number $\varepsilon$.

However, a finite number of coupled equations can be obtained if one expands the wave function ${\rm \Psi}^{\scr (+)}_{\scr C}({\bf R},\xi)$ over a finite set of states, 
$\phi_{n\beta}(\xi)$, instead of the infinite basis of scattering states. This is the basic idea of the continuum discretized coupled channel approximation (CDCC). 
The choice of these basis states is arbitrary, provided that it gives a good representation of the space spanned by the set of scattering states, truncated at 
some reasonable intrinsic energy, $\varepsilon_{\scr max}$. In the next sub-sections, we discuss the CDCC approximation in further detail.


\subsubsection{Three-body CDCC}


\begin{figure}
\begin{center}
\includegraphics*[width=8cm]{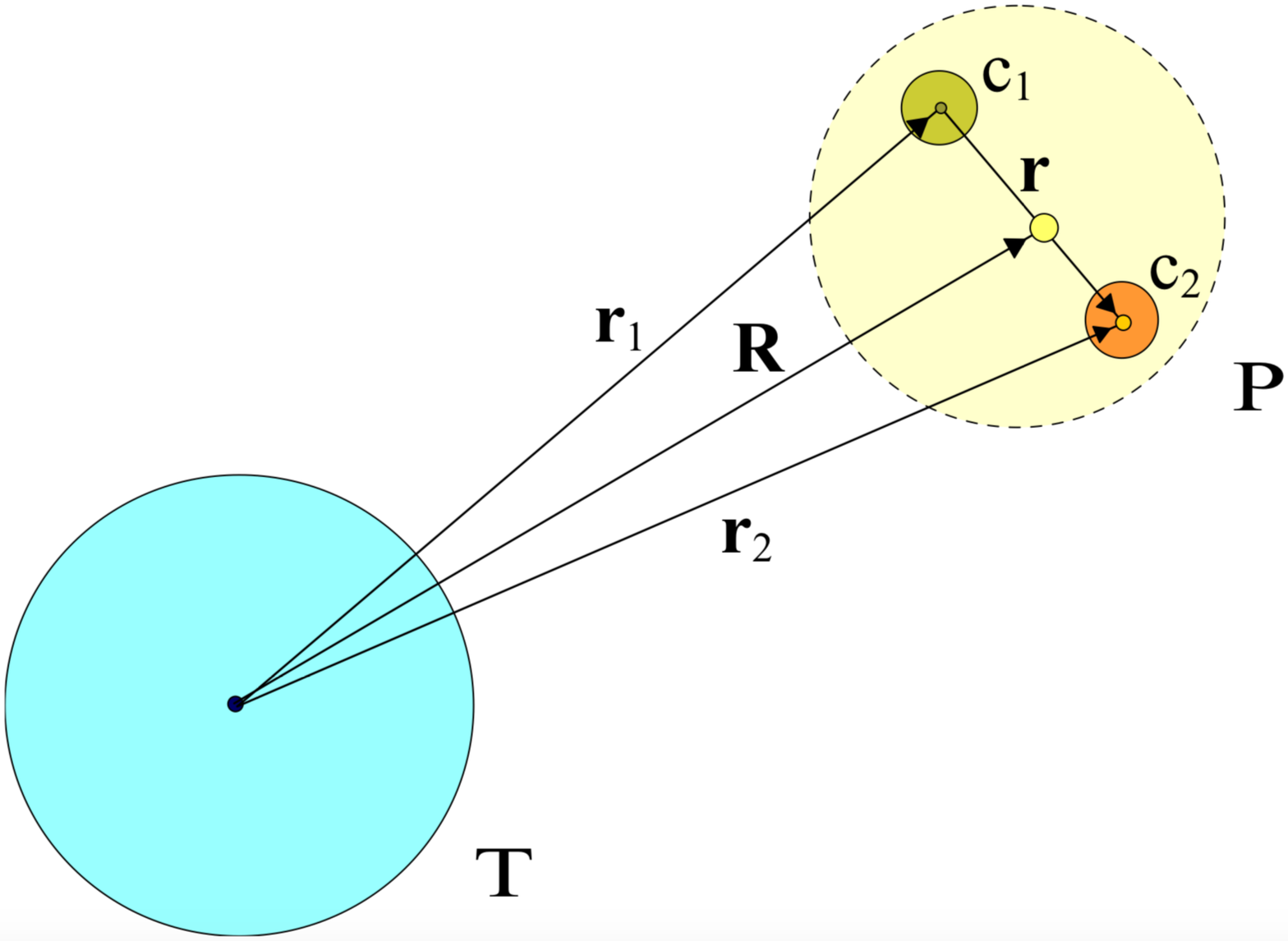}
\end{center}
\caption{(Color on line) Schematic representation of the colliding system in the 3-body CDCC calculations (figure taken from Ref.~\cite{KCD18}). 
For details, see the text.}
\label{3-B}U
\end{figure}
Let us consider the situation where, during the collision, the projectile breaks up into two fragments without internal structure, $c_1$ and $c_2$. 
In this case, the intrinsic coordinate (previously denoted by $\xi$) is the vector joining the centers of the two fragments, ${\bf r}$, as shown in Fig.~\ref{3-B}. 
Along the collision, the target interacts with the fragments through the complex potentials $U_1(r_1)$ and  $U_2(r_2)$, which are functions of the moduli of the vectors
${\bf r}_1 = {\bf R}-(A_2/A_{\scr P})\ {\bf r}$ and $  {\bf r}_2 = {\bf R}+(A_1/A_{\scr P})\ {\bf r}$, respectively. Thus, the projectile-target interaction is given by
\begin{equation}
{\mathbb U}({\bf R},{\bf r}) = U_1(r_1) + U_2(r_2).
\label{V1+V2}
\end{equation}
Then, the total Hamiltonian of the system is
\begin{equation}
{\mathbb H}({\bf R},{\bf r}) = h({\bf r})+\hat{K}+{\mathbb U}({\bf R},{\bf r}),
\label{H-2B}
\end{equation}
and the scattering state satisfies the Schr\"odinger equation,
\begin{equation}
\Big[ E - {\mathbb H} \Big]\, \left| {\rm \Psi^{\scr (+)}} \right> = 0.
\label{Sch_H-2B}
\end{equation}

Proceeding as in the case of tightly bound systems, we take scalar product of both sides of the above equation with each of the intrinsic states appearing
in the expansions of Eqs.~(\ref{CC-B}) and (\ref{CC-C}). In this way, we get the sets of coupled 
equations,
\begin{multline}
\big[E - H_{\alpha,\alpha}({\bf R})  \big] \,\psi_\alpha^{\scr (+)}({\bf R}) = \sum_{\alpha^\prime} H_{\alpha,\alpha^\prime}({\bf R})\,
\psi_{\alpha^\prime}^{\scr (+)}({\bf R}) \\
+\ \sum_\beta \int d\varepsilon\ H_{\alpha,\varepsilon \beta}({\bf R})\,\psi_{\varepsilon \beta}^{\scr (+)}({\bf R})
\label{CC-eq2}
\end{multline}
and
\begin{multline}
\big[E - H_{\varepsilon \beta,\varepsilon \beta}({\bf R})  \big] \,\psi_{\varepsilon \beta}^{\scr (+)}({\bf R}) = 
\sum_{\alpha} H_{\varepsilon \beta,\alpha}({\bf R})\,\psi_{\alpha}^{\scr (+)}({\bf R}) \\
+\ \sum_{\beta^\prime} \int d\varepsilon^\prime\ H_{\varepsilon \beta,\varepsilon^\prime \beta^\prime}({\bf R}) \,
\psi_{\varepsilon^\prime \beta^\prime}^{\scr (+)}({\bf R}) ,
\label{CC-eq3}
\end{multline}
where the matrix-elements of the Hamiltonian involving continuum states are defined analogously to Eq.~(\ref{Hmat-el}).  Now, however, the situation 
is different. This procedure lead to an infinite set of coupled equations, which cannot be solved. \\

The CDCC method deals with this problem by approximating the infinite dimensional space of scattering states by the reduced space spanned 
by a finite set of functions. In the simple case where the $c_1$ and $c_2$ have spin zero, $\beta$ stands for the total angular momentum of
the projectile, $j$, and its z-projection, $\nu$. Then, the wave function describing the scattering of the fragments with relative energy 
$\varepsilon$ can be written as:
\begin{equation}
\varphi_{\varepsilon\beta}({\bf r}) =  \frac{u_{\varepsilon \beta}(r)}{r}\ \ \mathcal{Y}_\beta(\hat{\bf r}) ,
\label{discrete}
\end{equation}
where $u_{\varepsilon \beta}(r)$ is the radial part of $\varphi_{\varepsilon \beta}({\bf r})$. Above, $\mathcal{Y}_{\beta}(\hat{\bf r})$ is a complementary
function of the orientation\footnote{For simplicity, we omit the spins of the collision partners. For non-zero spins, $\mathcal{Y}_\beta(\hat{\bf r})$ are
linear combinations of spin-dependent terms, involving  angular momentum coupling coefficients.} of ${\bf r}$. The radial wave functions must be normalized 
to satisfy the relations,
\begin{equation}
\int dr\  u^*_{\varepsilon \beta}(r) \ u_{\varepsilon^\prime \beta}(r) = \delta(\varepsilon- \varepsilon^\prime).
\label{ortho-scatt}
\end{equation}
In the particular case of fragments with spin zero, $\beta$ represents the quantum numbers $\{j,\nu\}$. Then, 
$\mathcal{Y}_\beta(\hat{\bf r}) \equiv \mathcal{Y}_{jm}(\hat{\bf r})$ are the usual spherical harmonics. In more general situations, 
it is a more complicated function~\cite{DBD10,ThN09,CaH13}, involving spherical harmonics, spin states and angular momentum coupling coefficients.\\

The CDCC approximation consists in replacing the infinite integral over $\varepsilon$ by the sum
\begin{equation}
\int \ d\varepsilon\  \frac{u_{\varepsilon \beta}(r)}{r}\ \longrightarrow\ \sum_{n=1}^{n_{\rm max}} \ \frac{\phi_{n\beta}(r)}{r} ,
\label{discrete-a}
\end{equation}
where the set of $n_{\rm max}$ functions $\phi_{n\beta}(r)$ must satisfy the orthonormality relations\footnote{Note that the angular part of the intrinsic wave functions 
guarantees the orthogonality for $\beta \ne \beta^\prime$.}
\begin{equation}
\int \ d\varepsilon\ \phi^*_{n\beta}(r)\ \phi_{n^\prime \beta}(r) = \delta_{n n^\prime}.
\end{equation}
Further, the intrinsic hamiltonian should be diagonal in the $n_{\scr max}$ dimensional space of the functions $\phi_{n\beta}(r)$. That is
\begin{equation}
\left\langle \phi_{n\beta}\, \right| \ h\ \left|\, \phi_{n^\prime \beta}\right\rangle = \bar{\varepsilon}_n\ \delta_{n n^\prime}.
\end{equation}
Otherwise, there would be channel couplings even without the interaction with the target.

\bigskip

With the continuum discretization of Eq.~(\ref{discrete-a}), bound and unbound states can be treated on the same grounds. Then, the infinite set of 
Eqs.~(\ref{CC-eq2}) and (\ref{CC-eq3}) reduces to a finite set of equations with the general form of Eq.~(\ref{CC-eq1}). Now, however, $\alpha$ stands for any 
state of the projectile, bound or unbound. The number of coupled equations, $N$, is given by 
\begin{equation}
N = N_{\scr B} + N_{\scr C},
\label{totalN}
\end{equation}
where $N_{\scr B}$ and $N_{\scr C}$ are respectively the number of bound and continuum-discretized states of the projectile. The latter is given by
\begin{equation}
N_{\scr C} =  n_{\scr max} \times \beta_{\scr max}.
\label{totalN-1}
\end{equation}
In practice, the number of coupled equation is much smaller than $N$. When proper angular momentum projections are carried out, and angular momentum,
and parity conservations are taken into account, the set of equations splits into decoupled sets of smaller dimensions, one for each invariant sub-space.\\ 

The discretization of the continuum may be performed by two methods: the {\it bin method} and the method of {\it pseudo-states}. These methods are
briefly discussed below.

\bigskip

\noindent {a) \it The bin method}

\medskip

In the bin method, the functions $\phi_{n\beta}(r)$ are generated by scattering states of the fragments through wave packets with the general form,
\begin{equation}
\phi_{n\beta}(r) = \int d\varepsilon\ \Gamma_n(\varepsilon)\, u_{\varepsilon \beta}(r),
\label{functions phi}
\end{equation}
with the weight function, $\Gamma_n(\varepsilon)$, concentrated around $\varepsilon_n$. \\

The most common weight functions used in nuclear physics are constant within some limited energy interval, and zero elsewhere~\cite{SYK86,AIK87,MKO03,ThN09}. 
The continuum is truncated at some energy $\varepsilon_{\scr max}$ and the interval from 0 to the cut-off energy is divided into a set of non-overlapping 
intervals $\Delta_m$, centered at the energy $\varepsilon_m$, such that the upper limit of each interval coincides with the lower limit of the subsequent one. 
That is,
\begin{eqnarray}
 \Gamma_m(\varepsilon) & = & \frac{1}{\sqrt{\Delta_m}}, \qquad {\rm if}\ \varepsilon_m^{\scr (+)} \ge \varepsilon_m \ge \varepsilon_m^{\scr (-)} 
\nonumber \\
                                         & = & 0, \qquad\qquad  {\rm otherwise},
\label{constant gamma}
\end{eqnarray}
where $\varepsilon_m^{\scr (\pm)} = \varepsilon_m \pm \Delta_m/2$ are the limits of the interval. The most common choice is to use bins of constant width in 
momentum space. In this case, the energy width increase linearly with $k$.  On the other hand, when there are sharp resonances in the scattering of the fragments, it is 
necessary to reduce the width and increase the density of bins around the resonance. 

\medskip

In most CDCC calculations the discretization is carried out with bins in momentum space, labeled by $k = \sqrt{2\mu_{\scr 12} \varepsilon}/\hbar$, where
$\mu_{\scr 12}$ is the reduced mass in the $c_1-c_2$ collision. In this case, the bins are given by
\begin{equation}
\phi_{m\beta}(r) = \int dk\ \Gamma_m(k)\, u_{k \beta}(r),
\label{functions phi-k}
\end{equation}
with the scattering states satisfying the orthonormality relations
\begin{equation}
\int dr\ u^*_{k\beta}(r) \ u_{k^\prime\beta}(r)= \frac{\pi}{2}\ \delta(k-k^\prime).
\label{ortho-n_k}
\end{equation}
The constant weight functions now are given by
\begin{eqnarray}
 \Gamma_m(k) & = & \frac{1}{\sqrt{\Delta_m}}, \qquad {\rm if}\ k_m^{\scr (+)} \ge k_m \ge k_m^{\scr (-)} 
\nonumber \\
                                         & = & 0, \qquad\qquad  {\rm otherwise},
\label{constant gamma-k}
\end{eqnarray}
with $k_m^{\scr (\pm)} =k_m \pm \Delta_m/2$, and $\Delta_m$ now has the dimension of $p/\hbar$. Since the energy is proportional to $k^2$, the energy width of a bin
with constant width in $k$-space increases linearly with $k$.\\

It can be easily proved that the functions generated by the weight functions of Eq.~(\ref{constant gamma}) form an orthonormal set. For this purpose,
we take the scalar product of two bins and use the orthonormality of the scattering states (Eq.~(\ref{ortho-scatt})). We get,
\begin{equation}
\left\langle \phi_{n\beta}\, | \, \phi_{n^\prime \beta}\right\rangle = \int d\varepsilon\ \Gamma_n(\varepsilon)\,  
\Gamma_{n^\prime}(\varepsilon) = \delta_{n,n^\prime}.
\end{equation}
The second equality in the above equation was obtained inserting the weight functions of Eq.~(\ref{constant gamma}) into the integral over $\varepsilon$. 
It is equally straightforward to show that the intrinsic Hamiltonian is diagonal in this bin space. Proceeding similarly, one gets
\begin{equation}
\left\langle \phi_{n\beta}\, |\,h\, |\, \phi_{n^\prime \beta}\right\rangle = \int d\varepsilon\ \Gamma_n(\varepsilon)\ \varepsilon \    
\Gamma_{n^\prime}(\varepsilon) = \bar{\varepsilon}_n\ \delta_{n,n^\prime}.
\end{equation}
\bigskip

The weight functions of Eq.~(\ref{constant gamma}) are very easy to handle but the abrupt change at the edges leads to bins with longer ranges and with {\it beats} at 
large radial distances~\cite{BeC92}. Although it is not a serious problem, it can be avoided with smooth weight functions, as the ones proposed in Ref.~\cite{BeC92}, and
used in Refs.~\cite{BeC92,MCD08,MCD14,KCD18}. However, there is a drawback with these weight functions: they do not diagonalize the intrinsic Hamiltonian. It is
then necessary to perform a unitary transformation in the bin space,
\begin{equation}
\phi_{n\beta}(r) \ \rightarrow\ \bar{\phi}_{n\beta}(r) = \sum_{n=1}^{n_{\scr max}} U_{n m}(\beta)\ \phi_{m \beta}(r),
\label{w-phi}
\end{equation}
with the operator $U(\beta)$ determined by the condition 
\begin{equation}
\left< \bar{\phi}_{nj} \right| h \left| \bar{\phi}_{n^\prime j} \right> = \sum_{m,m^\prime =1}^{n_{\scr max}}  U^\dagger_{nm}(\beta) \ h_{m m^\prime}\ 
U_{m^\prime n^\prime}(\beta)= \bar{\varepsilon}_n\ \delta_{n,n^\prime},
\label{eigen}
\end{equation}
where
\begin{equation}
h_{m m^\prime} = \left< \phi_{m\beta} \right| h \left| \phi_{m^\prime\beta} \right> .
\label{def hij}
\end{equation}

\bigskip

\noindent {b) \it The pseudo-states method}

\medskip

In the pseudo-states (PS) method, the space of scattering states is approximated by a finite dimensional space spanned by
a set of square-integrable functions. These functions are approximate eigenstates of $ h$ with positive
energy. The method is developed in two steps. First, one selects a set of square integrable functions $\phi_{m\beta}(r)$.
Then, the pseudo-states,  $\bar{\phi}_{n\beta}(r)$, are determined by diagonalizing $h$ in this space spanned by 
this set.  The diagonalization is performed as in the case of non-orthogonal bins, following the procedure of Eqs.~(\ref{w-phi}), 
(\ref{eigen}) and (\ref{def hij}). There is, however, a difference.  In the bin method, the functions $\phi_{m\beta}(r)$ are
wave packets of scattering states. Thus, the eigenvalues $\bar{\varepsilon}_n$ are all positive. Now the situation is different.
There are also negative eigenvalues, which represent the bound states of the projectile. In most cases, these states are calculated 
directly, without expansion in PC basis. In such cases, the eigenstates with negative energy obtained through the diagonalization
of $h$ should be discarded.\\

The choice of square-integrable set of states, $\phi_{n\beta}(r)$, is arbitrary, provided that they give a good description
of the radial wave functions, within the range of the coupling interactions. Choices based on Gaussian functions are
extensively discussed in Ref.~\cite{HKK03}. We mention two of them. The first is the set of $N$ real Gaussian functions with 
variable range~\cite{MKO03},
\begin{equation}
\phi_{n\beta} (r)= r^{l_ \beta}\ \exp\left[ -\,r^2 / a_n^2 \right],
\label{gaussians}
\end{equation}
where $l_\beta$ is the orbital angular momentum of the $c_1-c_2$ relative motion in the state $\phi_{n\beta}$. The range
increases from $a_1$ to $a_{\scr N}$, in the geometrical progression
\begin{equation}
a_n = a_1\ \left(\frac{a_{\scr N}}{a_1} \right)^{(n-1)/(N-1)},
\label{an}
\end{equation}
where $a_1$ and $a_{\scr N}$ are parameters of the set.
The second is a set of $2N$ functions obtained by the multiplication of the Gaussians of Eq.~(\ref{gaussians}) by oscillating functions, 
in the form~\cite{HKK03},
\begin{eqnarray}
\phi^{\scr C}_{n\beta} &=& \phi_{n\beta} \ \cos\left[ b\,(r/a_n) \right] \label{exp-cos}\\
\phi^{\scr S}_{n\beta} &=& \phi_{n\beta} \ \sin\left[ b\,(r/a_n) \right] \label{exp-sin},
\end{eqnarray}
where $b$ is an adjustable parameter. This set corresponds to taking the real and the imaginary parts of the functions of Eq.~(\ref{gaussians}) with 
complex widths. Calculations using this set of states converge more rapidly than the ones using the Gaussians of Eq.~(\ref{gaussians}).

\smallskip

Some CDCC calculations adopting the Lagrange-mesh method make a different choice of basis functions. In such cases, one frequently uses basis functions 
associated with a Gauss quadrature~\cite{HSR98,DBD10}. This greatly simplifies the numerical calculations. \\

A large number of three-body CDCC calculations neglecting excitations of the fragments and the target have been reported. For a review, see, e.g. 
Refs.~\cite{CGD15,SYK86,AIK87}. 


\subsubsection{Core and target excitations}\label{XCDCC}


Until recently, the available CDCC calculations ignored intrinsic structures of the projectile's fragments and of the target, treating them as {\it point particles}. 
Frequently, these approximations are reasonable. However, the neglected degrees of freedom may play an important role in the reaction dynamics of 
some colliding systems. Formally, the inclusion of excitations of the fragments or of the target is straightforward. However, from the computational point of 
view it is not a trivial task. Usually, the calculations are performed by standard computer codes available in the literature and the inclusion of fragment or 
target excitation involves modifications that requires considerable knowledge of the structure of the code. In addition, the inclusion of these excitations enlarges 
significantly the dimension of the matrices involved in the calculations, demanding much more computer power. Calculations with excitations of 
one of the fragments or of the target will be briefly discussed below (we follow Ref.~\cite{MoG16}).

\bigskip

\noindent a)  CDCC {\it with Core excitation} (XCDCC)

\medskip

\noindent If the intrinsic structure of a projectile fragment, say $c_1$, with coordinates $\xi_1$, is taken into account, the total Hamiltonian of the system
becomes
\begin{equation}
{\mathbb H}({\bf R};{\bf r}, \xi_1) = h({\bf r}, \xi_1)+\hat{K}+{\mathbb U}({\bf R};{\bf r}, \xi_1),
\label{H-2B-1}
\end{equation}
with the projectile-target potential\footnote{ If  $c_1$ is not spherical, $V_1$ depends both on the modulus and on the orientation of ${\bf r}_1$.}
\begin{equation}
{\mathbb U}({\bf R};{\bf r}, \xi_1) = U_1({\bf r}_1, \xi_1) + U_2(r_2).
\label{V1+V2-a}
\end{equation}
Above, $\xi_1$ stands for the intrinsic coordinates of fragment $c_1$. Then, the projectile's wave functions of Eq.~(\ref{discrete}) takes the form
\begin{equation}
\varphi_{\varepsilon\beta}({\bf r}) =  \frac{u_{\varepsilon \beta}(r)}{r}\ \ \mathcal{Y}_\beta(\hat{\bf r},\xi_1).
\label{discrete-1}
\end{equation}
Now the index $\beta$ stands for angular momenta of the projectile and the fragments, and also for the quantum numbers associated with the
degrees of freedom represented by $\xi_1$. \\

This generalization of the CDCC approximation, known as XCDCC, was introduced by Summers {\it et al.}~\cite{SNT06,SuN07}, to study the breakup
of $^{17}{\rm C}$ ($^{16}{\rm C}+p$) and $^{11}{\rm Be}$ ($^{10}{\rm Be}+n$) projectiles on $^9$Be targets. The calculations included excitation channels 
related to rotational bands of the deformed $^{16}$C and $^{10}$Be cores. \\

More recently, similar calculations have been carried out to study different reactions. Moro and Crespo~\cite{MoC12} studied the influence of rotational 
excitations of the deformed $^{10}$Be core in the breakup of $^{11}$Be ($^{10}{\rm Be} + n$) projectiles in collisions with a proton target. For this purpose, 
they proposed a simple reaction model using the DWBA.

\smallskip

De Diego {\it et al.}~\cite{DAL14} used a more realistic XCDCC model to evaluate quasi-elastic and breakup cross sections for the same system, at 
collision energies ranging from 10 to 200 MeV/nucleon. 

\smallskip

Chen  {\it et al.}~\cite{CLY16} measured elastic and breakup cross sections for the same $^{11}{\rm Be} + p$ system at 26.9 MeV/nucleon, and compared 
the data with results of CDCC and XCDCC calculations including rotational excitations of $^{10}$Be. They concluded that excitations of the core play 
a moderate role in the reaction dynamics.

Later on, De Diego {\it et al.}~\cite{DCM17} performed similar XCDCC calculations, to evaluate cross sections for different two- and three-body observables, at 
different collision energies, for which there are data available. In this way, they investigated the importance of core excitation in the reaction dynamics. 

\smallskip

\begin{figure}
\begin{center}
\includegraphics*[width=7.5cm]{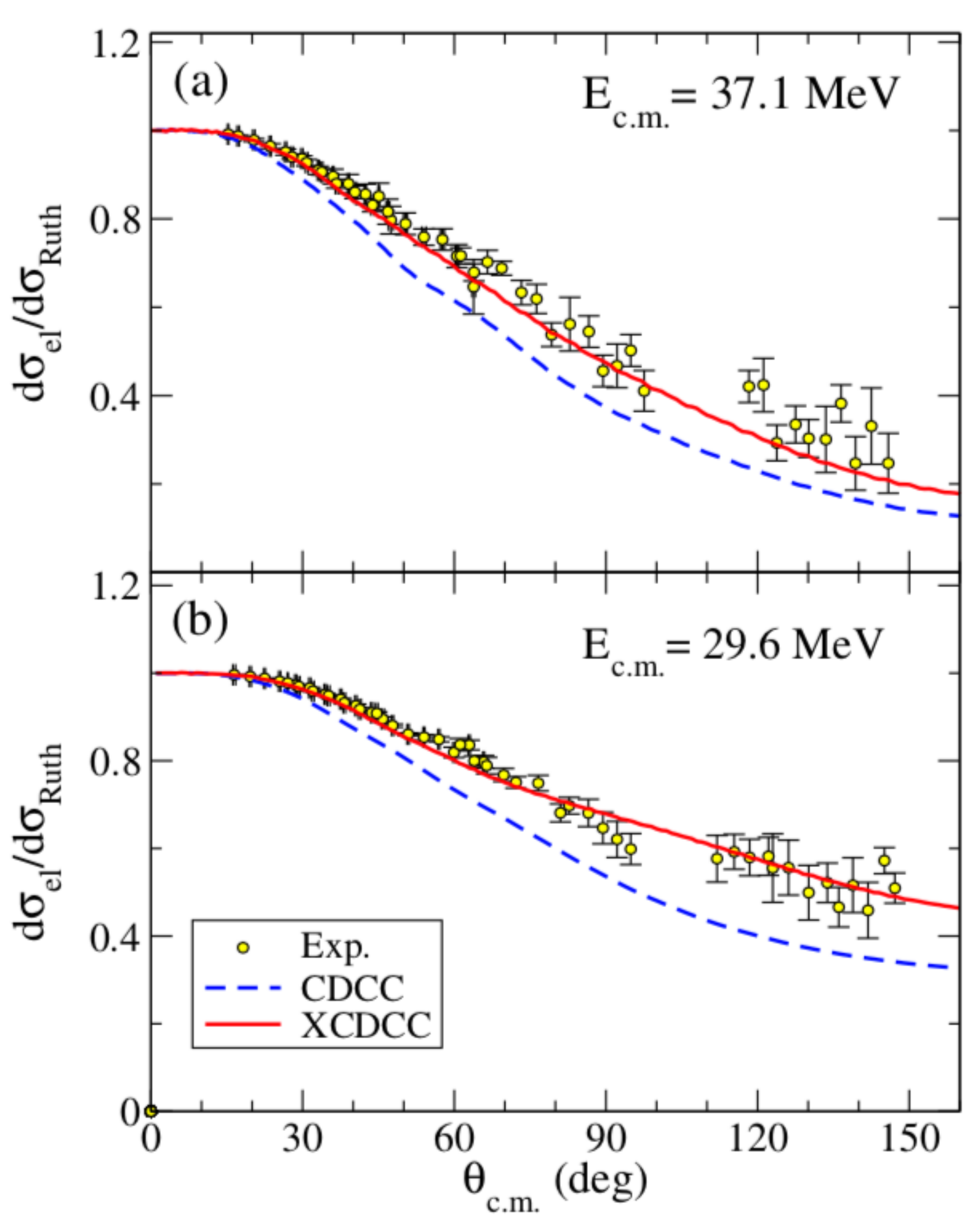}
\end{center}
\caption{(Color on line) Elastic angular distribution in the $^{11}$Be + $^{197}$Au collision at two collision energies, divided by the corresponding Rutherford
cross sections. Results of XCDCC (solid line) and CDCC (dashed line) are compared to the experimental data. The figure, the data and the calculations are
from Ref.~\cite{PBM17}.}
\label{XCDCC-el}
\end{figure}
\begin{figure}
\begin{center}
\includegraphics*[width=7.5cm]{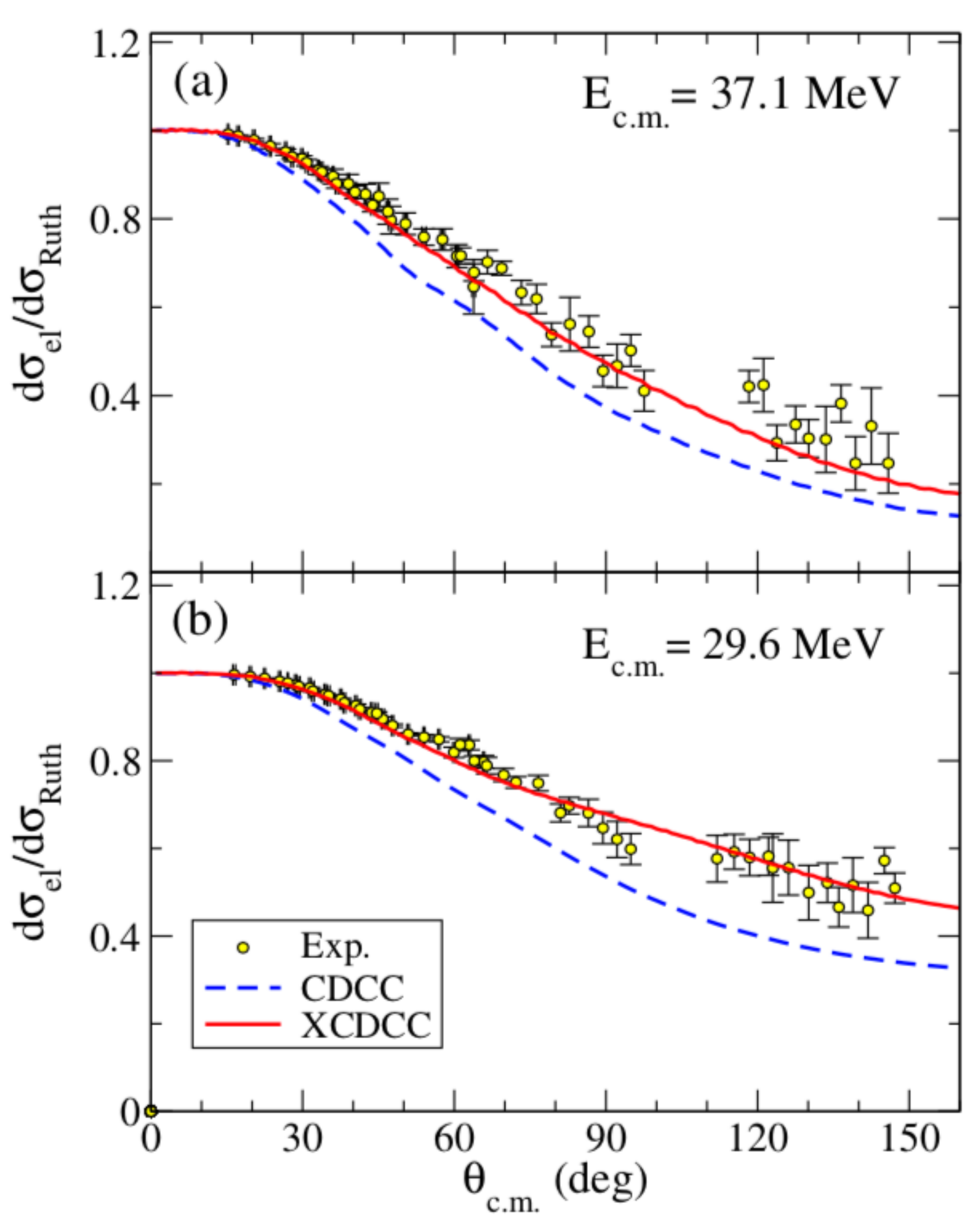}
\end{center}
\caption{(Color on line) Similar to the previous figure, but for breakup angular distributions (figure extracted from Ref.~\cite{PBM17}).}
\label{XCDCC-bu}
\end{figure}
Pesudo {\it et al.}~\cite{PBM17} measured elastic scattering, inelastic scattering and breakup in collisions of $^{11}$Be projectiles with a heavy target. They
studied the $^{11}{\rm Be}\, -\, ^{197}$Au collision at two energies, below and around the Coulomb barrier. The experimental cross sections were then
compared with predictions of CDCC and XCDCC calculations. Their elastic and breakup cross sections are shown in Figs.~\ref{XCDCC-el} and \ref{XCDCC-bu}, 
respectively. The elastic scattering data is well reproduced by the XCDCC calculations, whereas the predictions of the standard CDCC calculations fall 
systematically below the data. This clearly indicates the importance of the core excitation for a proper description of the reaction dynamics of this system. 
On the other hand, inspecting Fig.~\ref{XCDCC-bu}, one concludes that the inclusion of core excitations in the calculations of breakup cross sections is
not as important as in the case of elastic scattering.

\smallskip

Lay {\it et al.}~\cite{LDC16} performed XCDCC calculations for $^{19}{\rm C} + p$ system, to analyse the resonant breakup of $^{19}$C, which has been 
measured at RIKEN~\cite{SNF08}. The inclusion of core excitation was shown to be essential for a good description of the data. CDCC calculations with
an inert core largely underestimated the data.

\bigskip


\noindent b) {\it CDCC with excitations of the target}


\medskip

If the intrinsic degrees of freedom of the target, $\xi_{\scr T}$ are taken into account, the system's Hamiltonian becomes,
\begin{equation}
{\mathbb H}({\bf R};{\bf r}, \xi_{\scr T}) = h({\bf r}) + H_{\scr T}(\xi_{\scr T})+\hat{K}+{\mathbb V}({\bf R};{\bf r}; \xi_{\scr T}),
\label{H-Texc}
\end{equation}
with
\begin{equation}
{\mathbb V}({\bf R};{\bf r}; \xi_{\scr T}) = V_1({\bf r}_1, \xi_{\scr T}) + V_2({\bf r}_2, \xi_{\scr T}).
\label{V1T+V2T}
\end{equation}
In this case, the label $\beta$ stands also for quantum numbers of the target, and Eq.~(\ref{discrete}) becomes
\begin{equation}
\varphi_{\varepsilon\beta}({\bf r}) =  \frac{u_{\varepsilon \beta}(r)}{r}\ \ \mathcal{Y}_\beta(\hat{\bf r},\xi_{\scr T}).
\label{discrete-2}
\end{equation}

\bigskip

The importance of target excitations in weakly bound systems has been investigated by Lubian  {\it et al.}~\cite{LCA09}. They performed CDCC calculations 
for the $^9{\rm Be}\, +\, ^{58}{\rm Ni}$ system, including and not including excitations of the target. Comparing their results to the data of  Aguilera~{\it et al.}~\cite{EML09},
they concluded that the inclusion of continuum states was essential to describe the data, whereas the influence of target excitations was weak. 

\smallskip

Woodward {\it et al.}~\cite{WFO12} performed CDCC calculations for the $^6{\rm Li}+^{144}{\rm Sm}$ system. Besides the continuum space of the projectile,
the calculations took into account the excitation of the $2^+_1$ and $3^-_1$ states in $^{144}$Sm. Inelastic angular distributions populating the two excited
states of the target have been measured at near-barrier energies, and the results were compared with the predictions of standard coupled channel 
calculations and with their CDCC calculations with target excitation. This study lead to the conclusion that a full treatment of the continuum, including 
continuum-continuum couplings, is essential for a good description of the data. 

G\'omez-Ramos and Moro~\cite{GoM17} developed a comprehensive study of the influence of the breakup channel on excitations of the target, in collisions 
with weakly bound projectiles. They performed standard coupled channel and CDCC  calculations with target excitation for the following reactions:

\medskip

\noindent $^{58}{\rm Ni}(d,d)^{58}{\rm Ni}^*$, $^{24}{\rm Mg}(d,d)^{24}{\rm Mg}^*$, $^{144}{\rm Sm}(^6{\rm Li},^6{\rm Li})^{144}{\rm Sm}^*$,  
$^9{\rm Be} (^6{\rm Li},^6{\rm Li})^9{\rm Be}^*$.
\medskip

\noindent They obtained a satisfactory agreement with the data for both, standard and target excitation calculations, and concluded that the continuum had a moderate 
influence in the inelastic scattering. 


\subsubsection{Four-body CDCC}\label{4b-CDCC}


\begin{figure}
\begin{center}
\includegraphics*[width=8cm]{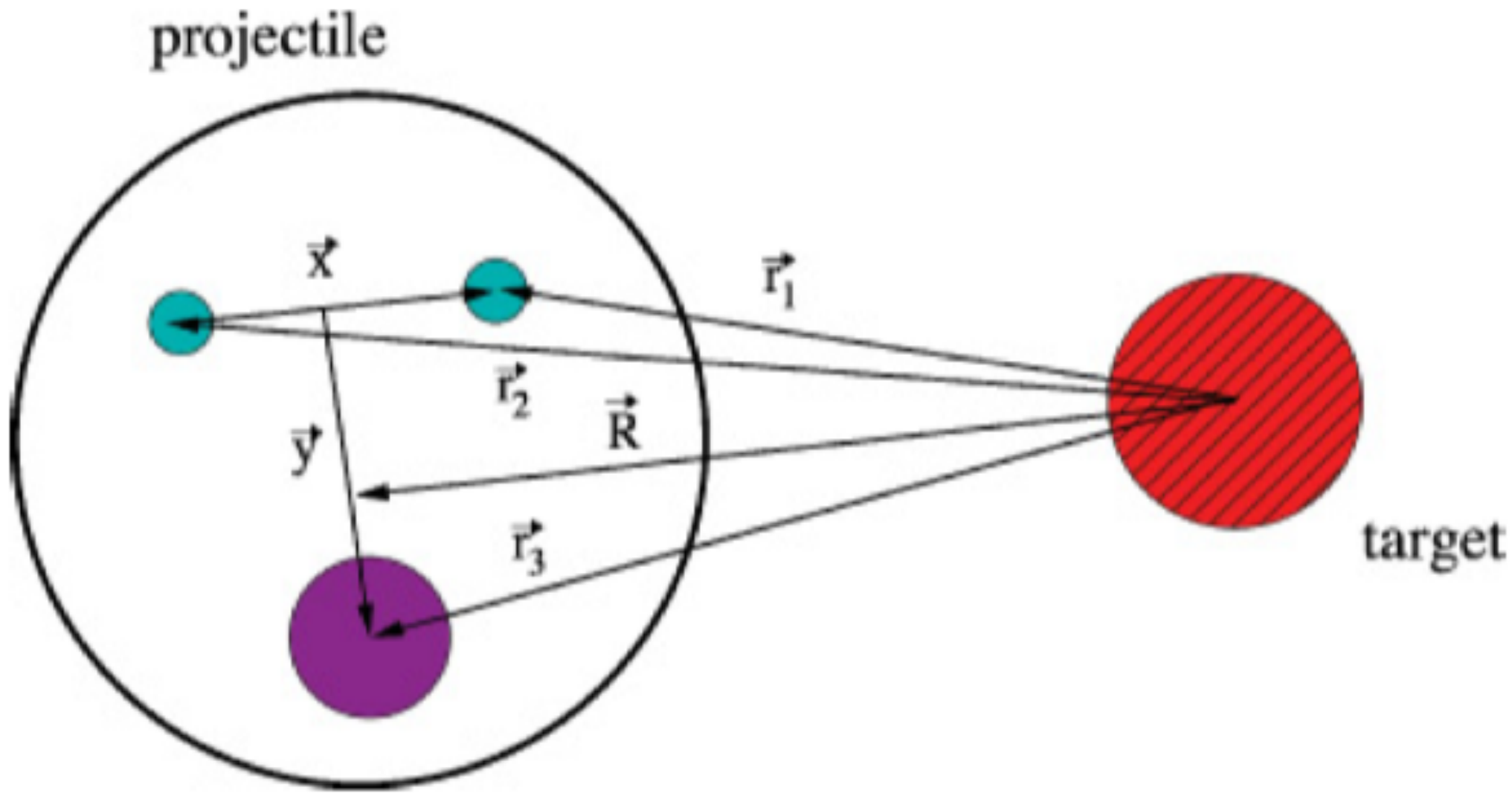}
\end{center}
\caption{(Color on line) Coordinates used in four-body CDCC calculations, where the projectile breaks up into three fragments 
(figure taken from Ref.~\cite{RAG08}).}
\label{4body}
\end{figure}
Owing to their cluster configuration, several weakly bound nuclei can break up into three fragments during a collision. Important examples are the stable $^9$Be 
($^4{\rm He}+^4{\rm He}+n$) and radioactive two neutron halo nuclei, like $^6$He ($^4{\rm He} + n + n$) and $^{11}$Li ($^9{\rm Li} + n + n$). 
Thus, in collisions of projectiles with this configuration, the reaction dynamics involves four particles: the three clusters of the projectile, and the target. This calls for a 
generalization of the CDCC method, usually called four-body CDCC (to distinguish these methods, we henceforth adopt the notations 3b-CDCC and 4b-CDCC).\\

A collision of a three-fragment projectile with a target is schematically represented in Fig.~\ref{4body}. The projectile-target interaction is the sum of interactions between 
the fragments and the target, $U_{c_i-T}\,( i=1,3)$, which depend on the fragment-target coordinates ${\bf r}_i$, shown in Fig.~\ref{4body}.  These coordinates 
are usually expressed in terms of the Jacobi coordinates\footnote{Alternatively, they can be expressed in terms of hyperspherical coordinates (see, e.g. Ref.~\cite{RAG08}).},
defined as 
\begin{eqnarray}
{\bf x} &=& \sqrt{\frac{A_{\scr 1}A_{\scr 2}}{A_{\scr 1}+A_{\scr 2}}}\ \left( {\bf r}_{\scr 2}-{\bf r}_{\scr 3} \right), \\
{\bf y} &=& \sqrt{ \frac{A_{\scr 3}\,\left(A_{\scr 1}+A_{\scr 2} \right)}{A_{\scr P}} }\ \left[ {\bf r}_{\scr 3} - \frac{A_{\scr 1}\,{\bf r}_{\scr 1}
+A_{\scr 2}\,{\bf r}_{\scr 2}}{A_{\scr 1}+A_{\scr 2}}\right].
\end{eqnarray}
The Hamiltonian of the projectile-target system then reads\
\begin{equation}
\mathbb{H}({\bf R};{\bf x},{\bf y}) = h({\bf x},{\bf y}) + \hat{K} + \mathbb{U}({\bf R};{\bf x},{\bf y}),
\end{equation}
where $\mathbb{U}({\bf R};{\bf x},{\bf y})$ is the sum of the three complex fragment-target interactions, expressed in terms of the Jacobi coordinates.\\

The system's wave functions still have the general form of Eq.~(\ref{discrete}), but with $\beta$ representing a larger number of intrinsic quantum
numbers, and with the replacement,
\[ 
 \mathcal{Y}_\beta(\hat{\bf r}) \rightarrow  \mathcal{Y}_\beta({\bf x}, {\bf y}). 
 \]
 
 \bigskip
 
4b-CDCC calculations have been performed to evaluate several observables in different collisions of weakly bound nuclei. Matsumoto {\it et al.}~\cite{MHO04} 
 performed 3b- and 4b-CDCC calculations of elastic angular distributions in $^6{\rm Li} - ^{12}{\rm C}$ scattering. 

\medskip

 Rodr\'\i guez-Gallardo  {\it et al.}~\cite{RAG08} performed 4b-CDCC calculations of elastic angular distributions for $^6$He projectiles on $^{12}$C, 
 $^{64}$Zn and $^{208}$Pb targets.
 
 \medskip
 
 Cubero {\it et al.}~\cite{CFR12} measured elastic angular distributions for $^{11}$Li projectiles on a $^{208}$Pb target, at two energies around the Coulomb barrier. 
 The data exhibited a strong damping at small angles, even at the energy below the barrier. This behaviour can be traced back to the breakup of the two-neutro-halo 
 projectile, under the action of the long-range dipole interaction. The authors performed  4b-CDCC calculations, and compared the resulting cross sections  with the data. 
 The agreement between theory and experiment was very good.

\medskip

\begin{figure}
\begin{center}
\includegraphics*[width=6cm]{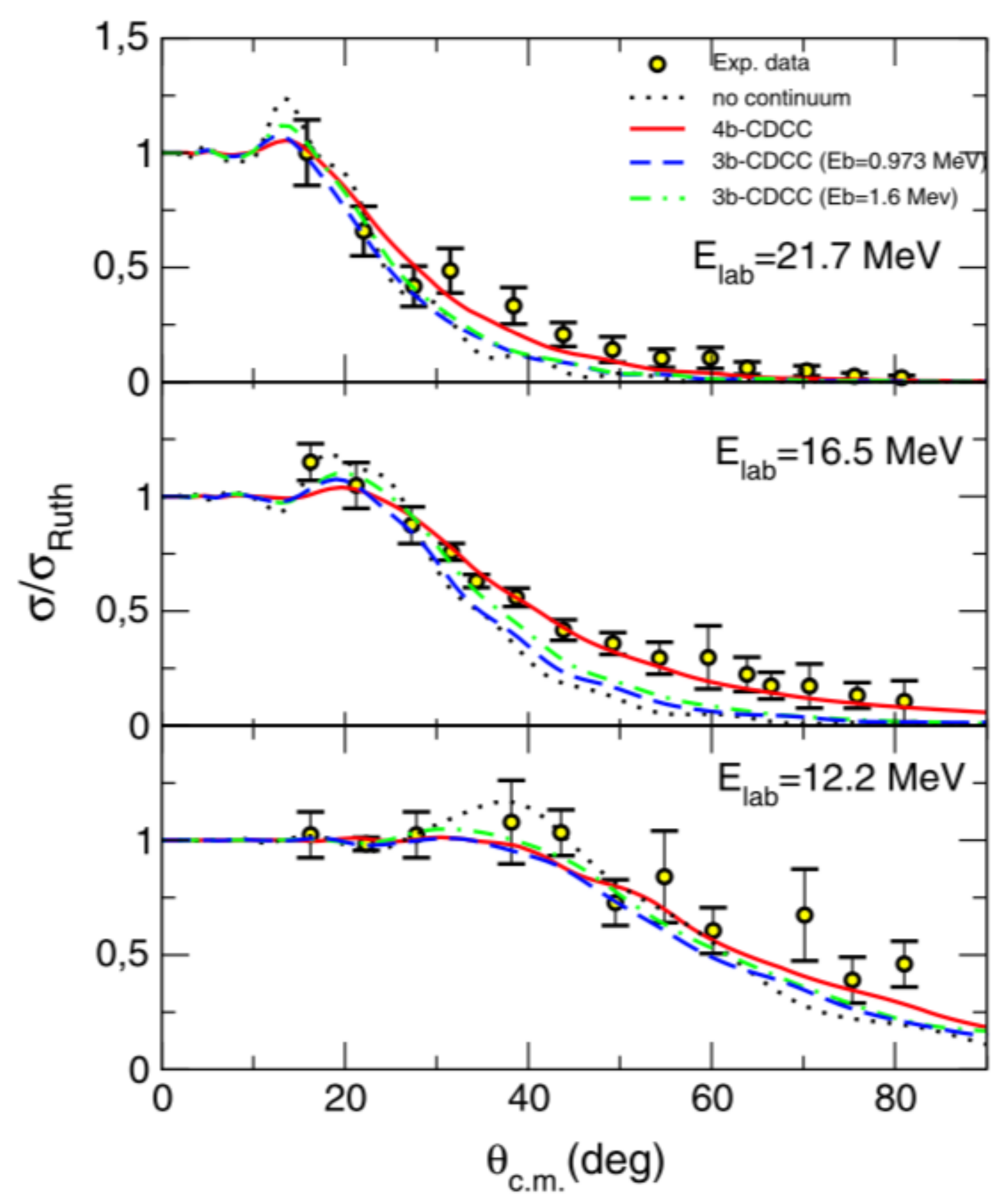}
\end{center}
\caption{(Color on line) Experimental elastic angular distribution in the $^{6}$He + $^{58}$Ni collision at three energies, in comparison with predictions
of 3-body and 4-body CDCC calculations~\cite{MPR14}.}
\label{F3b-4b}
\end{figure}
Morcelle  {\it et al.}~\cite{MPR14} measured elastic angular distributions in collisions of $^6$He projectiles with a $^{58}$Ni target, at three near-barrier energies.
The results are shown in Fig.~\ref{F3b-4b}, in comparison with predictions of 3b- and 4b-CDCC calculations. Clearly, the predictions of 4b-CDCC are much closer
to the data than those of 3b-CDCC, mainly at $E_{\rm lab} = 16.5$ MeV.

\medskip

Descouvemont {\it et al.}~\cite{DDC15} performed four-body CDCC calculations for the $^9{\rm Be}+^{209}{\rm Bi}$ system. Elastic scattering,
breakup and total fusion cross sections were evaluated simultaneously. The results were shown to be in good agreement with the data.

\medskip

Fern\'andez-Garc\'\i a {\it et al.}~\cite{FPF19} measured elastic angular distributions and cross sections for the production of $^{4}{\rm He}$,
in the $^{6}{\rm He}-^{64}$Zn collision. They performed coupled reaction channel (CRC), 3b-CDCC and 4b-CDCC calculations, and compared the 
resulting cross sections with the data. They found that the elastic cross sections of 3b-CDCC (using the di-neutron model) and 4b-CDCC are very 
similar, and close to the data. On the other hand, the contribution from elastic breakup to the $^{4}{\rm He}$ production cross section is very 
small. Although CRC calculations of 2n transfer gave a reasonable description of the data, inclusive cross section obtained with the Ichimura, Austern and 
Vincent (IAV) model~\cite{IAV85,AIK87} are closer to the experiment.

\medskip

A very nice example where 4b-CDCC reproduces the data much better than 3b-CDCC is shown in Fig.~\ref{Moro-private com}. Although the two CDCC calculations
reproduce equally well the data at backward angles, the 4b-CDCC cross section remains close to it at forward angles, whereas the 3b-CDCC cross section 
falls significantly below.
\begin{figure}
\begin{center}
\includegraphics*[width=8cm]{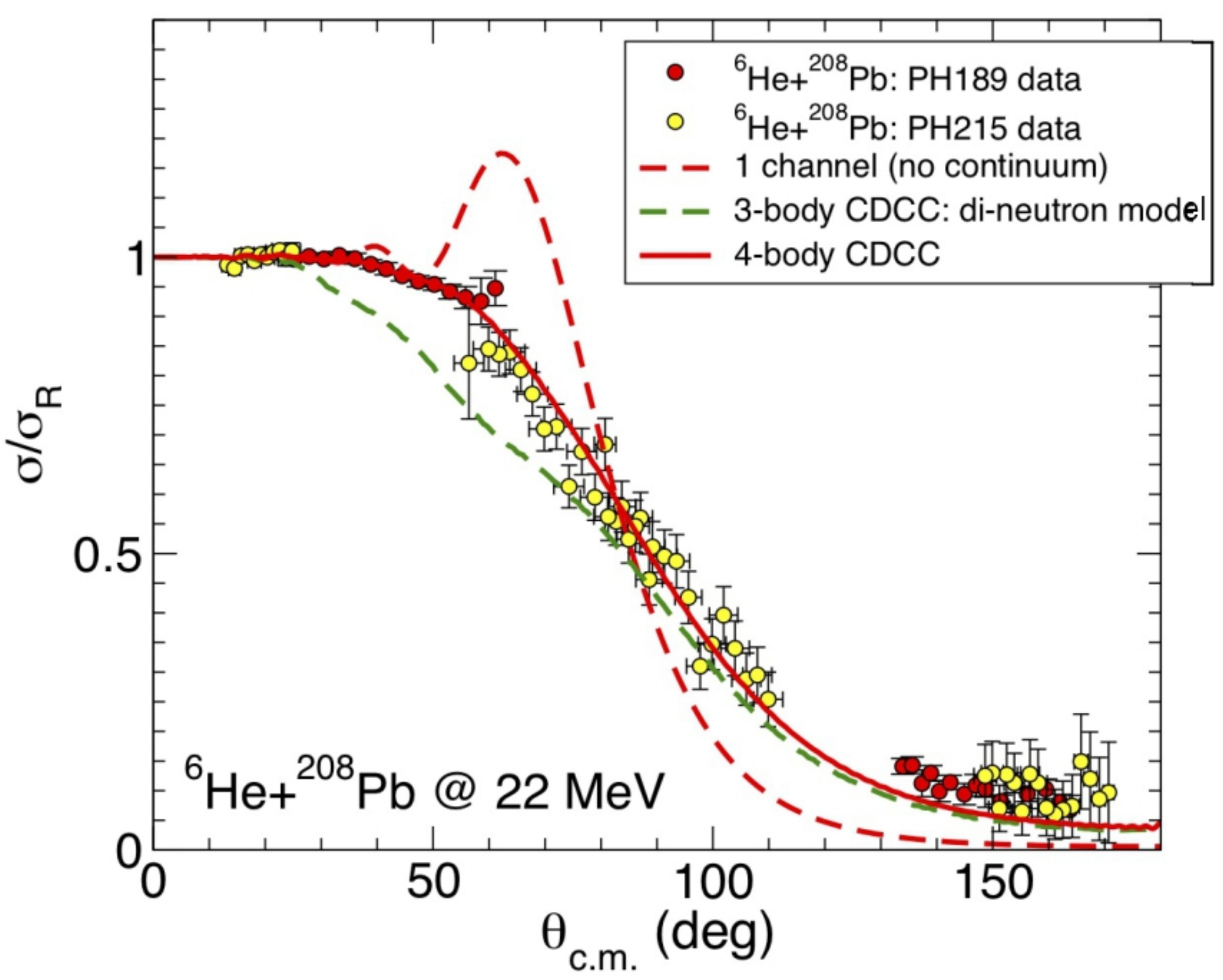}
\end{center}
\caption{(Color on line) Experimental angular distribution in $^6{\rm He}\, +\, ^{208}$Pb elastic scattering, in comparison with predictions of theoretical models. The data 
denoted by PH189 and PH215 are respectively from Refs.~\cite{SEA08} and \cite{ASG11}, the 1-channel and the 3b-CDCC calculations are from Ref.~\cite{MRA07},
and the 4b-CDCC calculations are from Ref.~\cite{RAG09} (figure taken from a private communication with A. Moro).}
\label{Moro-private com}
\end{figure}

 \subsubsection{Other generalizaions of the CDCC}
 
Recently, Pierre Descouvemont introduced two generalizations of the CDCC method. The first~\cite{DeI18} is to use microscopic wave functions in a multi-cluster model 
($\alpha + \alpha +n$) for the bound states of $^9$Be. This treatment has the nice feature of being based exclusively on nucleon-target interactions.  The model was 
applied to the $^9{\rm Be}\,+\,^{208}{\rm Pb}$ and  $^9{\rm Be}\,+\,^{27}{\rm Al}$ systems, and the results were shown to be in fair agreement with the data.\\

The second generalization~\cite{Des17,Des18} is an extension of the 4b-CDCC approach to deal with collisions between two weakly bound nuclei,
when each one can break up into two fragments.
The new reaction model was applied to the $^{11}{\rm Be}-d$ collision at $E_{\rm c.m.}=45.5$ MeV. It was shown that for a good description of the elastic
scattering data it is necessary to consider continuum states of the two collision partners simultaneously.  



\section{Hybrid reactions: The Surrogate Method}


With the advent of secondary beams of unstable nuclei one is bound to deal with reaction cross sections of only a piece of the projectile. Further, some of the desired reactions can not be 
measured in the laboratory even in the case of stable projectiles. Thus one has to find a way to extract the desired cross section from a measurement of the spectrum of the observed piece 
of the primary projectile. An example, is deuteron-induced reaction of the type $(d,p)$ where the neutron is captured by a target such as $^{238}$U or $^{232}$Th of importance
for nuclear energy generation in fast breeder reactions. The measured proton spectrum can then be used to extract cross sections for neutron capture reactions, like: $n + ^{238}{\rm U} \rightarrow\, 
^{239} {\rm U}$ or $n + ^{232}{\rm Th} \rightarrow\, ^{233}$Th. The method used to extract the desired neutron capture cross sections is referred to as the {\it Surrogate Method} (SM). Even in cases 
where the  primary projectile is a weakly bound two-cluster projectile such as $^6{\rm Li} =\, ^4{\rm He} + d$, the measurement of the deuteron spectrum will supply information on the alpha 
capture by the target. This represents the incomplete fusion of the projectile, which when added to the complete fusion supplies the total fusion. Therefore, as part of this review it is important 
to give an account of the theory which supplies the expression of the spectrum of the detected fragment, and exhibit how this spectrum is directly proportional to the secondary reaction cross
section (the desired cross section). The primary reaction cross section of the full projectile is of course extracted as discussed in the previous section, and relies on the careful measurement 
of the angular distribution of the elastic scattering. \\

The theory that we summarise below is the {\it inclusive nonelastic breakup theory} (NEB). Let us first deal with the reaction $A\, (d,p)\,B$, where $B = A + n$. 
The NEB cross section for the emerging proton at an angle $\theta_p$ and with energy $E_p$ is
\begin{equation}
\frac{d^{2}\sigma^{\scr NEB}_{p}}{dE_{p}\,d\Omega_p} = \rho_{p}(E_p)\ \bar{\sigma}_{\scr R}^{{\rm n}{\scr A}},
\end{equation}
where $\rho_{p}(E_p)$ is the density of proton states and $\bar{\sigma}_{{\scr R}}^{{\rm n}{\scr A}}$ is the medium-modified total reaction cross section in the collision between the 
neutron and the target.

\bigskip

The Surrogate Method (SM) purports to extract $\bar{\sigma}_{\scr R}^{n}$ through a measurement of $d^{2}\sigma_{p} / dE_{p}\,d\Omega_p$. 
In fact, what is done is a measurement of the protons in coincidence with one decay product of the secondary compound nucleus 
$B= A+n$~\cite{Hus17}. Several publications on the SM have recently appeared~\cite{EBD12}. Very recently this method was employed to populate the 
compound nucleus involving radioactive targets \cite{PPC17}.


\subsection{The inclusive nonelastic breakup cross section}\label{Sur EBU}


In the more general case of a two-cluster primary projectile, $a = b + x$, the NEB theory gives for the $a + A \rightarrow b + B$ (with $B = x + A$) reaction cross section,
\begin{equation}
\frac{d^{2}\sigma^{\scr NEB}_b}{dE_{b}\,d\Omega_{b}} = \rho_{b}(E_b)\ \bar{\sigma}_{\scr R}^{{\sc x}{\scr A}}  \ , 
\label{inebu1}
\end{equation}
where $\bar{\sigma}_{\scr R}^{{\sc x}{\scr A}}$ is the total reaction cross section of the interacting fragment, $x$, on the target and 
\begin{equation}
\rho_{b}(E_b)  \equiv \frac{1}{(2\pi)^3} \ \frac{d^3{\bf k}_{b}}{dE_{b}\,d\Omega_{b}}  = \frac{\mu_{b}\,k_{b}}{8 \pi^3\,\hbar^2}
\end{equation} 
is the density of states of the observed fragment, $b$. We consider the situation where $a$ is much lighter than $A$, so that one can assume the target
has infinite mass. \\

In this section we discuss the calculation of the NEB cross section within different participant-spectator models, following the work of Ichimura~\cite{Ich90} (a recent
review on this topic, can be found in Ref.~\cite{PPC17}). They all lead to an expression of the form
\begin{equation}
\frac{d^{2}\sigma^{\scr NEB}_b}{dE_{b}\,d\Omega_{b}} =-\, \frac{2}{v_a}\, \rho(E_b)\, \left\langle \varphi_{x} \right|\, W_{xA}\, \left|  \varphi_{x} \right\rangle,
\label{gen-abs}
\end{equation} 
where $W_{xA}$ is the imaginary part of an effective potential $U_{xA}$, which will be derived below, and $\varphi_x$ is is the so called source function, which varies according
to the particular implementation of the model.
The starting point is the system Hamiltonian,
\begin{equation}
H = K_b+K_x+h_{\sc A}(\xi)+V_{bx}+V_{xA}+U_{bA},
\label{H-IAV}
\end{equation}
where $K_b$ and $K_x$ are respectively the kinetic energy operators of particles $b$ and $x$, $h_{\sc A}$ is the intrinsic Hamiltonian of the target, $V_{bx}$ and 
$V_{xA}$ are respectively real potentials representing the interactions of $x$ with $b$ and $A$, and $U_{bA}$ is the complex optical potential between 
particle $b$ and the target. The intrinsic structures of $b$ and $x$ are neglected, whereas the target has a set of intrinsic states satisfying the equation,
\begin{equation}
 h_A \,\phi_\alpha (\xi) =  \varepsilon_\alpha  \,\phi_\alpha(\xi).
\label{intrinsic}
\end{equation}

\bigskip

We are interested in final states in the form
\begin{equation}
\left| 
\chi^{\scr (-)}_{{\bf k}_b}\,{\rm \Phi}_\alpha^{\scr (-)} 
 \right> \equiv
\left| \chi^{\scr (-)}_{{\bf k}_b}\right) \otimes \left| {\rm \Phi}_\alpha^{\scr (-)} \right),
\label{FS}
\end{equation}
where $\left| \chi^{\scr (-)}_{{\bf k}_b}\right) $ is a distorted wave with momentum $\hbar {\bf k}_b$ with ingoing wave boundary condition, satisfying the equation
\begin{equation}
\left[  K_b+U^\dagger_{bA} - E_b\right]  \left| \chi^{\scr (-)}_{{\bf k}_b}\right) = 0,
\label{chi_b}
\end{equation}
and $ \left| {\rm \Phi}_\alpha^{\scr (-)} \right)$ is a scattering state of the $x+A$ system with ingoing wave boundary condition, satisfying the equation
\begin{equation}
\left[  H_{xA}- E_\alpha\right]  \left| {\rm \Phi}_\alpha^{\scr (-)} \right) = 0.
\label{Phi_x}
\end{equation}
Above, $E_\alpha = E -E_b$ and
\begin{equation}
H_{xA} = h_{\scr A}+K_x+V_{xA}.
\label{HxA}
\end{equation}
is  the Hamiltonian of the $x+A$ system. Then, the corresponding T-matrix in the post representation is
\begin{equation}
T_{\alpha {\bf k}_b} = \left\langle  \chi^{\scr (-)}_{{\bf k}_b}\,{\rm \Phi}_\alpha^{\scr (-)}  \right| \, V_{bx}\, \left| {\rm \Psi}^{\scr (+)} \right\rangle,
\label{Tmat}
\end{equation}
where $\left| {\rm \Psi}^{\scr (+)} \right\rangle$ is the scattering wave function, which satisfies the Schr\"odinger equation with the full Hamiltonian of Eq.~(\ref{H-IAV}), namely
\begin{equation}
\left[  H - E \right]   \left| {\rm \Psi}^{\scr (+)} \right\rangle= 0.
\label{Psi+}
\end{equation}

\bigskip

The inclusive breakup cross section is given in terms of the T-matrix by the expression (see e.g. Eq.(1.31) of Ref.~\cite{RoT67})
\begin{equation}
\frac{d^{2}\sigma^{\rm inc}_b}{dE_{b}\,d\Omega_{b}} = \frac{2\pi}{v_a}\ \rho_{b}(E_b)\ \sum_\alpha \left| T_{\alpha {\bf k}_b} \right|^2\ 
\delta \left(E-E_b - E_\alpha \right),
\label{inebu2}
\end{equation}
where $v_a$ is the incident velocity of $a$. Using in Eq.~(\ref{inebu2}) the well known identity
\begin{equation}
\frac{1}{x+i\epsilon} = {\mathcal P}\left\{ \frac{1}{x} \right\} - i\pi\, \delta(x),
\label{identity-1}
\end{equation}
with ${\mathcal P}$ standing for the principal value, and replacing $T_{\alpha {\bf k}_b}$ by its explicit form ((Eq.~(\ref{Tmat})), the inclusive breakup cross section becomes
\begin{multline}
\frac{d^{2}\sigma^{\rm inc}_b}{dE_{b}\,d\Omega_{b}} = - \frac{2}{v_a}\ \rho_{b}(E_b)\ {\rm Im} \Big\{ \big\langle {\rm \Psi}^{\scr (+)} \big| \,V_{bx} \big|\chi_{{\bf k}_b}^{\scr (-)} \big) \\ 
\Big[ \sum_\alpha \big( {\rm \Phi}_\alpha^{\scr (-)} \big|\,\frac{1}{E-E_b-E_\alpha +\i \epsilon}
\big| {\rm \Phi}_\alpha^{\scr (-)} \big)\
\Big]\\
\big( \chi_{{\bf k}_b}^{\scr (-)}\big|\,V_{bx}\big| {\rm \Psi}^{\scr (+)} \big\rangle
\Big\}.
\label{inebu3}
\end{multline}
The quantity within square brackets in the above equation is the spectral representation of the Green's function associated with the Hamiltonian of Eq.~(\ref{HxA}),
$G^{\scr (+)}_{xA}(E-E_b)$. Then, we can write
\begin{multline}
\frac{d^{2}\sigma^{\rm inc}_b}{dE_{b}\,d\Omega_{b}} = - \frac{2}{v_a}\ \rho_{b}(E_b)\ {\rm Im} \Big\{ \big\langle {\rm \Psi}^{\scr (+)} \big| \,V_{bx} \big|\chi_{{\bf k}_b}^{\scr (-)} \big) \\ 
G^{\scr (+)}_{xA}(E-E_b)\ \big( \chi_{{\bf k}_b}^{\scr (-)}\big|\,V_{bx}\big| {\rm \Psi}^{\scr (+)} \big\rangle
\Big\}.
\label{inebu4}
\end{multline}

\bigskip

Different approximations have been adopted for the exact wave function $\left| {\rm \Psi} ^{(+)} \right\rangle$. Some of them are discussed below.


\subsubsection{The three-body model}


First, we consider the approximation of $\left| {\rm \Psi} ^{(+)} \right\rangle$ by the three-body wave function of Austern {\it et al.}~\cite{AIK87}, 
\begin{equation}
\big| {\rm \Psi} ^{\scr (+)} \big\rangle \sim \big| {\rm \Psi} ^{\scr (+)}_{3b} \big\rangle = \big| {\rm \psi} ^{\scr (+)}_{3b} \big) \otimes \big|\phi_0 \big),
\label{Psi-3b}
\end{equation}
where $ \left|\phi_0 \right) \equiv  \left|\phi_{\alpha = 0} \right)$ is the ground state of the target. Since the excitations of the target have been
neglected, it is necessary to replace the real interaction $V_{xA}$ by an optical potential, $U_{xA}$. Formally, this potential is the energy averaged
potential of Feshbach's theory~\cite{Fes58,Fes62,LeF73,Fes92}. However, for practical purposes, it is treated phenomenologically. The 
three-body wave function is then the solution of the Schr\"odinger equation
\begin{equation}
\Big[ K_b+K_x+V_{bx}+U_{xA}+U_{bA}  \Big]\,\big| {\rm \psi} ^{\scr (+)}_{3b} \big) = 
E\,\big| {\rm \psi} ^{\scr (+)}_{3b} \big).
\label{Sch-3b}
\end{equation}
Inserting Eq.~(\ref{Psi-3b}) into Eq.~(\ref{inebu4}), we get
\begin{multline}
\frac{d^{2}\sigma^{\rm inc}_b}{dE_{b}\,d\Omega_{b}} = - \frac{2}{v_a}\ \rho_{b}(E_b)\ {\rm Im} \Big\{ \big( {\rm \psi} ^{\scr (+)}_{3b} \big| \,V_{bx} \big|\chi_b^{\scr (-)} \big) \\ 
G^{\scr opt}_{xA}\ \big( \chi_b^{\scr (-)}\big|\,V_{bx}\big| {\rm \psi} ^{\scr (+)}_{3b} \big)
\Big\},
\label{inebu5}
\end{multline}
where 
\begin{equation}
G^{\scr opt}_{xA} = \big( \phi_0 \big|\
G^{\scr (+)}_{xA}(E-E_b)
\ \big| \phi_0 \big) .
\label{Gopt}
\end{equation}

\bigskip

Now, we split the inclusive BU cross section into its elastic (EBU) and inelastic (NBU) components. For this purpose, we use the identity~\cite{UdT81,KaI82}
\begin{multline}
{\rm Im}\left\{
G^{\scr opt}_{xA}\right\} = -\pi \int \left| \chi^{\scr (-)}_{{\bf k}_x} \right)
\delta\left(E-E_b-\frac{\hbar^2k_x^2}{2m_x}\right) \\
\times\left( \chi^{\scr (-)}_{{\bf k}_x} \right| \,d{\bf k}_x^3\ \ 
 + G^{{\scr opt}\dagger}_{xA}\, W_{xA}\, G^{{\scr opt}\dagger}_{xA} ,
 \label{ident-green}
\end{multline}
where $W_{xA}$ is the imaginary part of the optical potential $U_{xA}$. 
Inserting Eq.~(\ref{ident-green}) into Eq.~(\ref{inebu5}), the inclusive BU cross section can be put in the form
\begin{equation}
\frac{d^{2}\sigma^{\scr inc}_b}{dE_{b}\,d\Omega_{b}} = \frac{d^{2}\sigma^{\scr EBU}_b}{dE_{b}\,d\Omega_{b}} \, +\, 
\frac{d^{2}\sigma^{\scr NBU}_b}{dE_{b}\,d\Omega_{b}},
\label{EBU-NBU}
\end{equation}
where  
\begin{multline}
\frac{d^{2}\sigma^{\scr EBU}_b}{dE_{b}\,d\Omega_{b}} = \frac{2\pi}{v_a}\ \rho_{b}(E_b)\ \int \Big| \big( {\rm \psi} ^{\scr (+)}_{3b}\big| \,V_{bx} 
\big|\chi_{{\bf k}_b}^{\scr (-)} 
\chi^{\scr (-)}_{{\bf k}_x} \big) \Big|^2 \\
\delta\left(E-E_b-\frac{\hbar^2k_x^2}{2m_x}\right) \ d{\bf k}_x^3\ 
\label{sig-EBU}
\end{multline}
is identified with the inclusive elastic breakup cross section. The remaining part, which corresponds to the inclusive nonelastic breakup cross section, 
can be put in the general form of Eq.~(\ref{inebu1}), namely
\begin{equation}
\left[ \frac{d^{2}\sigma^{\scr NBU}_b}{dE_{b}\,d\Omega_{b}} \right]_{\scr 3b}= -\,\frac{2}{v_a}\ \rho_{b}(E_b)\  \big( \varphi^{\scr 3b}_x\,\big| \,W_{xA} \,\big| \, \varphi_x^{\scr 3b}
\,\big),
\label{sig-NBU}
\end{equation}
with
\begin{equation}
\big| \, \varphi_x^{\scr 3b}\,\big\rangle = G^{\scr opt}_{xA}\ \big( \chi_b^{\scr (-)}\big|\,V_{bx}\big| {\rm \psi}^{\scr (+)}_{3b} \big\rangle.
\label{varphi_3b}
\end{equation}

\medskip

Eq.~(\ref{varphi_3b}) can be simplified if one uses the identity
\begin{multline}
V_{bx} = \big[ V_{bx}+K_b+K_x+U_{xA}+U_{bA} \big]\\
-  \big[ K_b+K_x+U_{xA}+U_{bA} \big]
\end{multline}
in Eq.~(\ref{sig-NBU}), and take into account Eqs.~(\ref{chi_b}) and (\ref{Sch-3b}). One gets,
\begin{equation}
\big| \, \varphi_x^{\scr 3b}\,\big\rangle = \big( \chi_b^{\scr (-)}\big|{\rm \psi}^{\scr (+)}_{3b} \big\rangle.
\label{varphi_3b-1}
\end{equation}


\subsubsection{The IAV, the Hussein-McVoy and the Udagawa-Tamura formulae}


Now we consider the model of Ichimura, Austern and Vincent~\cite{IAV85,AuV81,KaI82}. These authors use the post representation but adopts the DWBA approximation. 
The exact wave function is replaced by
\begin{equation}
\big| {\rm \Psi}^{\scr (+)} \big\rangle \simeq  \big| {\rm \Psi}_{\scr IAV}^{\scr (+)} \big\rangle =  \big| \chi^{\scr (+)}_a\, \psi_a \big) \otimes \big| \phi_0 \big),
\label{Psi-IAV}
\end{equation} 
where $\psi_a$ is the ground state of the incident projectile and $\chi^{\scr (+)}_a$ is its distorted wave. They satisfy the equation,
\begin{equation}
\Big[ K_b+K_x+V_{bx}+U_{xA}+U_{bA}  \Big]\,\big| \chi^{\scr (+)}_a\, \psi_a \big) = 
E\,\big| \chi^{\scr (+)}_a\, \psi_a \big).
\label{Sch-IAV}
\end{equation}

To derive the NBU cross section one follows the same procedures as in the previous section, but replacing $\left| {\rm \Psi} ^{(+)}_{3b} \right\rangle$ by
$\big| {\rm \Psi}^{\scr (+)}_{\scr IAV} \big\rangle$. Then, Eq.~(\ref{inebu5}) becomes
\begin{multline}
\frac{d^{2}\sigma^{\rm inc}_b}{dE_{b}\,d\Omega_{b}} = - \frac{2}{v_a}\ \rho_{b}(E_b)\ {\rm Im} \Big\{ \big( \chi^{\scr (+)}_a\, \psi_a \big| \,V_{bx} \big|\chi_b^{\scr (-)} \big) \\ 
G^{\scr opt}_{xA}\ \big( \chi_b^{\scr (-)}\big|\,V_{bx}\big| \chi^{\scr (+)}_a\, \psi_a \big)
\Big\},
\label{inebu6}
\end{multline}

Next, we use Eq.~(\ref{ident-green}) to split the cross section into its EBU and NBU components, getting the DWBA version of Eq.~(\ref{EBU-NBU}).
The EBU cross section is given by Eq.~(\ref{sig-EBU}), with the replacement: $ {\rm \psi} ^{\scr (+)}_{3b}  \rightarrow  \chi^{\scr (+)}_a\, \psi_a$. The
NBU component reads
\begin{equation}
\left[ \frac{d^{2}\sigma^{\scr NBU}_b}{dE_{b}\,d\Omega_{b}} \right]_{\scr IAV} = -\,\frac{2}{v_a}\ \rho_{b}(E_b)\  \big( \varphi^{\scr IAV}_x\,\big| \,W_{xA} \,\big| \, \varphi_x^{\scr IAV}
\,\big),
\label{sig-NBU-IAV}
\end{equation}
with the source function
\begin{equation}
\big| \, \varphi_x^{\scr IAV}\,\big\rangle = G^{\scr opt}_{xA}\ \big( \chi_b^{\scr (-)}\big|\,V_{bx}\big| \chi^{\scr (+)}_a\, \psi_a \big\rangle.
\label{varphi_IAV}
\end{equation}

\bigskip

The Hussein-McVoy formula~\cite{HuM85} adopts also  the post representation and the approximation of Eq.~(\ref{Psi-IAV}) for the three-body wave function. However, the
source function is generated using this approximation in Eq.~(\ref{varphi_3b-1}) instead of in Eq.~(\ref{varphi_3b}). In this way, one gets the NEB cross section
\begin{equation}
\left[ \frac{d^{2}\sigma^{\scr NBU}_b}{dE_{b}\,d\Omega_{b}} \right]_{\scr HM} = -\,\frac{2}{v_a}\ \rho_{b}(E_b)\  \big( \varphi^{\scr HM}_x\,\big| \,W_{xA} \,\big| \, \varphi_x^{\scr HM}
\,\big),
\label{sig-NBU-HM}
\end{equation}
with the source function
\begin{equation}
\big| \, \varphi_x^{\scr HM}\,\big\rangle = \big( \chi_b^{\scr (-)}\big| \chi^{\scr (+)}_a\, \psi_a \big\rangle.
\label{varphi_HM}
\end{equation}

\bigskip

The IAV and the HM formulae are obtained using the same approximation for the three-body wave function in different equations for the source function. However, although 
these equations are formally equivalent when the three-body wave function is used, the IAV formula is more convenient for numerical 
calculations~\cite{Ich90}. The reason is that, owing to the short-range of the $V_{bx}$ interaction, one does not need a good approximation 
for the three-body wave function at large distances between particles $b$ and $x$. The calculation becomes considerably 
simpler if one adopts the zero-range approximation for $V_{bx}$~\cite{CCS17}. The accuracy of this approximation has been studied by Lei and 
Moro~\cite{LeM15a} and by Potel {\it et al.}~\cite{PPC17}. The situation is different in the HM formula, where the scalar product of 
Eq.~(\ref{varphi_HM}) is not constrained to small values of $r_{bx}$.\\

The Udagawa-Tamura~\cite{UdT81,LUT84} formula adopts the prior representation. Then, the T-matrix of Eq.~(\ref{Tmat}) will involve the coupling interaction  $U_{xa}+U_{bA}-U_{aA}$, instead of $V_{bx}$,
and the NEB cross section becomes
\begin{equation}
\left[ \frac{d^{2}\sigma^{\scr NBU}_b}{dE_{b}\,d\Omega_{b}} \right]_{\scr UT} = -\,\frac{2}{v_a}\ \rho_{b}(E_b)\  \big( \varphi^{\scr UT}_x\,\big| \,W_{xA} \,\big| \, \varphi_x^{\scr UT}
\,\big),
\label{sig-NBU-UT}
\end{equation}
with the source function
\begin{equation}
\big| \, \varphi_x^{\scr UT}\,\big\rangle = G^{\scr opt}_{xA}\ \big( \chi_b^{\scr (-)}\big|\,U_{xa}\,+\,U_{bA} \,-\,U_{aA}\big| \chi^{\scr (+)}_a\, \psi_a \big\rangle.
\label{varphi_UT}
\end{equation}

\bigskip

The relation between the IAV, the UT and the HM versions of the participant-spectator model has been discussed in several 
papers~\cite{Ich90,PPC17,LeM15a,LeM15b,PNT15}.
It has been shown that within  DWBA the IAV, the HM and the UT source functions satisfy the relation~\cite{HFM90,LUT84},
\begin{equation}
\big| \, \varphi_x^{\scr IAV}\,\big\rangle = \big| \, \varphi_x^{\scr UT}\,\big\rangle\,+\,\big| \, \varphi_x^{\scr HM}\,\big\rangle.
\label{source_IAV-HM-UT}
\end{equation}
Then, inserting the above equation into Eq.~(\ref{sig-NBU-IAV}), one gets
\begin{multline}
\left[ \frac{d^{2}\sigma^{\scr NEB}_b}{dE_{b}\,d\Omega_{b}} \right]_{\scr IAV} = \left[ \frac{d^{2}\sigma^{\scr NEB}_b}{dE_{b}\,d\Omega_{b}} 
\right]_{\scr UT} + \left[ \frac{d^{2}\sigma^{\scr NEB}_b}{dE_{b}\,d\Omega_{b}} \right]_{\scr HM}\\
+ \left[ \frac{d^{2}\sigma^{\scr NEB}_b}{dE_{b}\,d\Omega_{b}} \right]_{\rm int},
\label{sig_IAV-HM-UT}
\end{multline}
where
\begin{equation}
\left[ \frac{d^{2}\sigma^{\scr NBU}_b}{dE_{b}\,d\Omega_{b}} \right]_{\rm int} = -\,\frac{4}{v_a}\ \rho_{b}(E_b)\  {\rm Re}
\Bigg\{ \big( \varphi^{\scr IAV}_x\,\big| \,W_{xA} \,\big| \, \varphi_x^{\scr UT} \,\big) \Bigg\}
\label{sig-NBU-int}
\end{equation}
is the interference term.\\

\bigskip

\centerline{ \noindent {\it IAV vs. UT models}}

\medskip

There is some controversy about which model is more suitable to describe inclusive breakup experiments. Udagawa and Tamura~\cite{LUT84} argued that the last two terms
on the RHS of Eq.~(\ref{sig_IAV-HM-UT}) are {\it unphysical}, since they arise from a non-orthogonality component of the wave function. Later, Mastroleo, Udagawa,
and Tamura~\cite{MUT89} performed numerical calculations for $^{58,62}{\rm Ni} (\alpha,X)$ reactions, using both the UT and the IAV models. They showed that 
the UT model predicts accurately the experimental cross section of Kleinfeller {\it et al.}~\cite{KBE81}, whereas the cross section obtained with the IAV model 
overestimated the data.\\

\begin{figure}
\begin{center}
\includegraphics*[width=8cm]{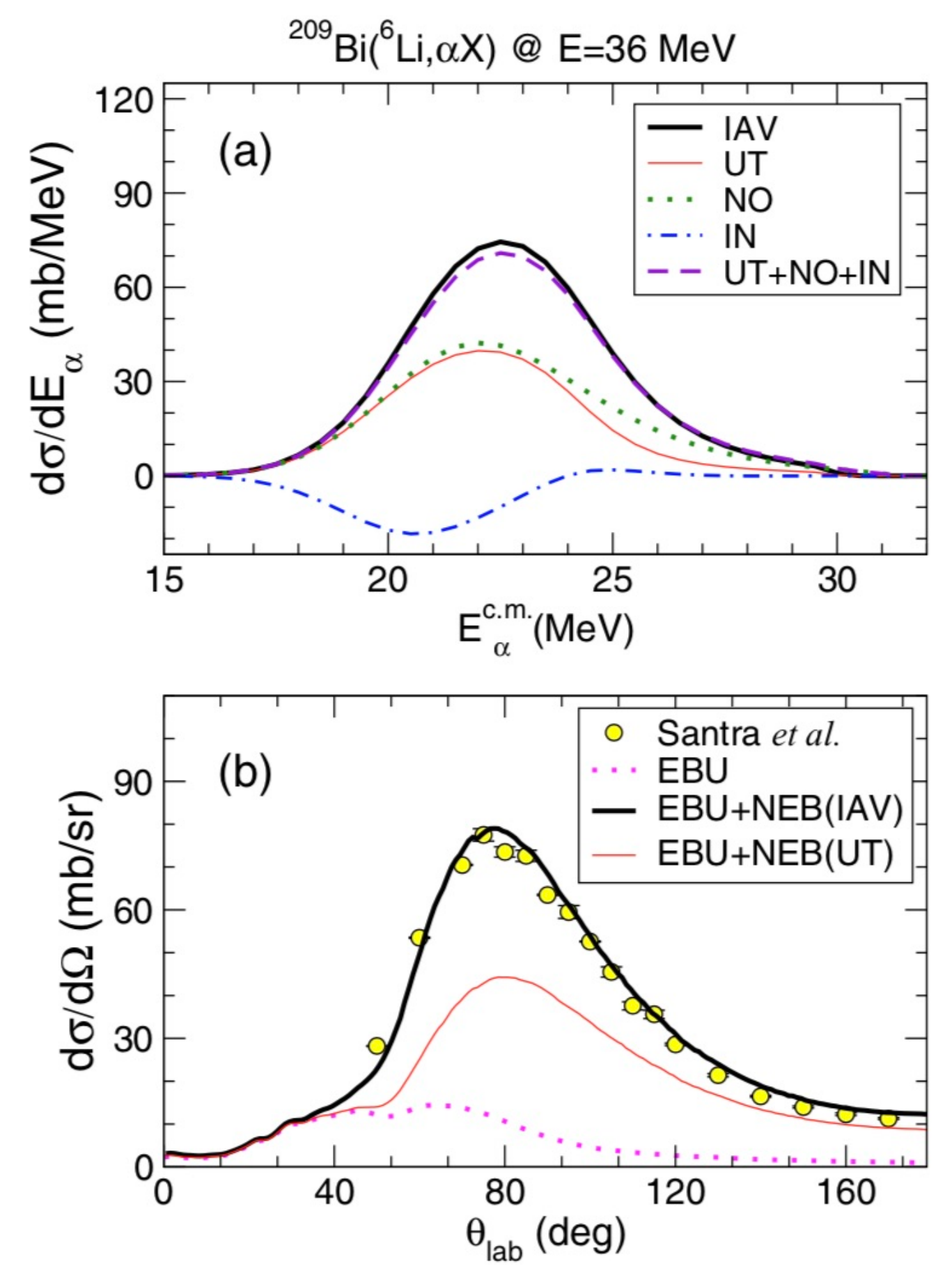} 
\end{center}
\caption{(Color on line) Panel (a): the IAV, UT, HM and IN cross sections; Panel (b): The inclusive BU cross sections predicted by the IAV and UT models, in comparison with 
data Ref.~\cite{SKR11}. See the text for details.}
\label{Fig13}
\end{figure}
On the other hand, more recently, Lei and Moro~\cite{LeM15b} studied  the accuracy of the IAV and the UT models, performing calculations of the $^{209}{\rm Bi} (^6{\rm Li}, \alpha X)$ 
reaction at $E = 36$ MeV. They obtained the EBU cross section by a CDCC calculation and evaluated the NBU cross section using both the UT and the IAV models. 
The IAV, UT, HM (denoted by NO) and the interference (IN) cross sections of Eq.~(\ref{sig_IAV-HM-UT}) are shown in panel (a) of Fig.~\ref{Fig13}.
The sum of the UT, HM and IN terms is represented by the long-dashed line. Note that this curve is very close to the one representing the IAV cross section, as predicted by
Eq.~(\ref{sig_IAV-HM-UT}).

\smallskip

Summing the NBE cross section of the IAV and UT models with the EBU cross section, Lei and Moro obtained the inclusive breakup cross section predicted by the
two models. The results are shown in panel (b) of Fig.~\ref{Fig13}, in comparison with the data of Santra {\it et al.}~\cite{SKR11}. Whereas the predictions of the IAV
model reproduces very well the experimental cross section, the results of the UT model fall well below the data.


\subsection{Applications}


The surrogate method allows evaluation of cross sections that cannot be directly measured, such as neutron capture by short-lived targets. In this section we illustrate the use of the 
SM in two examples.  


\subsubsection{Neutron capture by short-lived targets}

Neutron capture reactions are of great interest in nuclear physics, astrophysics and in theoretical models for generation of energy. In different 
situations, the direct measurement of these cross sections are very difficult to perform. One example is radioactive capture of neutrons by a 
short-lived target (more details on this topic can be found in Ref.~\cite{PPC17}). We will discuss below the
indirect measurement of the $(n,\gamma)$ cross section for the short-lived $^{87}{\rm Y}$ nucleus ($t_{\scr 1/2} = 79.8$ h), for which there are no data available. 
According to the Hauser-Feshbach theory~\cite{HaF52}, this cross section can be written as,
\begin{equation}
\sigma_{n\gamma} (E_n) = \sum_{J,\pi} \sigma_{\scr CN} \left( E_{\rm ex},J^\pi \right)\ G_\gamma\left( E_{\rm ex},J^\pi \right),
\label{HF}
\end{equation}
where $ \sigma_{\scr CN} \left( E_{\rm ex},J^\pi \right)$ is the formation probability of the $^{88}{\rm Y}^*$ compound nucleus with excitation energy
$E_{\rm ex}$, angular momentum $J$ and parity $\pi$, and $G_\gamma\left( E_{\rm ex},J^\pi \right)$ is the probability that the CN decays emitting one or
more $\gamma$-rays. The excitation energy is related to the energy of the incident neutron by the equation, $E_n = (1+1/A)(E_{\rm ex}-S_n)$, where
$S_n$ is the energy needed to remove the neutron from $^{88}$Y. The CN formation probability can be obtained theoretically, using a properly chosen
neutron-nucleus complex potential in a standard potential scattering calculation. The decay probability $G_\gamma$, which is very hard to calculate, can
be obtained by the surrogate method from the $^{89}{\rm Y}\,(p,d)\, ^{88}{\rm Y}^*$ reaction~\cite{EBH18}, through coincidence measurements
of the deuteron with the characteristic $\gamma$-rays emitted by $^{88}{\rm Y}^*$. The coincidence probability can be written as,
\begin{equation}
P(E_{\rm ex},\theta_d) = \sum_{J,\pi} F_{\scr CN} \left( E_{\rm ex},J^\pi,\theta_d \right)\ G_\gamma\left( E_{\rm ex},J^\pi \right),
\label{HF-1}
\end{equation}
where  $F_{\scr CN} \left( E_{\rm ex},J^\pi,\theta_d \right)$ is the formation probability of a $^{88}{\rm Y}$ CN with quantum numbers $\{E_{\rm ex},J^\pi\}$ in the $(p,d)$ reaction,
with the $d$ being emitted at an angle $\theta_d$, and $G_\gamma\left( E_{\rm ex},J^\pi \right)$ is the $\gamma$-decay probability that appears in Eq.~({\ref{HF}). The former
can be evaluated by standard nuclear reaction techniques. Then, $G_\gamma\left( E_{\rm ex},J^\pi \right)$ is expressed in terms of parameters, which were determined by fitting
the experimentally determined $P(E_{\rm ex},\theta_d)$ probability.  The cross section $\sigma_{n\gamma} (E_n)$ was then determined by inserting the resulting decay probability 
into Eq.~(\ref{HF}) and  using the calculated CN formation probability. The results are shown in Fig.~\ref{Fig14}, together with the TENDL 2015 (brown curves, with hatched 1 $\sigma$ 
uncertainty) and the Rosfond 2010 evaluations which are based on regional systematics (Nikolaev~\cite{nndc}, Koning~\cite{Kon15}).
\begin{figure}
\begin{center}
\includegraphics*[width=9cm]{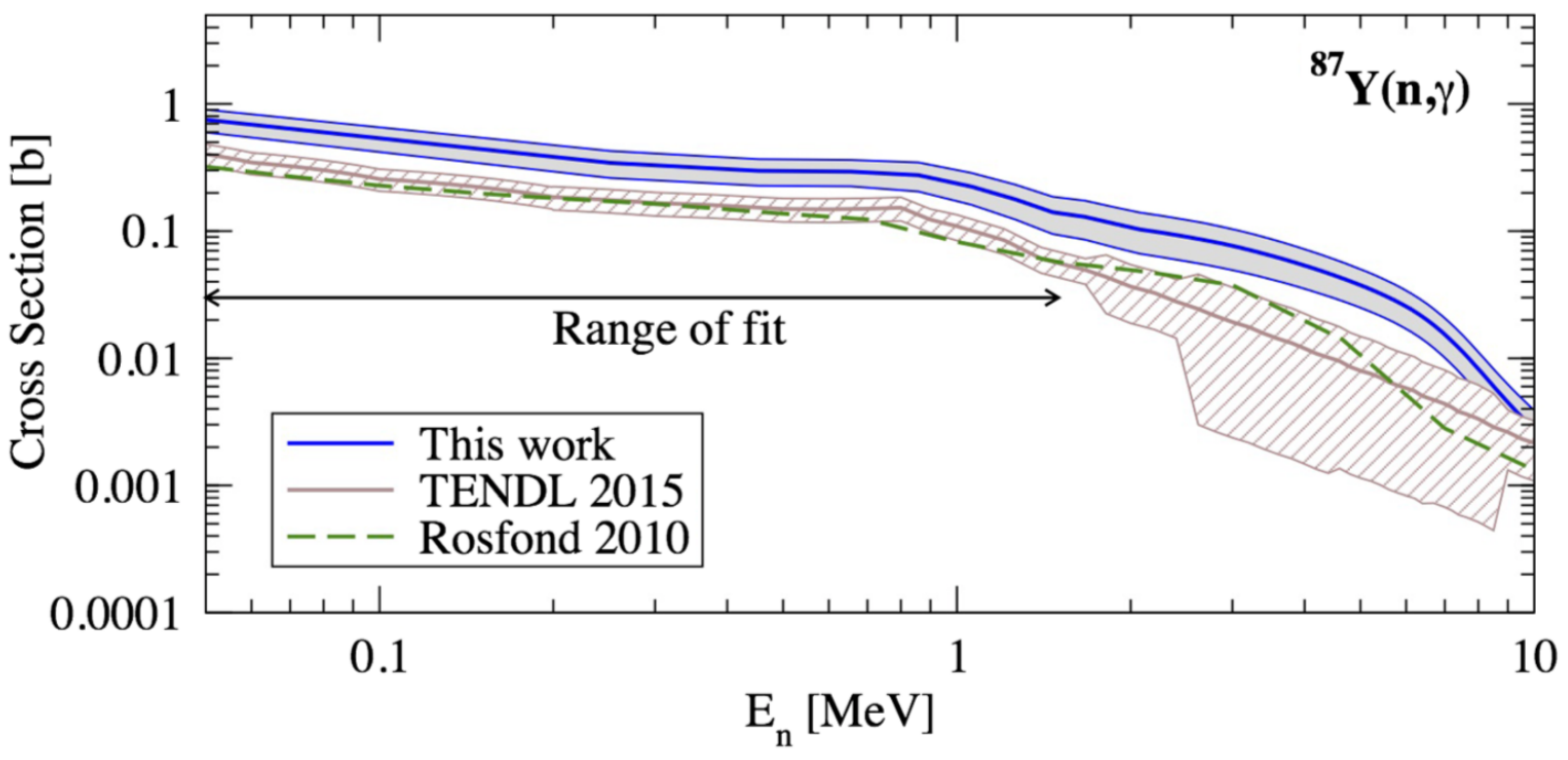} 
\end{center}
\caption{(Color on line) The $\sigma_{n\gamma} (E_n)$ cross section in the $p\,+\,^{89}{\rm Y} \rightarrow d\, +\, ^{88}{\rm Y}^*$ reaction, obtained by
the SM (blue solid curve), with 1 $\sigma$ uncertainty (gray band). The figure was taken from Ref.~\cite{EBH18}. For details see the text. }
\label{Fig14}
\end{figure}

\medskip

The validity of the SM could not be checked in the above discussed reaction because there is no direct measurement of $\sigma_{n\gamma} (E_n)$. However, 
Escher {\it et al.}~\cite{EBH18} assessed the accuracy of their surrogate approach using it for another target in the same Zr-Y region. They evaluated the $^{90}{\rm Zr}(n,\gamma)$ 
cross section through the surrogate reaction $^{92}{\rm Zr}\, (p,d)\, ^{91}{\rm Zr}^*$. The agreement was reasonable. The surrogate cross section and the one measured
directly exhibited similar energy dependences and the same order of magnitude.


\subsubsection{Complete fusion in collisions of weakly bound nuclei}


Determining complete fusion (CF) and incomplete fusion (ICF) cross sections in collisions of weakly bound nuclei is a great challenge both for experimentalists
and theoreticians~\cite{CGD15,CGD06,JPK20}. Owing to the influence of the breakup channel, the CF cross section is reduced in comparison to predictions of
barrier penetration models. In the $^ {6,7}$Li + $^{209}$Bi and $^9$Be + $^{208}$Pb collisions, for example, the reduction is $\sim 20-35$ \%. Recently, Lei and 
Moro evaluated the CF cross section using the inclusive breakup model~\cite{LeM19}. For this purpose, they wrote the total reaction cross section as,
\begin{equation}
\sigma_{\scr R} = \sigma_{\rm inel} + \sigma_{\scr EBU} + \sigma_{\scr NEB}^{\scr (b)} + \sigma_{\scr NEB}^{\scr (x)}+\sigma_{\scr CF},
\label{CF LeM}
\end{equation}
where $\sigma_{\scr R}, \sigma_{\rm inel}, \sigma_{\scr EBU}, \sigma_{\scr NEB}^{\scr (b)},\sigma_{\scr NEB}^{\scr (x)} $ and $\sigma_{\scr CF}$ are respectively the total 
reaction, inelastic scattering, elastic breakup, nonelastic breakup of fragments (b) and (x), and complete fusion cross sections. These cross sections were determined as follows.
The $\sigma_{\scr R}$, $\sigma_{\scr EBU}$ and $\sigma_{\rm inel}$ cross sections were obtained from standard theoretical approaches (CC or CDCC calculations),  and
$\sigma_{\scr NEB}^{\scr (b,x)}$ were obtained from inclusive breakup model calculations as described in section~\ref{Sur EBU}. The remaining cross section, 
$\sigma_{\scr CF}$ was extracted from Eq.~(\ref{CF LeM}).\\

The method described above was used to study CF in the $^{6,7}$Li + $^{209}$Bi collisions and the theoretical cross sections were compared with the data of 
Dasgupta {\it et. al}~\cite{DHH02,DHB99,DGH04}. The comparison is reproduced in Fig.~\ref{Fig15}.
\begin{figure}
\begin{center}
\includegraphics*[width=7cm]{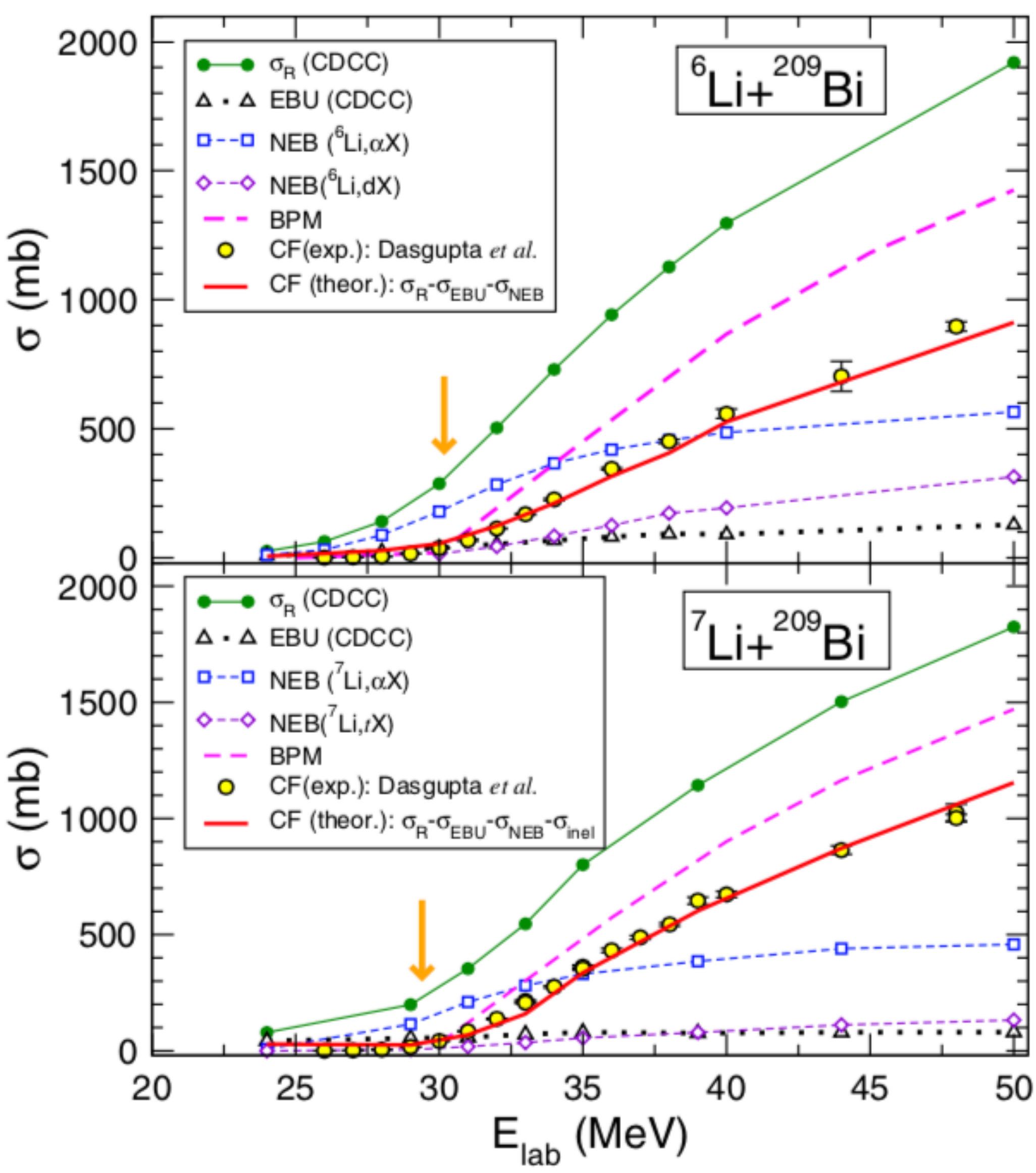}
\end{center}
\caption{(Color on line) The cross sections of Eq.~(\ref{CF LeM}) together with the fusion cross section of barrier penetration models and the CF data. The figure was taken
from Ref.~\cite{LeM19} and the data are from Refs.~\cite{DHH02,DGH04}. }
\label{Fig15}
\end{figure}
Clearly, the surrogate approach of Lei and Moro~\cite{LeM19} can describe the CF cross section very well, at leat for collision energies above the Coulomb barrier.  On the other hand,
it is unable to predict ICF cross section individually. This cross section is included in the NBE cross section, together with that for target excitations.

\vskip 1cm

\section{The total reaction cross section data}


The total reaction cross section, $\sigma_{\scr R}$, is a very important quantity in collisions of heavy nuclei, since it contains information about the size 
of the collision partners and on the open channels. Although very desirable, determining $\sigma_{\scr R}$ as the sum of  individual 
measurement of each channel is a great challenge for experimentalists. Thus, most of the available total reaction data have been obtained from 
optical model analyses of experimental angular distributions in elastic scattering. In this way, the angular momentum components of the nuclear S-matrix
are determined, and the total reaction cross section is evaluated.\\

In the past two decades or so, with the development and modernization of experimental techniques, several new experiments have been performed with 
the purpose of determining total reaction cross sections, and investigating the reaction mechanisms involved in heavy nuclei collisions. In particular, 
the possibility of using radioactive beams out of short-lived nuclei has offered unique opportunities for research in the frontier field of nuclear physics. Several 
experiments with radioactive nuclei as projectiles have been performed, and previous reviews on these measurements can be found in 
Refs.~\cite{KGA16,CGD15,KAK09}. The results of the analysis of these experiments have provided very interesting information, improving our understanding 
of the interaction mechanism between heavy nuclei and their structures. 
The strong synergy between reaction mechanism and structure of the nuclei involved 
in the collision allows for both static and dynamics effects to be investigated in the experiments related to the elastic scattering and direct measurements. It is 
important to mention that, from its beginning, nuclear physics is a science impelled by experiments. The development of nuclear physics is strongly dependent on 
the continuous advance of experimental techniques and instrumentation. A review on the thousands of 
interesting experiments related to elastic scattering and total reaction cross section would be unpractical. Thus, in this section, we are going to review and 
discuss some basic features of these experiments with a focus on reactions with projectile in the mass range o
$6 < A < 20$ on medium to heavy mass target, and at energies close and above the Coulomb barrier, performed in the last 10 years or so. 
Special attention will be given to experiments with radioactive ion beams.


\subsection{Elastic scattering measurements and total reaction cross section}


As mentioned, total reaction cross sections can be obtained from optical model analyses of elastic angular distributions. Elastic scattering is the simplest process 
which can occur in collisions of two nuclei, since it involves only a few degrees of freedom. At low energy, the cross sections are quite large and a 
phenomenological approach, in which the interaction between the colliding nuclei is represented by an appropriate optical-model potential, is still a very 
reliable and practical method to analyze angular distribution data. Complex potentials are used to fit the experimental angular distribution and with these
potentials the total reaction cross section can be determined. Investigations of elastic scattering is a very fruitful area and thousands of experiments have 
been carried out with very interesting results. In the past, most of the experiments were related to reactions with different combinations of complex 
projectile-target nuclear systems where the projectiles were stable nuclei. Elastic scattering cross sections of a quite large number of heavy-ion systems 
have been measured at several energies and the optical model has been found to be successful in extracting total reaction cross sections, as well as 
interaction radii. A large amount of nuclear reaction data can be found in the EXFOR database~\cite{EXFOR}. A compilation of elastic scattering data can 
also be found in the webpage of the NRV project~\cite{KDN17,ZDK99}.\\

The important requirement in experiments related to elastic scattering is the identification and separation of the scattered particles from the large number of 
reaction products which can be produced in collisions between two heavy nuclei. Although it might seem to be a simple requirement, it can be more challenging 
for experiments with radioactive ion beam, which, sometimes, is produced as a cocktail of beams. To fulfil this requirement, a detection system 
with good energy, mass, and charge resolution, as well as a large dynamic range, is necessary. Over these several years of elastic scattering measurements, 
different experimental techniques have been developed to meet these requirements, all of which have its advantages and disadvantages. In the earlier times, 
the experiments with stable nuclei as projectiles, had a quite simple setup, which consisted of just a combination of thin and thick standard planar Silicon (Si) 
surface-barrier detectors, forming $\Delta E-E$ telescopes. This kind of detectors was largely used in experiments and they were enough for particle 
identification and to cover the complete angular range for precise angular distribution measurements. A recent paper discussing improvements on particle
identification when using this kind of detector configuration can be found in Ref.~\cite{SGC20}. Although the planar silicon detectors are still being used, the recent generation of silicon detectors, with a more complex configuration, as double-sided silicon strip detectors (DSSSD), as well as gas ionizing detectors
and active targets, have been developed and are being used.\\

The possibility of using radioactive beams out of short-lived nuclei has offered unique opportunities for research in the frontier field of nuclear physics. Many 
laboratories have installed or upgraded their facilities, and/or developed new techniques to produce radioactive nuclear beams. The idea of these facilities is 
to investigate nuclei at extreme conditions in terms of density, temperature, angular momentum and isospin. Reactions induced by these beams have 
been performed and exotic nuclear structures, such as halo properties, have been investigated~\cite{KGA16}. Several laboratories are pushing to produce 
all kinds of exotic and very energetic species of nuclei as beams. Some laboratories use fragmentation reactions and in-flight technique to produce radioactive ion beams at intermediate energies: NSCL at MSU (USA) ~\cite{NSCL}, RIBLL at HIRFL (China) ~\cite{SZG03}, FAIR at GSI (Germany) \cite{FAIR}, RIBF at RIKEN 
(Japan) \cite{Yan07}, LISE at GANIL (France) \cite{ABM87}, EXOTIC at INFN-LNL (Italy) \cite{PBB17}, MARS at Texas A\&M (USA) \cite{TBG89}, 
and ACCULINNA at Dubna (Russia) \cite{TOI04}. Other laboratories are using the ISOL technique: SPIRAL2 at GANIL (France) \cite{GANIL2}, 
SPES at INFN-LNL (Italy) \cite{BPA16}, REX-ISOLDE at CERN (Switzerland) \cite{VWP08}  and ISAC at TRIUMF (Canada) \cite{TRIUMF}. There are also 
several laboratories producing radioactive nuclei at the lower energy regime such as Twinsol at University of Notre Dame (USA) \cite{BLO03}, RIBRAS (Brazil) 
\cite{LLG05}, ATLAS and CARIBU at ANL (USA) \cite{HPR00}, CRIB-RIKEN (Japan) \cite{YKT05}  and SOLEROO (Australia) \cite{RHD11}. 
These laboratories are constantly improving their capabilities in terms of beam intensity, and development of new devices to increase the detection efficiency. 
Reviews on these facilities can be found in Refs. \cite{BND13} and \cite{Raa16}. Some new laboratories, such as VECC-RIB (India) \cite{Cha07}, BRIF 
\cite{ZCL13} and HIAF (China) \cite{WXX13}, which are under construction, will broaden even further the scope of the studies on nuclear reactions.\\

Most of these laboratories have scientific programs \break based on measurements of nuclear reactions, and several elastic scattering experiments, 
at energies close and above the Coulomb barrier, have been performed. These measurements allowed the extraction of total reaction cross sections and to 
investigate the influence of other mechanisms in the elastic scattering. The basic idea of these experiments is to obtain angular distributions 
for a wide range of angles and energies and extract the total reaction cross section. The angular distributions for elastic scattering can exhibit different 
features depending on the incident energy and on the structure of the colliding nuclei. For instance, angular distributions for tightly bound projectiles 
at near-barrier energies divided by the corresponding Rutherford cross section exhibit typical Fresnel oscillatory diffraction patterns~\cite{SMC86}. For light 
projectiles, and at higher energies, the Coulomb force is weaker and the diffractive pattern changes from Fresnel to Fraunhofer oscillations. 

On the other hand, angular distributions for weakly bound nuclei may deviate from the diffractive patterns. When the binding energy of the projectile 
fragments is very low, their motion decouples as they approach the target, leading to nonelastic processes. Owing to the long ranges of the Coulomb and 
nuclear couplings, this can occur even in distant collisions, giving rise to strong absorption at forward angles. As a result, the diffractive oscillations and the 
Fresnel peak are damped, or completely disappear.\\

Some weakly-bound nuclei are called exotic, as they can exhibit peculiar properties such as halo and borromean\footnote{Borromean nuclei are 
nuclei with a bound three cluster configuration, which becomes unbound when one of the clusters is removed.} configurations, where the valence nucleon(s) 
orbits around a compact core, forming an extended matter distribution~\cite{THH85,TSR13}. Typical examples of exotic nuclei with two valence 
nucleons (borromean) are $^6$He, and $^{11}$Li, and with one valence nucleon (halo nuclei) are $^8$B, $^{11}$Be and $^{15}$C. In  
collisions of these nuclei with medium to heavy mass target, the breakup process is favoured during the interaction, due to the Coulomb field of 
the target and/or possibly a long-range component of the nuclear potential. During the collision, the charged core is decelerated by the repulsive 
Coulomb forces, while the valence neutrons are not affected. This phenomenon is called {\it electric dipole polarization} 
or {\it Coulomb dipole excitation} (CDE).\\

\begin{figure}
\begin{center}
\includegraphics*[width=6cm]{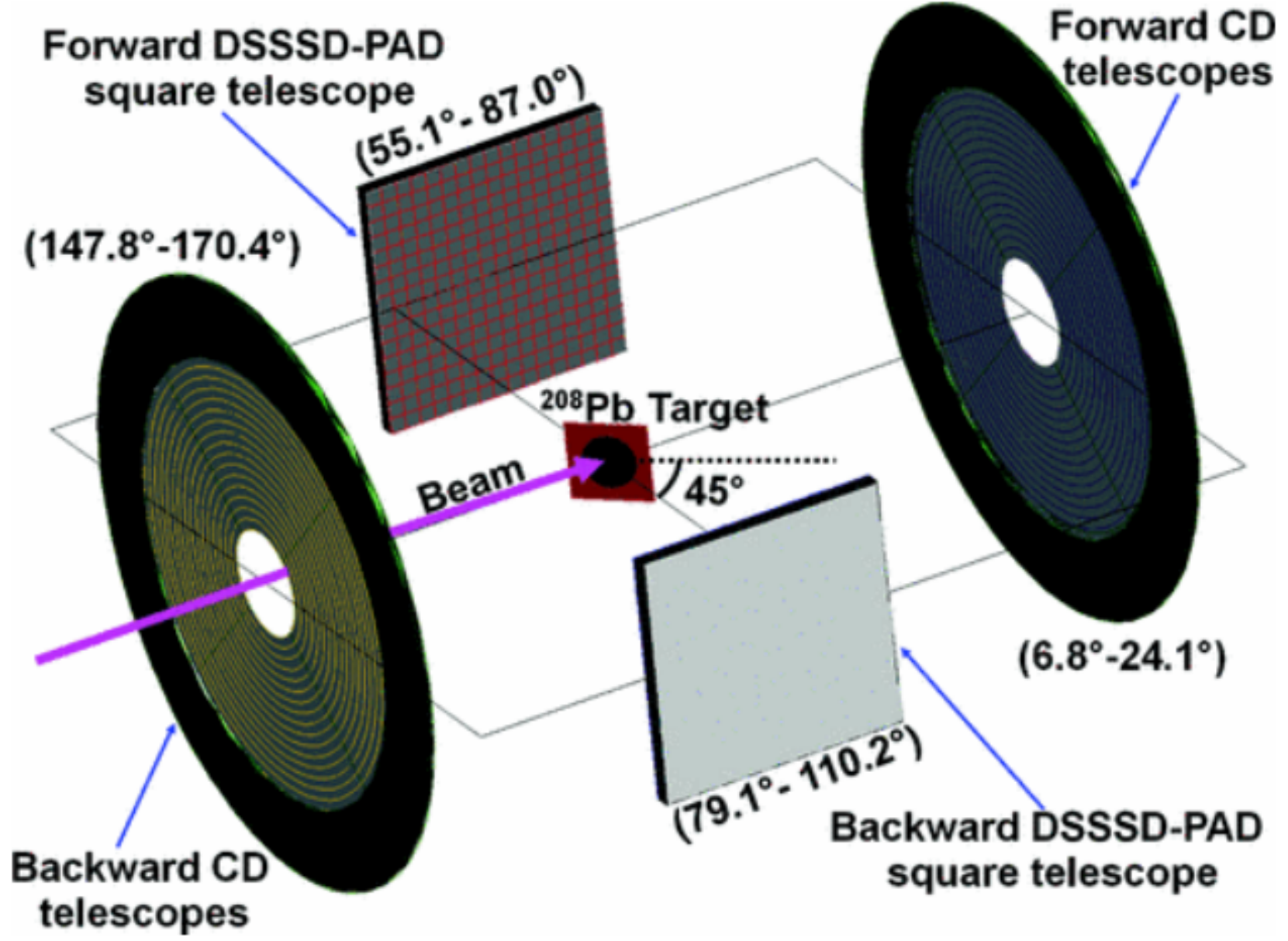}
\end{center}
\caption{(Color online) Detector setup (DINEX) used in the experiment for the $^6$He+$^{208}$Pb \cite{ABG11}.  Figure extracted from ref. \cite{ABG11}.}
\label{F5.1}
\end{figure}
Radioactive ion beams produced as secondary beams have the limitation of much lower intensity (in the order of $10^3$ up to $10^6$ pps), when compared
to beams of stable nuclei ($\sim 10^{12}$ pps). For experiments with these beams, larger solid angles and better detection efficiency of the setup are required 
to compensate for the low beam intensity. State-of-art experiments, using a combination of double-sided silicon strip detectors (DSSSD) have been performed 
for the $^6$He, $^8$He and $^{11}$Li nuclei on $^{208}$Pb targets. Experiments with $^6$He beams were performed at \break Louvain-la-Neuve (Belgium), 
before the facility was decommissioned, using an array of DSSSD telescopes called DINEX~\cite{ABG11}. For the $^8$He beam, an experiment was 
performed at GANIL, France, using an arrangement called GLORIA (Global Reaction Array) detection system~\cite{MAB14,MMS16}, while for the 
$^{11}$Li beam an experiment at the ISAC-II line at TRIUMF, Canada, using an array of four DSSSD telescopes, was performed~\cite{CFR12}. A sketch 
of the DINEX setup used in the elastic scattering of the $^6{\rm He}\,+\,^{208}{\rm Pb}$ system is shown in Fig. \ref{F5.1}. With this kind of arrangement, 
a quite large total solid angle (about 20\% of 4$\pi$) can be reached and a wide angular range (10$^{\circ}$ to 150$^{\circ}$) can be covered, allowing 
precise angular distribution measurements. It is also important to mention that these experiments, using low-intensity radioactive ion beams, are quite challenging 
from the experimental point of view. For instance, the use of detection setups with several DSSSDs demands a considerable 
effort in the analysis to assign the scattering and solid angles of each pixel of the array. However, all the efforts paid off, since covering a wide angular range 
and obtaining cross sections at very backward angles, is quite important for a better theoretical analysis and better 
interpretation of the influence of the several mechanisms such as breakup and transfer in the elastic scattering. The measured angular distributions obtained 
from these experiments have been analyzed with different approaches, such as optical model, coupled channel calculations and phenomenological models 
\cite{KKM19,MKK17,SKC14,GLK18}. From the optical model analysis reported in Ref. \cite{KKM19},  the derived total reaction cross section for  
the $^6$He+$^{208}$Pb and $^8$He+$^{208}$Pb systems are very similar.
The decoupling of the single-particle motion of the halo with respect to the core has also motivated the application of new 
reactions models such as the CDCC, discussed in section \ref{CDCC}. In particular, the analysis performed for the $^6{\rm He}\,+\,^{208}{\rm Pb}$ data 
is a good example of the success of 4b-CDCC applications, as discussed in section \ref{4b-CDCC}.\\

Besides the previously discussed experiments with the neutron-rich borromean beams, several experiment have been performed with 
$^{11}$Be~\cite{PSM12,AAA09,PBM17,MSR06} and $^{12}$B beams. $^{11}$Be is a typical one-neutron halo nucleus, where the valence 
neutron moves around the $^{10}$Be core in an s-orbit, with the low separation energy ($S_n$=0.503 MeV). Differently from the borromean nuclei, 
 $^{11}$Be has the peculiarity of having one bound excited state, with excitation energy 320 keV ($j^\pi = 1/2^Ã¢ÂÂ$). 

\begin{figure}
\begin{center}
\includegraphics*[width=6cm]{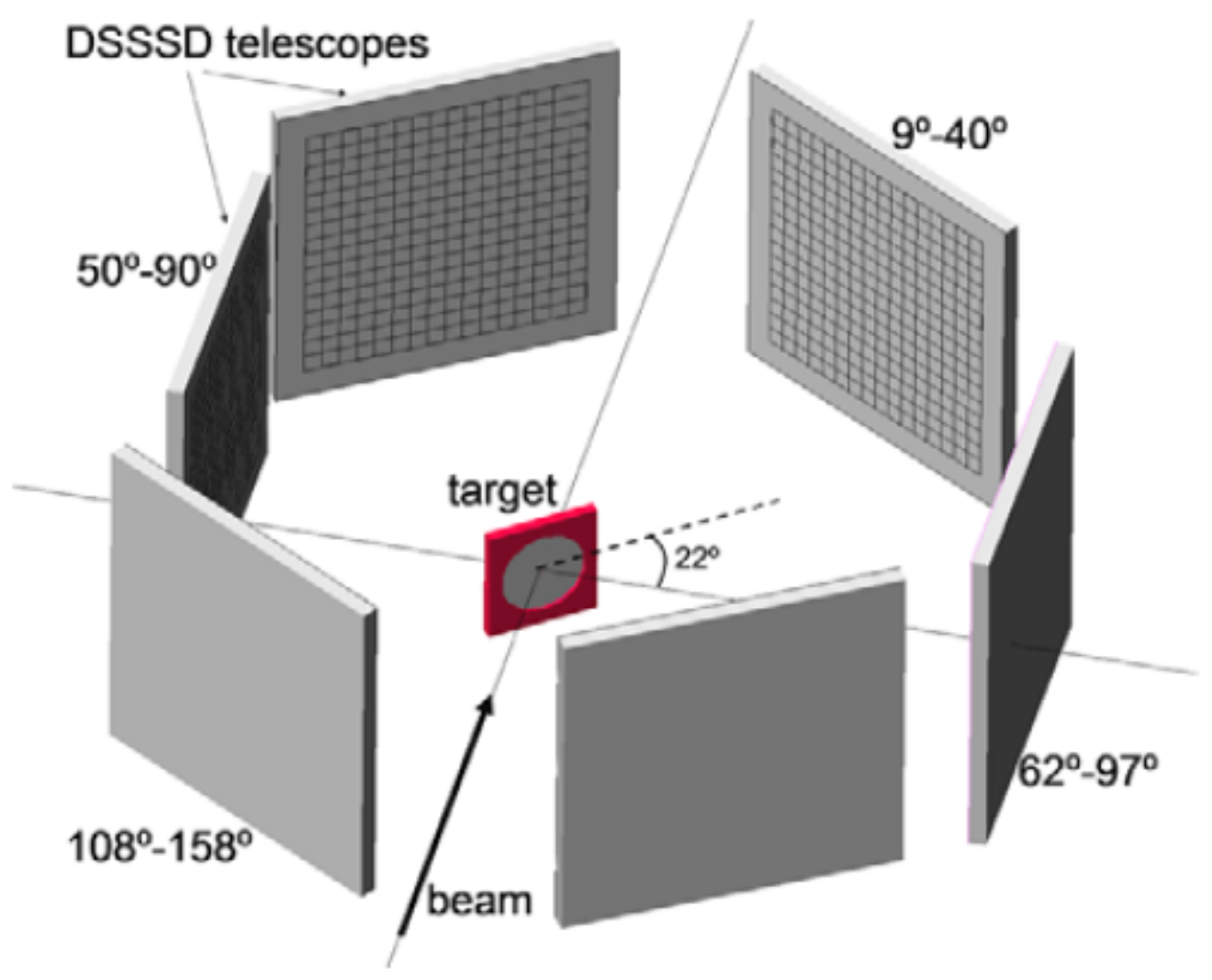}
\end{center}
\caption{(Color online) T-Rex experimental setup used for the $^{10}$Be, $^{11}$Be+$^{64}$Zn and  $^{11}$Be+$^{120}$Sn experiments at REX-ISOLDE, 
at CERN. Figure extracted from Ref. \cite{AAA09}.}
\label{F5.2}
\end{figure}
Di Pietro {\it et al.} \cite{PSM12}  obtained high quality quasi-elastic scattering angular distribution data for collisions of $^{11}$Be with $^{64}$Zn, at 
energies around the Coulomb barrier. The experiment was performed at the REX-ISOLDE facility of CERN, using the array of six DSSSDs telescopes, 
shown in Fig. \ref{F5.2}. Since the energy resolution was not good enough to distinguish between elastic scattering and inelastic scattering to the 
excited state of $^{11}$Be, the angular distributions were considered quasi-elastic in nature. They also measured breakup cross sections but, due to low 
statistics at the backward angles, they considered only data from the three DSSSDs telescopes. The data clearly indicates the suppression of the Coulomb$-$nuclear interference peaks due to the halo 
structure of the $^{11}$Be nucleus, which is an indication of strong absorption, even at forward angles. Indeed, the obtained total reaction cross section 
was $\sigma_{\scr R} =2730$ mb for $^{11}$Be, which is more than twice those obtained for $^{10}$Be ($\sigma_{\scr R} =1.260 $mb) and $^9$Be 
($\sigma_{\scr R}  =1.090$ mb).

Another quasi-elastic scattering experiment at the 
REX-ISOLDE facility was performed by Acosta {\it et al.}~\cite{AAA09}. They studied the $^{11}$Be+$^{120}$Sn system, using the same setup 
as in the experiment of Di Pietro {\it et al.}~\cite{PSM12}, for the  $^{11}{\rm Be} \,+\,  ^{64}{\rm Zn}$ system. Despite their impressive detection system, 
allowing measurements of 1536 pixels and covering an angular range of 10$^{\circ}$ to 150$^{\circ}$, they could only obtain reliable data at forward angles, 
between 15$^{\circ}$ and 38$^{\circ}$. This was mainly due to the use of a too thick target, which made it impossible to separate $^{11}$Be from $^{10}$Be contamination at more backward angles. 

Mazzoco {\it et al.}~\cite{MSR07a} measured elastic cross sections in collisions of $^{11}$Be projectiles on a $^{209}$Bi target, at near-barrier energies. 
The experiment was performed at RIKEN, using eight two-stage $\Delta  E - E$ telescopes placed along the lateral faces of two cubes closely-packed around the 
target, called EXODET array~\cite{MSR07a}. Despite the large fluctuations in the cross sections, the results of the optical model analysis indicate that direct 
processes related to the $^{11}$Be halo structure and smaller binding energies are more important at near-barrier energies.

More recently, Pesudo {\it et al.}~\cite{PBM17} measured elastic, inelastic and breakup angular distributions for the $^{11}{\rm Be}\,+\,^{197}{\rm Au}$
system. The experiment was performed at TRIUMF, Canada, with an array of 4 DSSSD telescopes surrounded by twelve high-purity germanium clovers of 
TIGRESS. This setup allowed measurements of elastic and inelastic scattering events at angles ranging from  $14^{\circ}\,{\rm to} 157^{\circ}$. The inelastic channel was identified by gating the $^{11}$Be events on the 320 keV gamma-ray peak. The data were 
analyzed with an extended continuum-discretized coupled-channels (XCDCC), discussed in section \ref{XCDCC}, where the $^{10}$Be core excitation was 
considered. Theory and experiment were shown to be in excellent agreement (see Figs. \ref{XCDCC-el} and \ref{XCDCC-bu}).\\

Experiments on elastic scattering induced by proton-rich nuclei, such as $^7$Be, $^8$B, $^9$C, $^{10}$C and $^{17}$F, on medium to heavy target, 
have also been performed in several laboratories, as described below.
One of the first experiment with $^7$Be and $^8$B secondary beams on medium to heavy target was performed at the University of Notre Dame, USA \cite{AML09}. 
In this experiment, angular distributions for the elastic scattering of the 
$^7$Be, $^8$B + $^{58}$Ni systems were measured simultaneously, at several energies near and above the coulomb barrier. The large derived 
total reaction cross sections for $^8$B were considered an evidence of the halo effect in this nucleus. The CDCC calculations performed for the 
$^8$B + $^{58}$Ni system showed the strong influence of the breakup channel in the elastic scattering \cite{LCG10,Kee12}. 

For heavier targets, the breakup is expected to be more prominent, since the long-range Coulomb interaction is predominant over the nuclear 
potential. Elastic scattering measurements for the $^8{\rm B}\, +\, ^{208}{\rm Pb}$ system were performed at $E_{\scr Lab} =50$ MeV \cite{MKB19} and 
$E_{\scr Lab} =170$ MeV  \cite{YLP18}. The measurement at the lower energy, close to the barrier, was performed at the
CRIB facility in RIKEN, Japan, using the EXPADES detector array \cite{PBB16}, which consisted of six DSSSD telescopes covering the angular range of 
8$^{\circ}$ to 166$^{\circ}$. The obtained angular distribution shows a quite strong damping of the Fresnel peak, although the differential cross sections data
in the corresponding angular region show large fluctuations and uncertainties. The extracted total reaction cross section from an optical model 
analysis was $\sigma_{\scr R}=1112$ mb, which is much larger than the one for the $^7$Be core on the same target, measured in the same experiment
($\sigma_{\scr R}=253$ mb). Full CDCC calculations performed for the $^8{\rm B}\,+\, ^{208}{\rm Pb}$ system poorly reproduces the experimental 
angular distribution, indicating that some other effect, such as core excitation, may be playing an important role for this system. For the data obtained at the 
higher energy, three times that of the Coulomb barrier, at the Radioactive Ion Beam Line in Lanzhou (RIBLL), the Fresnel peak is still present and the full 
CDCC calculation describes quite well the data \cite{YLP18}.\\

Besides the data on the $^8{\rm B}\,+\,^{58}{\rm Ni}$ system reported in Ref. \cite{AQL09}, elastic scattering experiments involving other boron isotopes 
have been performed, studying the $^{10}{\rm B}\,+\,^{58}{\rm Ni}$ \cite{SCG17}, $^{11}{\rm  B}\,+\,^{58}{\rm Ni}$ \cite{DGC15} and 
$^{12}{\rm B}\,+\,^{58}{\rm Ni}$ \cite{ZGC19} systems. All these data were analyzed with coupled channel calculations, and the analyses indicated that the 
elastic cross sections are sensitive to the cluster configuration of the projectiles. Different configurations of the boron isotopes were shown to produce 
different effects on the elastic angular distributions~\cite{GZC20}.\\

Elastic scattering experiments on collisions of $^9$C, $^{10}$C, $^{11}$C, $^{12}$C, $^{13}$C, $^{14}$C, and $^{15}$C beams on heavy targets have also been
performed, as described below. 
Santra {\it et al.} \cite{SSK01} studied the $^{12}$C+$^{208}$Pb system, measuring elastic and inelastic scattering, transfer, fission, and 
evaporation residue cross sections, at energies around and above the Coulomb barrier, in the range: $E_{\scr Lab}= 58.9 - 84.9$ MeV. Simultaneous coupled 
reaction channel (CRC) calculations involving the elastic and all the nonelastic channels were performed. Optical-model analyses of the elastic data have 
been also carried out. The resulting optical potential showed a strong energy dependence, mainly at near-barrier energies, and some long-range absorption.
Although elastic scattering and various nonelastic channels are studied in Ref.~\cite{SSK01}, the total reaction cross section has not been given in this work.

Several measurements of elastic cross sections in collisions of $^{13}$C and $^{14}$C on heavy targets have been performed (see, for instance, the database 
in the website given in Ref.~\cite{KDN17,ZDK99}). Most of these experiments investigated the influence of one- or two-neutron transfer reactions on the elastic 
process. In particular, the study of Landowne and Wolter for the $^{13}$C+$^{208}$Pb system~\cite{LaW81} at sub-barrier energies is one of the pioneering 
works in the early days on the use of single-folding potential models and on the importance of valence nucleon in elastic scattering. \\

Experiments with secondary beams of radioactive carbon isotopes have also been performed. 
Yang {\it et al.}\cite{YLP18,YWW14} measured elastic angular distributions for collisions of $^{9,10,11}$C on $^{208}$Pb. The experiments were performed at the RIBLL 
in Lanzhou, China, at collision energies three times the height of the Coulomb barrier.
The measured angular distributions did not show any suppression of the Fresnel peak, resembling those for stable projectile. On other hand, a strong absorption
has been observed for the elastic scattering of $^{10}$C+$^{58}$Ni system at energies closer to the barrier, most probability due to the combination of 
deformation and cluster configuration \cite{GCS19}. The $^{10}$C isotope is assumed to be the only nucleus to have a {\it brunnian} (super-borromean) structure, where the four interconnected rings are associated to the four-body interactions ($\alpha+\alpha+p+p$) \cite{CAA08,CWM09}. \\

An interesting experiment was performed with the ATLAS accelerator, at ANL, to measure fusion reactions for the $^{12,13,14,15}{\rm C}\,+\,^{232}{\rm Th}$ 
systems, at near-barrier energies~\cite{ARB11}. The $^{15}$C nucleus is a good candidate for an exotic halo nucleus. The valence neutron is orbiting 
around the $^{14}$C core, in an $s_{1/2}$ orbit with a binding energy of S$_n$=1.218 MeV. The experiment showed that, at the lowest energies, 
the fusion-fission cross sections for $^{15}$C+$^{232}$Th is enhanced by a factor of 5, in comparison to those for $^{12,13,14}$C projectiles.
Considering that, at this very low energy, the fusion-fission would exhaust the total reaction cross section, this is an indication of large total reaction cross section, 
typical of an exotic nucleus. 

Very recently an elastic scattering experiment for the $^{15}$C+$^{208}$Pb system was performed at the HIE-ISOLDE facility, at CERN~ \cite{OKM20}. Although they used the GLORIA detector array \cite{MAB14}, which covers a large solid angle and angular range of 15$^{\circ}$ to 165$^{\circ}$, the cross sections were 
determined only at the most forward angles. The preliminary data indicated a strong absorption and strong damping of the Fresnel peak.

Laboratories as GANIL (France), NSCL-MSU (USA), ANL (USA), RIKEN (Japan), TRIUMF (Canada), RIBLL (China), Dubna (Russia)  etc. are producing
several other radioactive ion beams, with $Z>6$, such as $^{12}$N, $^{13}$O, $^{14}$O, $^{17}$F, and $^{17}$Ne etc. However, most of the experiments are 
related to the investigation of the structure of these nuclei,  using reactions on light targets (H and He) at high 
energies ($ > 50 {\rm MeV}/A$). Very few experiments are being performed to measure reactions induced by radioactive nuclei with $Z>6$ on medium to 
heavy target. The exception is the $^{17}$F  nucleus, where some efforts have been devoted to measure elastic scattering, breakup and fusion 
reaction. Fluorine-17 is a weakly-bound proton drip line nucleus, where the  valence proton has the binding energy $S_p= 0.601$ MeV. 
 One of the first experiment involving $^{17}$F was the fusion-fission measurement of $^{17}$F+$^{208}$Pb performed at energies in the vicinity of the 
 Coulomb barrier at ANL, USA  \cite{HPR00}. 
 For proton drip-line nuclei, the breakup process produces a residual nucleus (the core),  with one less proton, and thus with a lower Coulomb barrier. This 
 is expected to lead to appreciable incomplete fusion, and to large  total fusion cross sections. However, no enhancement of fusion-fission yields due 
 to breakup or to a large interaction radius was observed in this experiment. 
 
 Later, some other experiments were performed~\cite{BBG11} at the Holifield Radioactive Ion  Beam Facility (HRIBF) at Oak Ridge, USA, to  measure 
 elastic scattering and breakup of the same $^{17}$F+$^{208}$Pb system  \cite{LBE02,LBG03}. In these works, a discussion on the diffraction and 
 stripping breakup was presented, where the late is related to incomplete fusion of the proton fragment. Their results suggested that,  for proton-rich 
 nuclei, sub-barrier  fusion would be suppressed. This is in sharp contrast with the large sub-barrier fusion enhancements observed  for the exotic 
 neutron-rich nucleus $^6$He \cite{KGP98a}.  

\begin{figure}
\begin{center}
\includegraphics*[width=8.5 cm]{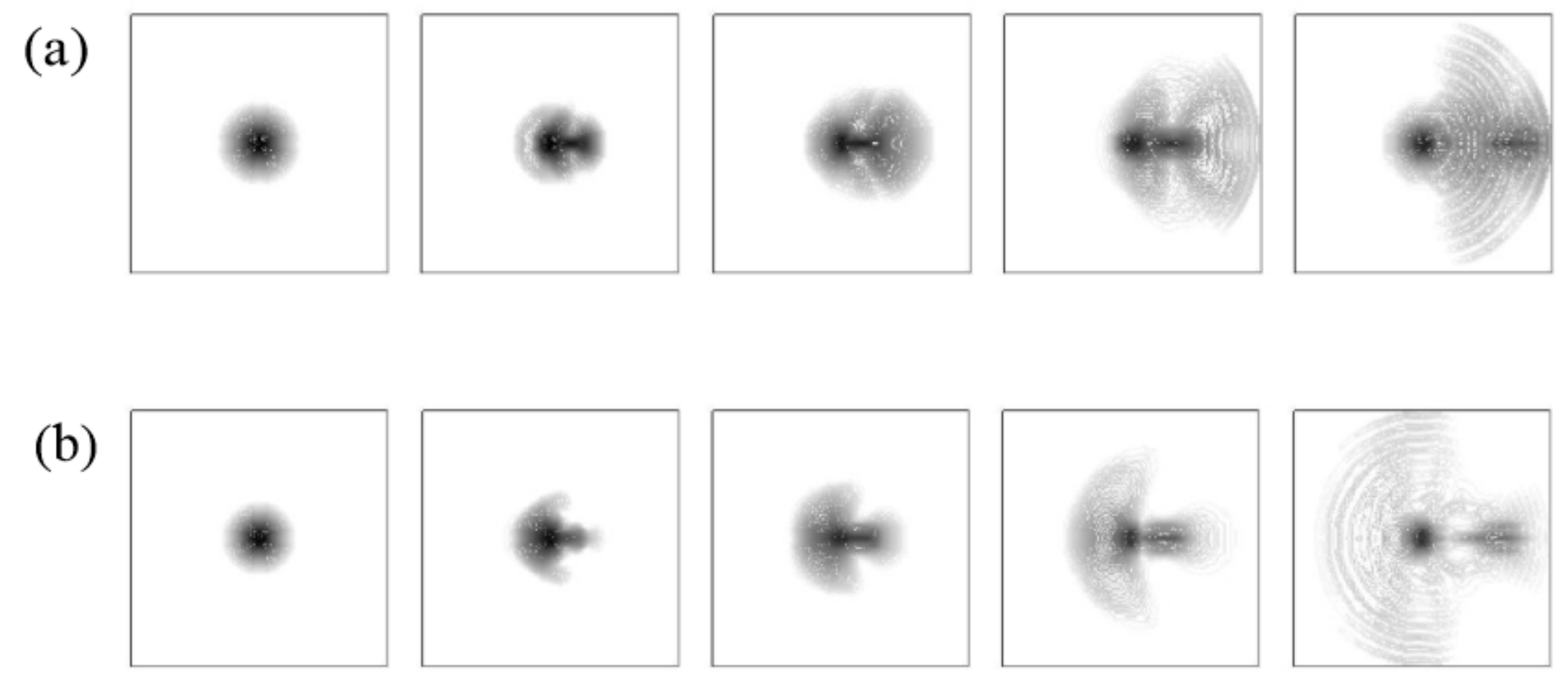}
\end{center}
\caption{(Color online) Time-dependent density distribution of: (a) the neutron;  (b) the proton valence particle (figure extracted from Ref.~\cite{IYN06}).}
\label{F5.3}
\end{figure}
There has been some speculation about the possible polarizability of the proton- and neutron-rich exotic nuclei, as they approach  a heavy target. In the 
case of  a proton halo nucleus, both the valence proton and the core would be scattered to the backward direction by the Coulomb field of the target.
On the other hand,  for neutron-rich nuclei, the charged core is repelled by the Coulomb field, whereas the neutron halo is not affected. In this way, the 
neutron-rich projectile would be polarized along the collision. This idea is corroborated by the time-dependent dynamic calculation of the fusion process 
by Ito {\it et al.} \cite{IYN06}, illustrated in Fig.\ref{F5.3}. The figure shows time evolutions of the projectile's density in a collision of a neutron-rich (panel 
(a)) and of a proton-rich (panel (b)) projectile with a heavy target, at a sub-barrier energy.\\

Due to the high interest in investigations of breakup effects in fusion, elastic and total reaction cross sections for proton-rich nuclei, other elastic 
scattering experiments were carried out, for the $^{17}$F+$^{208}$Pb \cite{SPM10} and $^{17}$F+$^{58}$Ni \cite{MSP10} systems at near-barrier energies.  
The experiments  were performed at INFN-LNL, Italy, using the EXOTIC facility \cite{PBB17}. In these experiments the charged particles coming out from the 
reaction were detected by the EXODET detector array, which consisted of eight $\Delta E$-$E$ telescopes arranged as faces of two cubes closely packed 
in the front and at the back side of the target \cite{RPV05}. This setup covered a solid angle of about 70\% of $4\pi$ sr and allowed the coincidence detection 
of both $^{16}$O and $p$ fragments from the $^{17}$F breakup (exclusive breakup). 
\begin{figure}
\begin{center}
\includegraphics*[width=8cm]{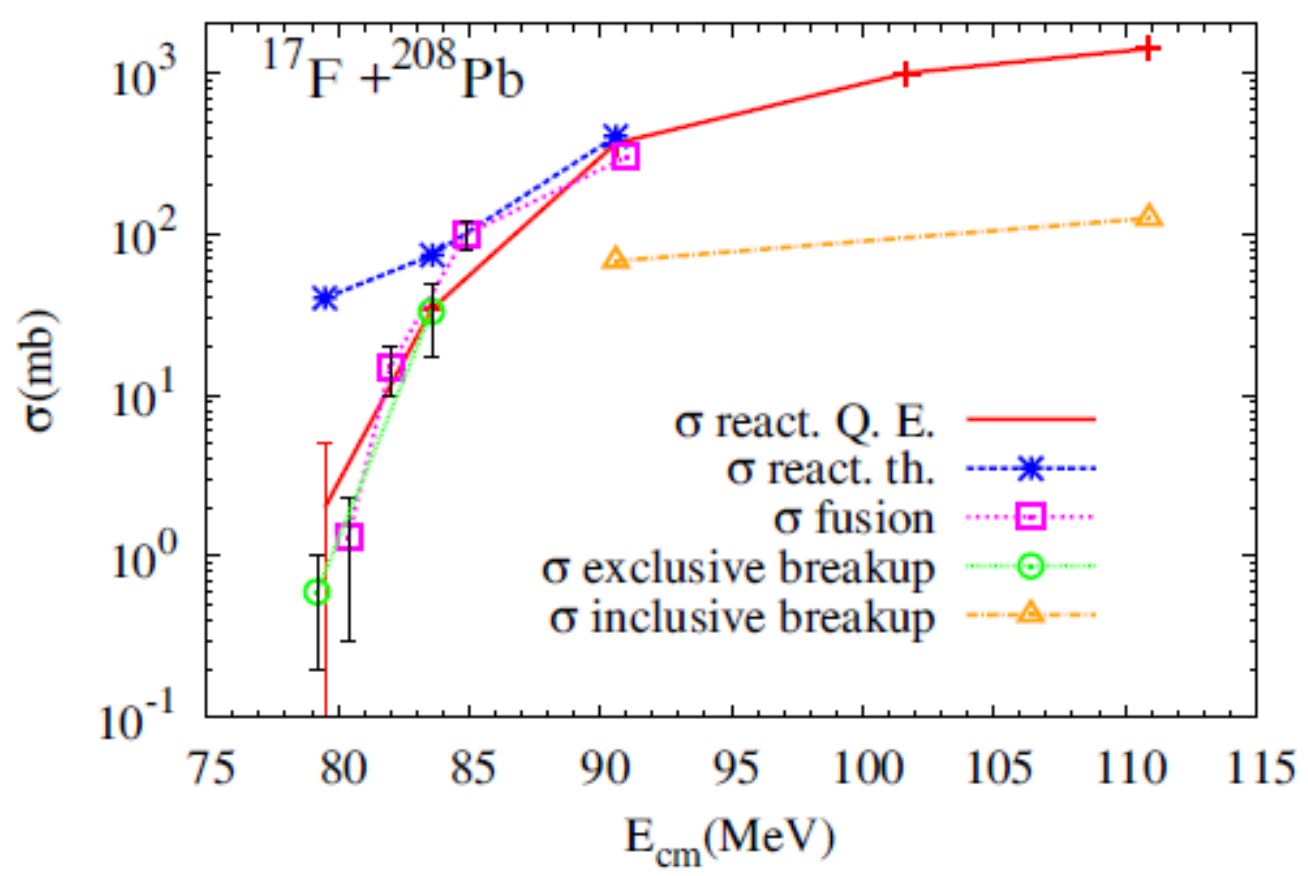}
\end{center}
\caption{(Color online) Cross section for total reaction, inclusive and exclusive breakup and fusion. Figure taken from Ref. \cite{SPM10}. }
\label{F5.4}
\end{figure}
The results for the $^{17}$F+$^{208}$Pb system are summarized in Fig.~\ref{F5.4}. At energies below the coulomb barrier (V$_{\scr B}=92.0$ MeV), the sum 
of exclusive breakup and fusion exhausted the total reaction cross section. Above the barrier, the inclusive breakup, where only the $^{16}$O fragment is detected, was 
quite strong. The conclusion is that the reactivity of the proton-rich $^{17}$F nucleus is not very large. Thus, the small binding energy of the valence proton plays 
a minor role in the reaction dynamics. 

\begin{figure}
\begin{center}
\includegraphics*[width=5cm]{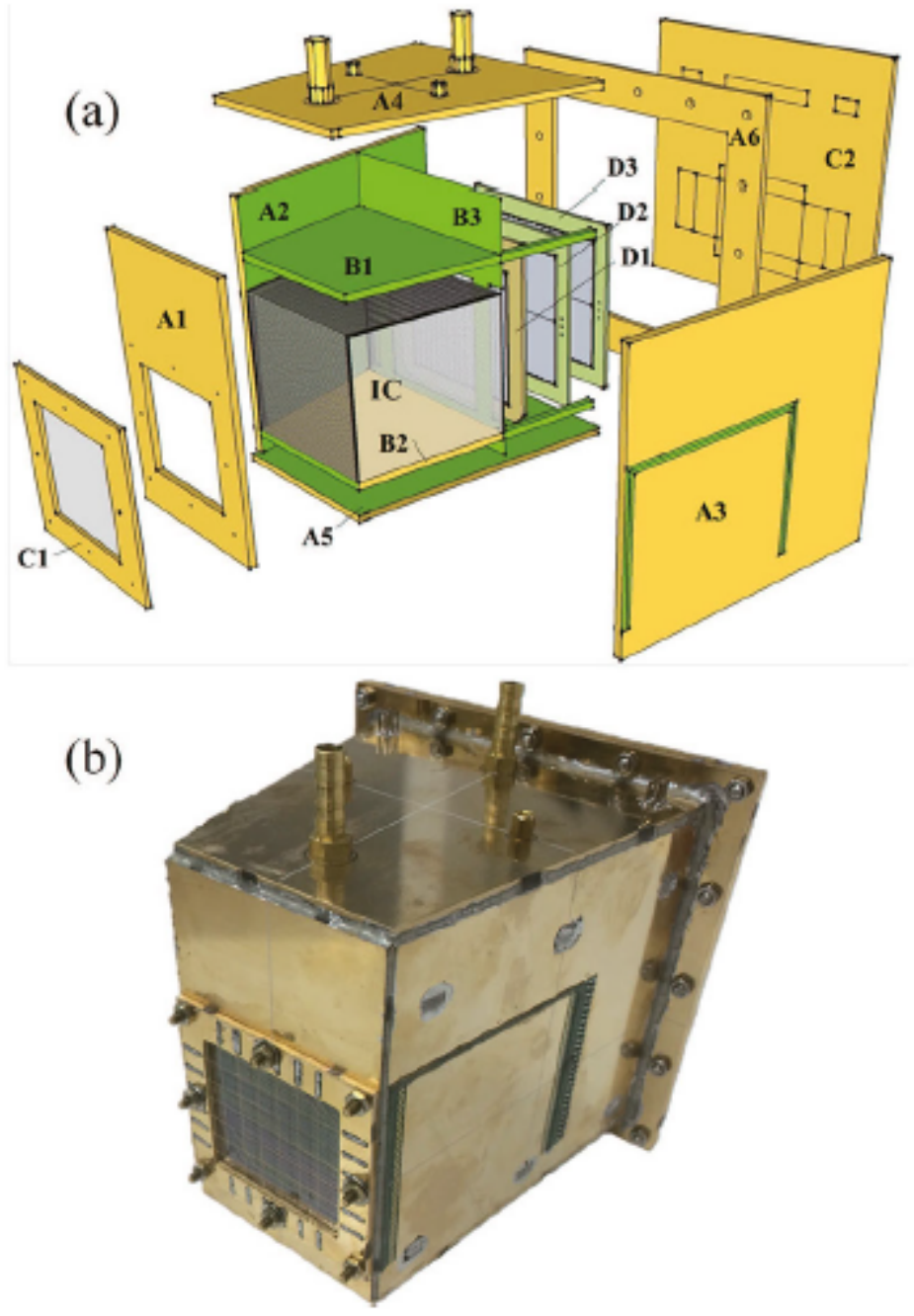}
\end{center}
\caption{(Color online) View of general structure and the photograph of the assembled multilayer ionization-chamber telescope array MITA. Figure extracted 
from Ref. \cite{MYL19}.  }
\label{F5.5}
\end{figure}
Very recently, another experiment for the $^{17}$F+$^{58}$Ni system was performed at the CRIB, RIKEN, Japan. In this experiment, elastic, breakup and 
fusion cross sections were measured using a multilayer ionization-chamber telescope Array (MITA) \cite{MYL19}, shown in Fig.~\ref{F5.5}. This device is a 
combination of 10 independent telescopes, where each of them consists of four stages detectors: one ionization chamber, followed by thin (40 or 
60 $\mu$m) DSSSD and two layers of thick (300 and 1000 $\mu$m) quadrant silicon detectors (QSDs). The possibility of detecting, distinguishing and 
separating high Z particle as well as light particles enables the simultaneous measurements of all the important reactions (elastic, breakup and fusion) 
produced in the collision, where the ionization chamber is important to measure energy loss of high Z particles. A single angular distribution for 
the elastic scattering is presented in Ref. \cite{MYL19}. The full set of data of this experiment is been submitted for publication and the results is promising 
and very much awaited. \\

Usually different experimental setups, not always available at the same laboratory, would be required to measure the different reaction channels produced in a collision 
between heavy nuclei. In most cases, it would demand different measurements at different times and, consequently, it would take several years to get information on all 
channels for a given system.

An example of such an effort is the work developed at University of Notre Dame to investigate the channels involved in the collision of $^{6}$He on a $^{209}$Bi target. The 
different reactions were measured in different experiments, using different setups, such as sub-fusion measurement ($1n,2n,3n$ evaporation) \cite{KGP98a}, 
fusion-fission \cite{KGP98b}, fusion-evaporation ($4n$ evaporation) \cite{DHJ98}, transfer and breakup \cite{AKN00}, elastic scattering \cite{AKB01}, $2n$-transfer \cite{DMK05} 
and breakup \cite{KAB07}. The discussion of these mechanisms, in terms of total reaction cross section and evidence for core-halo decoupling, was reported in Ref. \cite{AKA10}. 
A similar effort was also devoted to the $^{8}$B+$^{58}$Ni system, where elastic \cite{AQL09}, breakup \cite{GKP00,AQB08} and fusion cross sections~\cite{AVQ11} were 
measured in different experiments, and with different setups. The discussion of the total reaction cross section for this system was reported in Ref. \cite{KAG17}. The sum of the 
total fusion plus breakup cross sections gives about the same large total cross section as those extracted from elastic scattering, see Fig.~\ref{F5.6}. 
\begin{figure}
\begin{center}
\includegraphics*[width=6cm]{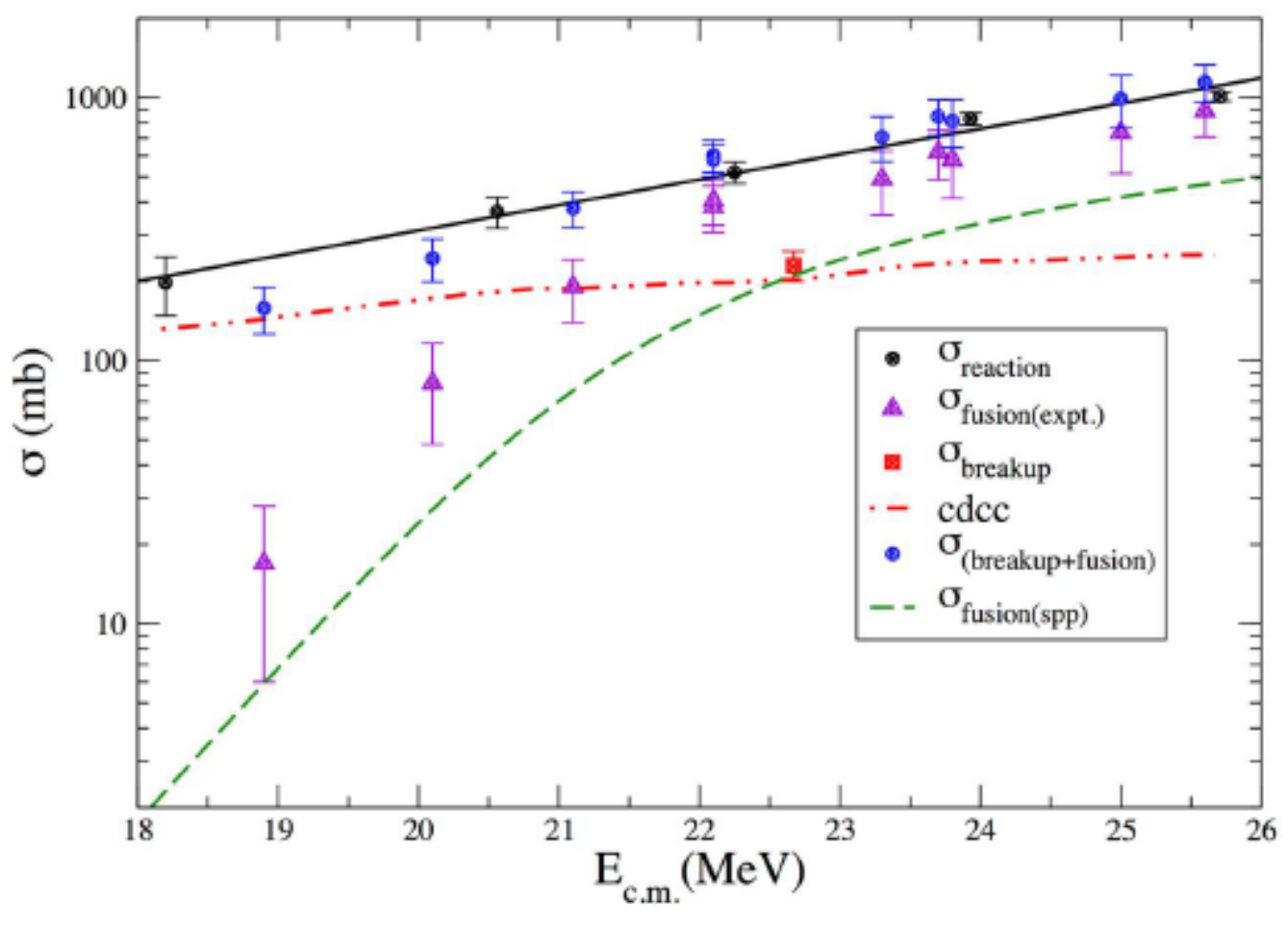}
\end{center}
\caption{(Color online) The several cross sections for $^8$B+$^{58}$Ni system, as indicated. The fusion curve was calculated using the S\~aÃÂ£o Paulo potential. Figure extracted from Ref. \cite{KAG17}.}
\label{F5.6}
\end{figure}
%


\subsection{Direct measurements and total reaction cross section}


In the previous section, some experiments on the elastic scattering of radioactive ion beams were highlighted. The total reaction cross sections were extracted from optical model 
(OM) analyses of the scattering data. 

\medskip

However, this indirect method has a limitation. It relies on a precise determination of the optical potential, which requires accurate elastic scattering data over a fine angular mesh.
Frequently, this condition is not met, mainly in collisions of exotic nuclei. 
On the other hand, some direct methods to measure the total reaction cross section have been developed, and reported in the literature. A brief discussion of these methods
is presented below.


\subsubsection{Sum of differences}\label{SOD tech}


The sum of difference method, described in section \ref{SOD}, is a very convenient way to obtain the total reaction cross section  and
information on the nuclear interaction from the elastic angular distribution at very forward angles. In this angular region, heavy-ion scattering exhibits oscillations, 
known as {\it forward glory effect}. 
These oscillations result from the interference between the nuclear and Coulomb scattering amplitudes. As the Coulomb amplitude is known analytically, it is possible 
to extract $f{\scr N}(\theta=0)$ from the data, and then obtain the total reaction cross section (see section \ref{SOD}).\\

This method was used to investigate forward nuclear glory in the scattering of the $^{12}$C+$^{12}$C system of identical nuclei~\cite{OTV91}. Total reaction cross sections were extracted and a clear evidence of nuclear glory in heavy-ion collisions was observed. In another experiment, this method was used to study collisions of $^{12}$C, $^{13}$C, 
$^{15}$N and $^{16}$O projectiles on $^{28}$Si targets~\cite{YIY98}. The elastic scattering data were used to extract the total reaction cross section and to 
investigate the importance of the neutron transfer in the reaction. 

Ueda and Takigawa studied the forward glory effect in collisions of the two-neutron halo $^{11}$Li nucleus on $^{12}$C, within a semiclassical 
approximation~\cite{UeT96}. They found that the halo of $^{11}$Li leads to a higher glory angular momentum and shifts the first glory minimum towards a smaller 
angle. Further, they concluded that these features enhance the oscillations of the nuclear amplitude in the sum of difference analysis. 

To investigate these effects, Ostrowski {\it et al.}~\cite{OSG99} measured elastic angular distributions for the $^6$He+$^{12}$C system at forward angles. 
The experiment was performed at Louvain-la-Neuve, Belgium. They concluded that the low neutron binding energy of $^6$He reduces the nuclear forward glory 
effect due to flux losses from the elastic channel to breakup or transfer channel, and that the total reaction cross section for this weakly bound 
system is twice as high at that for for $^6$Li + $^{12}$C at the same Coulomb parameter value.


\subsubsection{Backscattering cross section}


A simple method to extract the total reaction cross section has been proposed by Sargsyan {\it et al.} \cite{SAA13b}, which consists of using only the cross 
section measured at very backward angles. \\

The starting point is the usual expression for the total reaction cross section,
\begin{equation}
\sigma_{\scr R}(E) = \frac{\pi}{K^2} \sum_{J=0}^\infty\ (2J+1) \ \mathcal{P}_{\scr R}(J,E),
\label{sigR-Jsum}
\end{equation}
where $\mathcal{P}_{\scr R}(J,E)$ is the total reaction probability in a collision with energy $E$ and angular momentum $J$. Then, 
$\mathcal{P}_{\scr R}(0,E)$ is written in terms of the elastic scattering probability for $J=0$, 
\begin{equation}
\mathcal{P}_{\scr R}(0,E) = 1\,-\,\mathcal{P}_{\scr el}(0,E),
\label{PR}
\end{equation}
and $\mathcal{P}_{\scr el}(0,E)$ is approximated by the ratio between the elastic and the Rutherford cross sections at $\theta = 180^\circ$, namely
\begin{equation}
\mathcal{P}_{\scr el}(0,E) \simeq \left[\frac{d\sigma_{\scr el}(\theta){\big /} d\Omega}{d\sigma_{\scr Ruth}(\theta){\big /}
d\Omega} \right]_{\theta=180^\circ}.
\label{Pel}
\end{equation}
Next, the total reaction probabilities for $J\ne 0$ are approximated by the probability at $J=0$, but at the lower energy 
\begin{equation}
E^\prime = E\,-\, \frac{\hbar^2}{2\mu R_{J}^2}\ J(J+1),
\label{E-prime}
\end{equation}
where $R_J$ is the radius of the barrier of the effective potential (including the centrifugal term), $V_J(R)$. \\

The sum over $J$ in Eq.~(\ref{sigR-Jsum}) is then transformed into an integral over $E^\prime$ involving $\mathcal{P}_{\scr R}(0,E^\prime)$, which
is related to the backscattering scattering data ($\theta_{\rm Lab} > 150^{\circ}$), through Eqs.~(\ref{PR}) and (\ref{Pel}).\\

To assess the validity of the method, Sargsyan {\it et al.} \cite{SAA13b} used it to evaluate cross sections for several systems: $^4{\rm He}\,+
\,^{92}{\rm Mo}$,  $^4{\rm He}\,+\,^{110}{\rm Cd}$, $^4{\rm He}\,+\,^{110}{\rm Cd}$, $^4{\rm He}\,+\,^{116}{\rm Cd}$, $^4{\rm He}\,+\,^{112}{\rm Sn}$, 
$^4{\rm He}\,+\,^{120}{\rm Sn}$, $^{16}{\rm O}\,+\,^{208}{\rm Pb}$ and $^{6,7}{\rm Li}\,+\,^{64}{\rm Zn}$. In each case, the obtained cross section
was compared with total reaction data obtained by the standard method, fitting elastic angular distributions. The overall agreement was good,
except for the $^{16}{\rm O}\,+\,^{208}{\rm Pb}$ and $^{6}{\rm Li}\,+\,^{64}{\rm Zn}$ systems. There was no clear justification for these discrepancies.
The authors suggested that it might be related to the uncertainties in the elastic data at backward angles. 

 An analogous procedure can be used to determine the capture component of the total reaction cross section. In this case, one replaces in 
 Eq.~(\ref{PR}) the  elastic  cross section by the quasi-elastic one, and truncate the $J$-sum of Eq.~(\ref{sigR-Jsum}) at the critical angular momentum
 (the limit of integration over $E^\prime$ is modified accordingly). Using this method, capture cross sections for the $^{6,7}{\rm Li}\,+\,^{64}{\rm Zn}$
 systems were determined in Ref.~\cite{SAA13b}.\\
  
Guimar\~aes {\it et al.}~\cite{GKA16} performed a dedicated backscattering experiment to determine the elastic cross section for the $^6$He+$^{209}$Bi 
system, at sub-barrier energies. They measured cross sections at 150$^{\circ}$, with uncertainties better than 10\%. Several other elastic 
scattering data sets for collisions of exotic beams at back angles (150$^{\circ}$ or higher) are available in the literature. However, usually they have 
uncertainties of not less than 20\%. Thus, it would be interesting to employ the backscattering method to determine the total reaction cross section for the
$^6$He+$^{209}$Bi system,  using the data of Ref.~\cite{GKA16}.\\
%


\subsubsection{Total cross section from beam attenuation}


The {\it beam attenuation} or {\it transmission} method is an old and quite simple direct method to obtain the total reaction cross section~\cite{PKL82,KGC84}. 
In this method, one measures the number of particles in the incident beam, $N_{\scr B}$, and the number of unscattered particles plus particles scattered elastically, 
after the beam traverses a thick target, $N_{\scr T}$. The difference of the two is proportional to the total reaction cross section. That is,
\begin{equation}
\sigma_{\scr R} = k \times\ \left[\frac{N_{\scr B} - N_{\scr T}}{N_{\scr B}}\right],
\end{equation}
where $k$ is a constant related to the target thickness. This is a reliable method to be applied to nucleus-nucleus collisions at medium to high energies (10-50 MeV/A). 
At first, it was used to measure the total reaction cross section in collisions of light particles ($p, d, t$ and $\alpha$) with several targets, mostly for 
practical applications in reactors and space science~\cite{MEC60}. \\

This method was first applied in an experiment performed at CERN, to measure the total reaction cross section for the $^{12}$C+$^{12}$C system~\cite{PKL82}. 
An improvement of the method was achieved when TOF was added to the $\Delta E-E$ signals for a better identification of the particles. Details can be found in the 
work of Zheng {\it et al.}~\cite{ZYO02}, reporting their experiment to study of the halo structure of $^{16}$C, performed at RIKEN.   \\

Using a simular approach, Erdemshimeg {\it et al.}~ \cite{EAD19} performed experiments at the Flerov Laboratory of Nuclear Reactions, JINR, Russia. They measured 
total reaction cross sections in collisions of  $^{4,6,8}$He, $^{7,9,10,11,12}$Be, $^{7,8,9}$Li and $^{8,10,11,12}$B projectiles on $^{28}$Si targets, at energies in the 
range $(10-50)\, {\rm MeV}\cdot A$. 
A smooth mass dependence was observed for the extracted radius, but with fluctuations for some particular projectiles. These fluctuations might be an indication of 
halo structure. \\

Another interesting experiment was performed at the Radioactive Ion Beam Line in Lanzhou (RIBLL) at the HIRFL, China. Li {\it et al.}~ \cite{LLW10} measured total
reaction cross sections in collisions of the mirror nuclei  $^{12}$N and $^{12}$B  on  $^{28}$Si. The data were analyzed by Glauber model calculations, using 
gaussian-gaussian distributions~\cite{ADE02} for the nuclear densities of $^{12}$N and $^{12}$B. The parameters of the densities were then fitted to reproduce the 
total reaction data. The density of $^{12}$N obtained in this way is consistent with the $^{11}{\rm C}+p$ cluster configuration.
%


\subsubsection{Total reaction cross section by solid active target method}


The pioneering work of Tanihata {\it et al.}~\cite{THH85} on the interaction cross section\footnote{The interaction cross section, $\sigma_{\scr I}$, is defined as the sum of all 
nonelastic cross sections in which at least one nucleon is removed from the projectile. Since the total reaction cross section is the sum of all nonelastic processes, they are 
related by the equation: $\sigma_{\scr R} = \sigma_{\scr I} + \sigma_{\rm inel}$.} in collisions of the radioactive lithium and beryllium isotopes triggered a series of experiments  
to measure total reaction cross sections for halo nuclei. \\

A simple method has been developed for this purpose, using solid active targets. In this method, the incident particles are stopped within the solid-state detector, 
which has the double function of detector and target. Owing to the $Q$-value of the non-reacting events ($Q=0$), it is possible to distinguish them from reacting events. The
former produce a sharp peak, with  large intensity, at the beam energy, $E_0$, whereas reactive events give much weaker contributions to the spectrum at different energies
($E=E_0+Q$).} The energy signal in coincidence with $\gamma$-ray signals from a surrounding 4$\pi$ array of NaI(Tl) detectors can help to clean the energy spectra from 
spurious events, mainly in the energy region of $Q \sim 0$. 
The reaction probability, $\mathcal{P}_{\scr R}$, is then determined by dividing the number of reactive events (outside of the sharp elastic peak) 
by the total number of counts in this spectrum. \\

This method was first applied at GANIL, for silicon target-detectors  and light radioactive projectiles at intermediate
energies (20-50 ${\rm MeV}/A$)  \cite{VMP91}.
In this experiment, the mean (energy-integrated) total reaction cross section and the associated reduced radius, $r_0^2$, were obtained for several neutron-rich 
radioactive nuclei in the mass range $8<A<40$, such as $^8$He, $^{12}$Be, $^{15}$B, $^{22}$O, $^{26}$Ne, $^{37}$Si. The authors observed a 
strong isospin-dependence in the total reaction cross section and in the  reduced radius for all isobars with $A=10$ to 18. 

Applying this technique and using a stack of silicon detectors, Warner {\it et al.}~\cite{WPV96,WCB06} performed several experiments with proton- and neutron-rich
radioactive beams at the NSCL-MSU laboratory. They performed simultaneous measurements of total reaction and multi-nucleon removal cross sections at intermediate energies.
The radioactive nuclei were separated by the A1200 analyzing system of this laboratory \cite{SMN92} and impinged on a stack of silicon detectors.  Reaction events were 
then identified by the total energy loss in the telescope, which is different from that of non-reacting projectiles. 
Since the projectile's energy decreases as it travels through the telescopes, the cross sections were obtained at different energies by identifying in which detector 
the reaction occurred.
The obtained data were interpreted with the phenomenological strong absorption models and the optical limit of Glauber multiple-scattering theory. The 
later reproduced quite well the data.  

The above technique has also been applied to low energy incident projectiles~\cite{MFL10}. In this case, the short-range of the particles within the detector is an important issue 
to be considered.
Besides, the reaction probability is substantially reduced by the Coulomb barrier, and this reduction leads to a low signal-to-noise ratio. Further, particles may backscatter and then
leave the detector without loosing completely its energy. This would yield a small pulse that would be misinterpreted as a reaction event~\cite{WBD88}. 
Using this technique, total reaction cross section were obtained for the $^{6,7}$Li+$^{28}$Si systems at near-barrier energies \cite{PMP06}. The measured excitation function 
for the total reaction cross section agrees with the data obtained in previous standard experiments, where the total reaction cross section is determined from optical model 
analysis of angular distributions. 

The active silicon target technique has also been used to study fusion of exotic nuclei. An experiment to measure fusion cross sections for the $^8$B + $^{28}$Si system was 
performed at LNL, Italy \cite{PSP13}. The $^8$B secondary beam was produced at the EXOTIC facility and, from the detected evaporated alpha particles, the authors determined 
complete fusion cross sections at several energies above the Coulomb barrier. This measurement was based on the assumption that the observed alpha particles
resulted exclusively from the evaporation of the compound nucleus formed in the collision. Their fusion excitation function was compared to data for several other weakly
and tightly bound systems, measured in different experiments. To eliminate the influence of trivial factors, like charges and masses of the collision partners, the data were 
reduced by the fusion function method~\cite{CGL09a,CGL09b} (see section \ref{fusfun}). They concluded that their reduced complete fusion cross section was similar to the 
data of most other systems, and to the predictions of barrier penetration models. The exception was the $^8$B+$^{58}$Ni data of Ref.~\cite{AVQ11}, obtained from 
the proton evaporation multiplicity. 
These data were much larger that those for the other systems, above and below the barrier. The $^8$B + $^{58}$Ni and $^8$B + $^{28}$Si data were
later analyzed on the same basis and potentials assumptions and they were shown to be consistent with each other and in the fusion enhancement, 
despite the small overlap of the collision energies \cite{AVQ16}. 
To clarify this issue, a new experiment was performed at the Cyclotron Institute of the Texas A\&M University, USA, involving a similar system. Direct 
 measurements of the fusion cross sections for the $^8{\rm B}\,+\,^{40}{\rm Ar}$ system were performed, using the gaseous active target TexAT \cite{KRP20} 
 (this technique is discussed in the next 
 section). The final results of this study are expected to be soon available.


\subsubsection{Promising future methods}


%
\begin{figure}
\begin{center}
\includegraphics*[width=6cm]{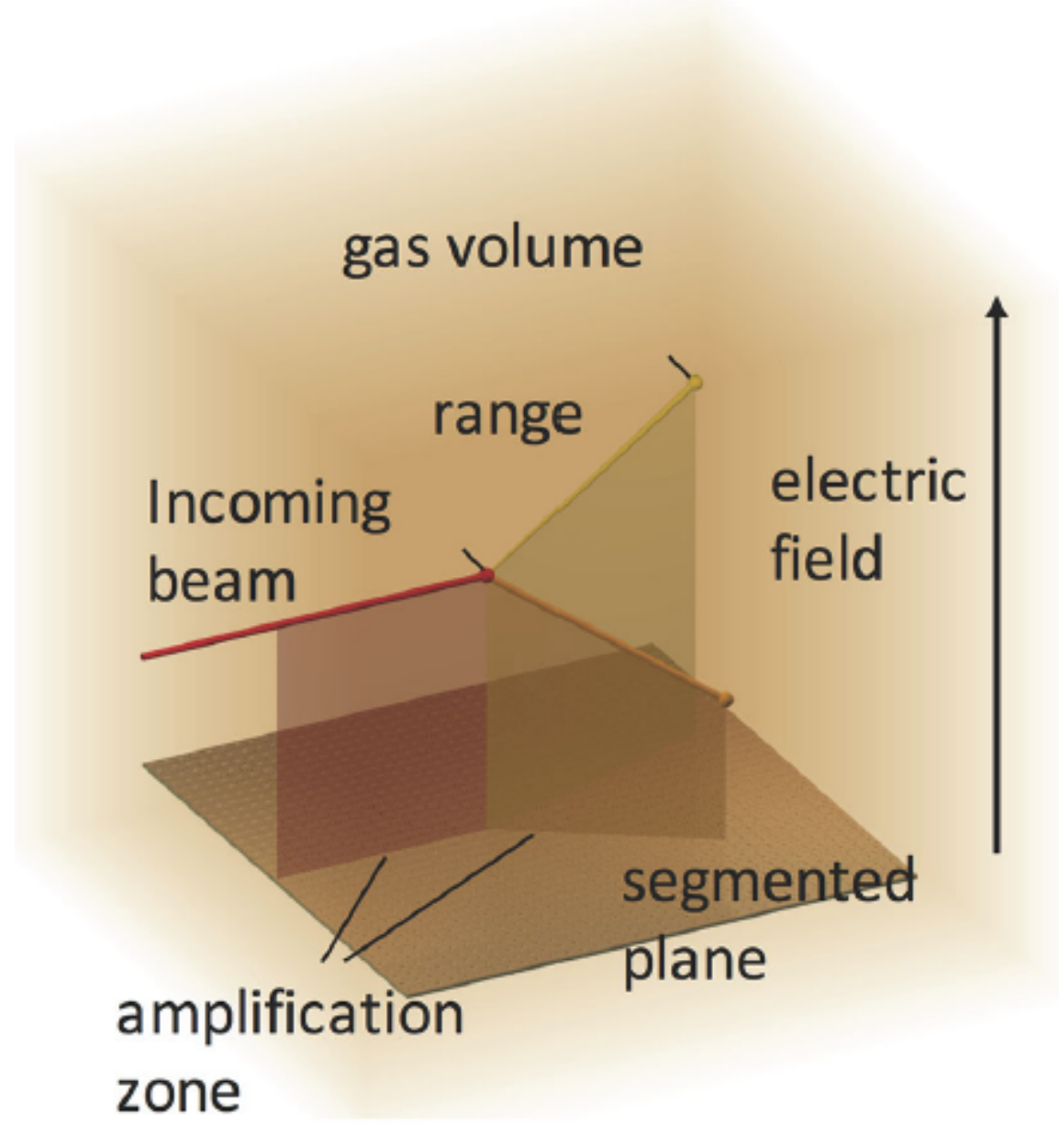}
\end{center}
\caption{(Color online) Working principle of an Active target. Figure extracted from Ref. \cite{Raa16}.}
\label{F5.7}
\end{figure}
As described in the previous sections, facilities offering radioactive ion beams are constantly improving, and developing new techniques to overcome the limitations 
due to the low beam intensity. One of these techniques is the use of active gaseous targets. They consist of an ionizing gas, which plays the double role of target 
and detector.  Their geometry allows a high solid angle detection (4$\pi$), they have variable thickness, and great detection efficiency. In addition, they have good 
resolution, since the locations of the reaction events within the target are accurately measured. 

The working principle of an active target  is schematically represented in Fig.~\ref{F5.7}. The trajectory of the incident projectile (thick red line) ends at the point 
where the reaction takes place (red circle), giving rise to reaction products - two in this example - that follow new trajectories (thin lines). Along the trajectories, the 
particles ionize the medium, producing free electrons. These electrons move under the action of the vertical electric field, until they are detected on the segmented 
horizontal plane (bottom of the figure). In this way, the projections of the three trajectories on the horizontal ($x-y$) plane are determined. Then, three-dimensional 
trajectories are constructed using the information of the drift time of the pulses detected on the horizontal plane.
The high luminosities achieved in gaseous active targets (due to their large dimensions in comparison with a target foil) are of great importance in experiments involving 
low intensity radioactive ion beams. \\

Active targets as the IKAR, at GSI \cite{IAA12}, and the MAYA, at GANIL \cite{DCW07}, have been used for a long time, to investigate resonant scattering and transfer 
reactions on hydrogen and helium targets in inverse kinematics.
A review on the existing active target and its corresponding working principles can be found in Ref. \cite{NAB15,BAA20}.\\

 A new generation of active targets, based on a time projection chamber, are becoming operational. The electrons produced by the particles inside the gaseous 
 target drift towards a micromegas plane, which multiplies them and provide position and time information of the particle tracks. This allows a complete kinematical 
 reconstruction of the reaction in the 3-dimensional  space. The ACTAR-TPC, at GANIL \cite{GPP20}, the TexAT, at Texas A \& M University \cite{KRP20}, 
and the AT-TPC, at NSCL-MSU/University of Notre dame \cite{MNF15} are examples of such new generation devices. More recently, the prototype AT-TPC was 
used to measure fusion cross sections for the $^{10}$Be+$^{40}$Ar system \cite{KHM16}, indicating that this new technique can be extended to experiments with 
heavy targets (at least in the case of $^{40}$Ar).
This kind of device is quite promising to allow the simultaneous measurement of complete fusion, incomplete fusion, breakup, scattering, and transfer 
reactions, in one experiment, which is a dream for experimentalist.\\

Another promising experimental technique, not yet fully explored for reaction measurements, is based on the stored ion beams. The use of  a heavy-ion storage ring 
with internal gas-jet targets, such as the ESR, at GSI \cite{Fra87}, or the HIRFL-CSR, at Lanzhou \cite{XZW02}, allows a great increase of the beam luminosity, since 
the beam revolutions in a closed trajectory with a very high frequency ($\sim 10^6$  MHz). However, a major challenge of this technique is to install a detector setup compatible with the ultra-high vacuum regime 
(in the order of 10$^{-10}$ mbar or below) needed to the operation of a storage ring. At this level any type of outgassing material can significantly deteriorate 
the vacuum conditions \cite{MEE15}. Recently, nuclear reaction experiments were successfully performed at the heavy-ion storage ring at the GSI facility, using 
stored $^{56,58}$Ni beams, and hydrogen and helium internal gas-jet targets. The recoiled particles were detected by in-ring DSSSDs and angular distributions 
were measured, allowing studies of nuclear matter \cite{SBB15,ZAB17} and giant resonances \cite{ZAB16}. In the future this type of technique may be applied 
to investigate nuclear reactions using heavy jet gas targets.


\subsection{Comparative studies of cross sections}
  

It is well known that total reaction cross sections in heavy-ion collisions at near-barrier energies have important contributions from fusion and also from direct reactions 
(inelastic scattering, transfer, and breakup). Besides,  couplings with direct reaction channels may exert a strong influence on the fusion cross section itself 
\cite{BBE81,BSS82,RHH97}.  An important example is the collision of a heavy or medium mass projectile with a strongly deformed target~\cite{StG81} at sub-barrier
energies. The fusion cross section is several orders of magnitude larger than those for the same projectile on a spherical target with similar mass. 
In this way, the enhancement of the fusion cross section may be a signature of the deformation.  Nevertheless, despite this enhancement, 
the total reaction cross section is still dominated by direct reactions in this energy region.
Another example, which has attracted considerable interest in the last few decades, is the suppression of complete fusion in collisions of weakly bound nuclei (see e.g. 
Ref.~\cite{CGD15} and references therein). \\

Clearly, comparisons of fusion and/or reaction data for different systems is a very useful tool in studies of nuclear structure and reaction 
mechanisms. However, direct comparisons of data for different systems may be misleading. The reason is that the cross sections also depend 
on trivial properties of the system, like the charges and sizes of the collision partners. We illustrate this point with an example. 
We consider the cross sections for collisions of $^{16}$O projectiles with energy $E_{\rm lab} = 50$ MeV on two different targets: the spherical
$^{40}$Ca and the highly deformed $^{154}$Sm nuclei.
Estimating the cross sections by potential scattering calculations with the Aky\"uz-Winther potential~\cite{BrW04,AkW81} and short-range absorption, one finds 
$\sigma_{\scr F} = 764$ mb for $^{40}$Ca and $ \sigma_{\scr F}= 0.6 \times10^{-10}$ mb for $^{154}$Sm.  Clearly, the 
comparison of these cross sections at the same beam energy is meaningless. In the case of the $^{40}$Ca target, the beam energy corresponds 
to $E_{\rm c.m.} = 35.7$ MeV and the barrier is $V_{\scr B} = 24$ MeV, whereas for $^{154}$Sm the energy is $E_{\rm c.m.} = 45.3$ MeV and the 
barrier $V_{\scr B} = 60$ MeV. Thus, the vanishingly small cross section for the heavier target is a trivial consequence of the fact that the collision 
energy is almost 15 MeV below the Coulomb barrier. At this very low energy, all nonelastic cross sections are extremely low. 
The importance of the barrier height in comparisons of cross sections for different systems emerges more clearly in Fig.~\ref{Lub1}, which shows fusion cross sections for 
systems in different mass regions, as functions of the collision energy in the center of mass frame. One observes that this kind of comparison does not reveal
any physical property of these systems. It only shows trivial differences in the Coulomb barriers.
\begin{figure}
\begin{center}
\includegraphics*[width=8cm]{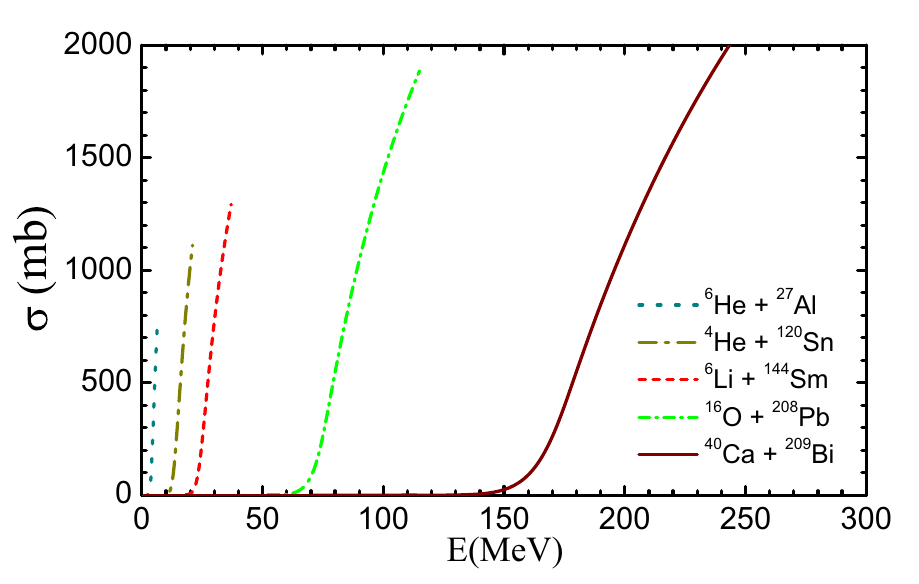}
\end{center}
\caption{(Color online) Comparison of the total reaction cross sections for several systems, obtained through single-channel calculations. (Figure taken from Ref.~\cite{GMC16}).}
\label{Lub1}
\end{figure}
The examples discussed above clearly show that static properties, like the barrier height and geometric effects arising from the size of the system, must be eliminated from 
the analysis. For this purpose, it is necessary to make transformations on the collision energy and on the cross section to get rid of these trivial factors. This procedure is 
known as {\it reduction}. 
Several reduction methods have been proposed over the last two decades. Detailed discussions of this subject, pointing out their successes and shortcomings, can be found, e.g., 
in Refs.~\cite{CGD15,CMG15,GMC16}. Some of the main reduction methods available in the literature are discussed below.\\

We begin with the {\it traditional} reduction method, which consists in the  transformations,
 \begin{eqnarray}
 E\ \rightarrow\ E_{\rm red} &= & \frac{E}{V_{\scr B}} \label{traditional-1} ,\\
 \sigma\ \rightarrow\ \sigma_{\rm red} &=& 
 \frac{\sigma}{\pi R{\scr B}^2} .
 \label{traditional-2}
 \end{eqnarray}
Qualitatively, the above transformations go in the right direction. They modify the scales of the x- and y-axes of Fig. \ref{Lub1}, bringing the curves of the different systems to the same 
region\footnote{A variant of the traditional method is to replace Eq. (\ref{traditional-1}) by the transformation $E \rightarrow E_{\rm red} = E - V_{\scr B}$.}.
 
 The second method considered here, proposed by Gomes {\it et al.}~\cite{GLP05}, adopts the transformations,
 \begin{eqnarray}
 E\ \rightarrow\ E_{\rm red} &= & E \ \times\  \left[
 \frac{Z_{\scr P} Z_{\scr T}}{A_{\scr P}^{\scr 1/3} + A_{\scr T}^{\scr 1/3}}
 \right]^{-1} \nonumber\\
 \sigma\ \rightarrow\ \sigma_{\rm red} &=& 
 \frac{\sigma}{\left( A_{\scr P}^{\scr 1/3} + A_{\scr T}^{\scr 1/3}\right)^2} .
 \label{gomes}
 \end{eqnarray}
This method, inspired by the previous one, is based on the assumptions that the barrier radius scales like $A_{\scr P}^{\scr 1/3} + A_{\scr T}^{\scr 1/3}$ and that the potential barrier 
can be approximated by the Coulomb potential at the barrier radius. Thus, it scales like $Z_{\scr P} Z_{\scr T} /(A_{\scr P}^{\scr 1/3} + A_{\scr T}^{\scr 1/3})$. This method has the
advantage of being independent of the parameters of the interaction potential, $V_{\scr B}$ and $R_{\scr B}$.\\

The above  methods have been widely used to reduce reaction and fusion data. However, a comprehensive study of reduction methods for fusion cross sections, considering a large number of 
systems in different mass ranges~\cite{CGL09a,CGL09b}, concluded that the best results were achieved by the {\it fusion function} method, which is discussed below.\\

The fusion function method hinges on the use of the Wong's formula for the fusion cross section (Eq.(\ref{sigWong})). The collision energy
and the cross section are replaced by the dimensionless quantities~\cite{CGL09b,CGL09a},
\begin{equation}
E\ \rightarrow\ x = \frac{E-V_{\scr B}}{\hbar\omega} ;\ \ \  \sigma(E) \ \rightarrow\ 
F(x) = \frac{2\,E}{\hbar\omega\,R_{\scr B}^2}\times \sigma(E).
\label{x-Fx}
\end{equation}
It can be immediately checked that if one applies the reduction method to the Wong's cross section itself, one gets
\begin{equation}
F_0(x) = \ln{[1 + e^{2\pi\,x}]} .
\label{UFF}
\end{equation}
The above function, being basically independent on the system, is called \textit{Universal Fusion Function} (UFF). 
The UFF would be a good representation of the reduced fusion data when they can be described by a one-channel calculations with a standard interaction,
like the S\~ao Paulo~\cite{CPH97,CCG02} or the Aky\"uz-Winther~\cite{BrW04,AkW81} potential, with short-range absorption. The Wong's formula is a reasonable
approximation in heavy-ion fusion at near-barrier energies.  For heavy systems ($Z_{\scr P}\,Z_{\scr P} \gtrsim 250$), it remains valid down to several MeV below
$V_{\scr B}$. However, for lighter systems it deteriorates more rapidly as the energy decreases below $V_{\scr B}$. A detailed discussion of the reduction of fusion
data is presented in the next section.

\subsubsection{Reduction of fusion cross sections}\label{fusfun}

The reduction of fusion cross sections by different methods has been systematically investigated in Ref.~\cite{CMG15}. The methods were submitted to a simple test. 
They were used to reduce theoretical cross sections obtained by one-channel calculations using the Aky\"uz-Winther potential and strong short-range absorption. This study 
involved several systems in different mass regions. Since there are no nuclear structure effects in these calculations, a successful reduction method should lead to 
reduced cross section with very weak system dependence. This study has shown that only the fusion function method meets this criterium. The other methods kept 
important system dependence, mainly in comparisons of reduced cross sections for systems in different mass ranges. \\

The reduction of fusion cross sections by the fusion function method is carried out in two steps, as follows.

\begin{enumerate}
\item  The Coulomb barrier calculated with some version of the double-folding potential, like the AW or the SPP, is fitted by a parabola, and the barrier parameters 
$R_{\scr B}$, $V_{\scr B}$, and $\hbar\omega$ are determined. 

\medskip

\item The data points, $\{E_i,\sigma (E_i) \}$, are converted to reduced data points, $\{ x_i,F_{\scr exp}(x_i) \}$, according to Eq.~(\ref{x-Fx}). 

\end{enumerate}

If there are no relevant channel coupling effects and the Wong's formula gives an accurate description of the optical model cross section, the experimental fusion function, 
$F_{\scr exp}(x)$, is expected to be very close to the UFF. In this way, the function $F_0(x)$ can be used as a benchmark, to which $F_{\scr exp}(x)$ should be compared. 
Deviations from  the UFF in a given energy region indicates that the Wong's formula is inaccurate there, 
or it is an evidence of relevant channel coupling effects. In the latter case, the strength of the deviation measures the importance of the couplings. 
In most collisions, couplings with direct reaction channels play an important role in the reaction dynamics. Then, the fusion data can not be accurately described by one-channel 
calculations with typical heavy-ion potentials.  \\

Usually, one wants to assess the influence of some specific process on fusion, like the breakup. Then, there may be a difficulty. The deviations of the experimental fusion function from the UFF may arise from breakup couplings and couplings with other channels. Besides, deviations may also arise from inaccuracies of the Wong's formula. Then, the 
influence of such undesired effects should be minimized. This can be done introducing the renormalized experimental fusion function, denoted by $\overline{\rm F}_{\scr exp}(x)$, 
defined as,
\begin{equation}
F_{\scr exp}(x)\ \rightarrow\ \overline{\rm F}_{\scr exp}(x) = F_{\scr exp}(x) \times \frac{\sigma^{\scr W}(E)}{\sigma^{\scr CC}_{\scr F}(E)}.
\label{sigmared2}
\end{equation}
Above, $\sigma^{\scr W}(E)$ is the Wong's cross section of Eq.~(\ref{sigWong}), and $\sigma_{\scr F}^{\scr CC}(E)$ is the theoretical fusion cross section obtained through a 
CC calculation that includes all relevant channels, but the ones we are investigating (usually the breakup channel). 
Using the reduction procedure of Eq.~(\ref{sigmared2}), one estimates the influence of the channels left out of the CC calculations. In Refs.~\cite{CGL09a,CGL09b}, where
this method was introduced, the channels left out were breakup and transfer.  
Of course, if all relevant channels are included in the CC calculations, the renormalized experimental fusion function is expected to coincide with the UFF.
The fusion function method has been widely used in comparative studies of fusion cross sections in collisions of weakly-bound and tightly-bound projectiles \cite{AAA20,KGT20,TAG20,AKG19,GhZ19,MKB19,RSH19,SAA17}. \\

\noindent {\it Detailed discussion of breakup effects on fusion}

\bigskip

Significant channel coupling effects have been observed in collisions of weakly bound nuclei. In this case, the relevant couplings are with the breakup channel. 
These couplings give rise to different fusion processes. In addition to complete fusion (CF), where the whole projectile is absorbed by the target, there is incomplete fusion 
(ICF). The latter takes place in two steps. First, the weakly bound projectile breaks up into fragments, say two, as it interacts with the target. Then, one fragment is absorbed 
by the target, while to other is not. Compared with the fusion cross section of a typical one-channel calculation, the experimental CF cross section is suppressed above the 
Coulomb barrier, and the TF (the sum of CF and ICF) cross section is enhanced at sub-barrier energies, due to the ICF of the lighter fragment. 
The determination of individual CF and ICF cross sections in collisions of weakly bound projectiles is a great challenge for both experimentalists and theorists. This  subject 
has been reviewed in several works \cite{CGD06,KRA07,KAK09,CGD15,KGA16,JPK20,RCL20}. Most fusion experiments measure only the TF cross section. However, the 
available measurements of CF cross sections for weakly bound systems systematically find suppression at above-barrier energies. Further, the strongest suppressions were 
observed in collisions of the most weakly bound projectiles on heavy targets. \\

Another kind of reduction method has been used to investigate the influence of breakup couplings in the fusion of weakly bound systems. It is well established that there is
 suppression of a CF in collisions of weakly bound projectiles on heavy targets at above-barrier energies compared to predictions of barrier penetration models or fusion 
 cross sections of tightly bound projectiles of similar masses. Further, the suppression is not very sensitive to the collision energy, provided that it is higher enough than the 
 Coulomb barrier ($E_{\rm c.m.} \gtrsim V_{\scr B}+2\,\hbar\omega$). For comparative studies of this phenomenon, it is convenient to measure the suppression by the single 
 number:
\begin{equation}
{\rm F}_{\scr CF} = \frac{\sigma_{\scr CF}}{\sigma_{\scr TF}},
\label{FCF}
\end{equation}
where $\sigma_{\scr CF}$ and $\sigma_{\scr TF}$ are respectively the complete and total fusion cross sections, above the Coulomb barrier.
This procedure was used in studies $^{6,7}$Li collisions with several medium-heavy and heavy targets, with the aim of correlating CF suppression with the mass and charge of the 
targets \cite{WLL03,RSS09,DHH02,DGH04,MRP06,KJP12,PMB11,SNL09,GVD20}. They concluded that F$_{\rm{CF}}$ has a weak target dependence. It is between 0.60 and 
0.67 for $^6$Li, and between 0.70 and 0.75 for $^7$Li. It is not surprising that the largest suppressions (lowest F$_{\rm{CF}}$) were found for $^6$Li, which has a lower breakup threshold.\\

Jha \textit{et al.}~\cite{JPK14} carried out a study along the same line, for collisions of $^9$Be on several targets. They investigated the correlation of the fusion processes with 
the charge of the target. Instead of ${\rm F}_{\scr CF}$, they used an ICF probability, $P_{\scr ICF}$, defined as 
\begin{equation}
P_{\scr ICF} = 1\,-\,{\rm F}_{\scr CF}.
\end{equation}
They concluded that the dependence of the CF/ICF processes on the target is more pronounced than that observed for the $^{6,7}$Li projectiles.\\

\begin{figure}
\begin{center}
\includegraphics*[width=8cm]{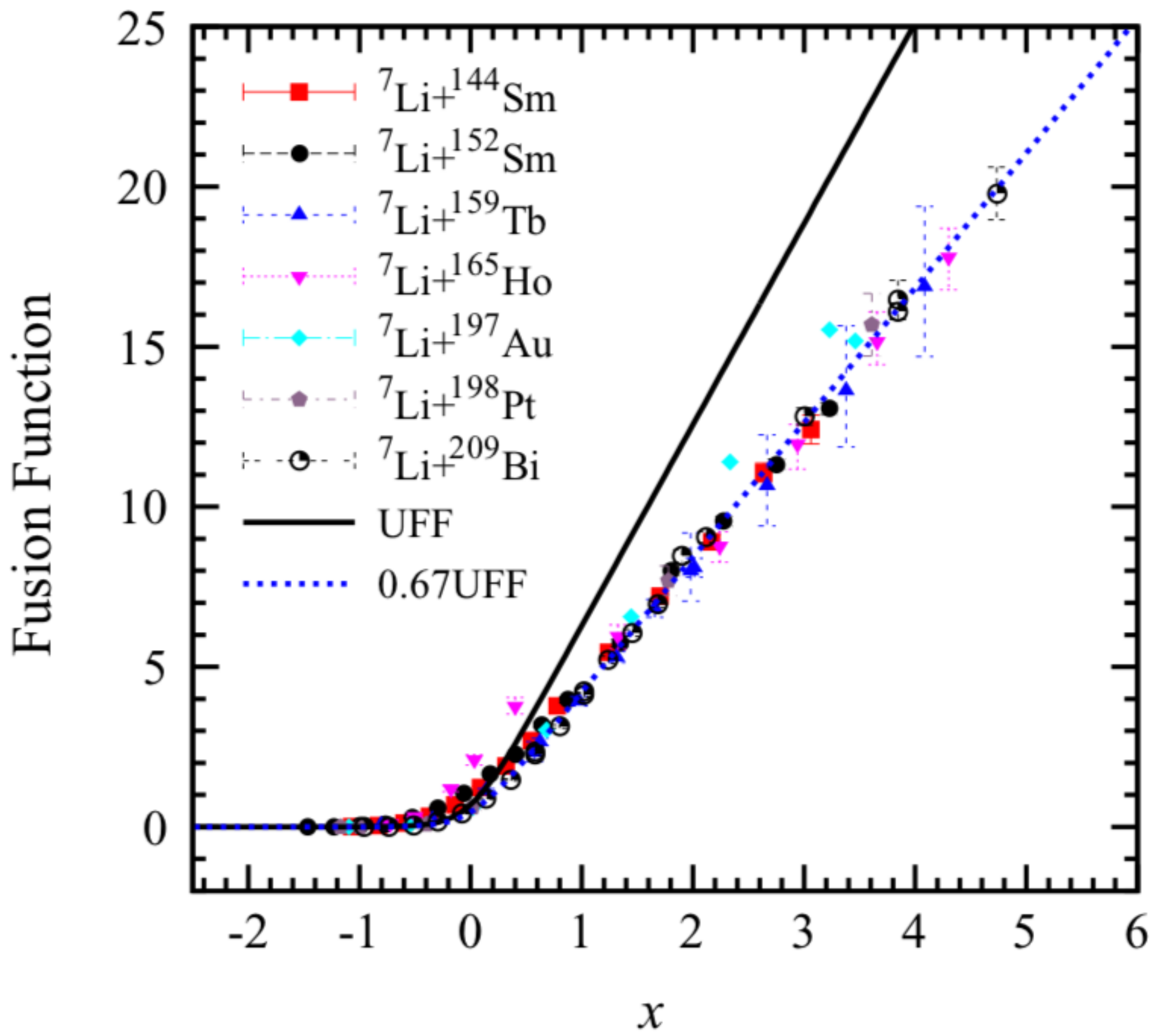}
\end{center}
\caption{(Color online) Experimental fusion function for collisions of $^4$Li with different heavy targets (figure taken from Ref.~\cite{WZG14}).}
\label{F24}
\end{figure}
A similar reduction method has been used to study the influence of the breakup threshold of the projectile on the CF cross section \cite{WZG14,SSY18}. These studies 
considered CF data of several weakly bound projectiles on different targets, with mass numbers in the range $89 \le A_{\scr T} \le 209$. They found that the experimental 
fusion functions were essentially independent or the target. This is illustrated in Fig. \ref{F24}, which shows experimental fusion functions for collisions of $^7$Li on 
different heavy targets. One observes that the data points follow the dotted line, corresponding to $0.67\times F_{\scr 0}(x)$. Then, the authors adopt the suppression 
factor
\begin{equation}
F_{\scr{B.U.} }= \frac{F(x)}{F_{\scr 0}(x)}
\label{FBU}
\end{equation}
to express the influence of the breakup channel on CF.

On the other hand, the suppression factor has an appreciable dependence on the breakup threshold of the projectile. This can be seen in Fig. \ref{Lub5}, which shows
the logarithm of $1 - F_{\scr{B.U.}}$, as a function of the breakup threshold of the projectile, $E_{\scr B.U.}$.
\begin{figure}
\begin{center}
\includegraphics*[width=8cm]{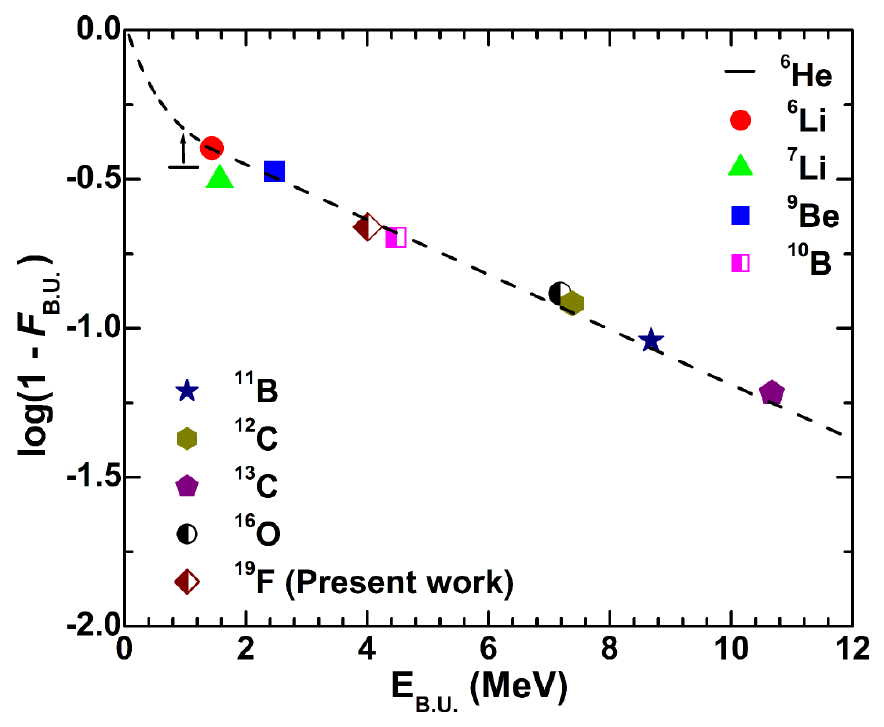}
\end{center}
\caption{(Color online) Dependence of the suppression factor on the breakup threshold of the projectile. (Figure taken from Ref.~\cite{SSY18}).}
\label{Lub5}
\end{figure}


\subsubsection{Reduction of the reaction cross section data}\label{red sigR}


The reduction of total reaction (TR) data is much more complicated because they are sums of contributions from fusion and direct reactions, which have very different natures.
Fusion is absorption by a strong imaginary potential acting within the inner region of the Coulomb barrier. In this way, 
it is equivalent to a barrier penetration problem. For this reason, the cross section scales with the barrier parameters, as predicted by the Wong formula. This explains the
success of the fusion function method to reduce fusion data. On the other hand, direct reactions are processes taking place in peripheral collisions. In 1-channel calculations, 
they can be simulated by a long-range imaginary potential, reaching beyond the barrier radius. These processes do not 
depend on the barrier parameters as predicted by the Wong formula. \\

Owing to the different characteristics of fusion and direct reactions, the available reduction methods have poor performances when applied to total reaction data.  A detailed study 
of this problem was carried out in Refs.~\cite{CMG15,GMC16}. Different reduction methods were applied to total reaction cross sections  obtained by 1-channel calculations using 
the Aky\"uz-Winther interaction~\cite{AkW81,BrW04}, $V_{\scr N}(R)$, and the long-range imaginary potential,  $W(R) = 0.78\,V_{\scr N}(R)$.  This procedure was applied to a 
large number of systems in different mass regions. Since the calculations did not take into account nuclear structure properties of the collision partners, differences among the
reduced cross sections should be attributed to the gross properties of the systems, like charges and masses. Such differences should not be found in a successful reduction method. 
This study has shown that the reduced cross sections kept a strong system dependence, independently of the method used. Nevertheless, the situation was much better when 
the comparison involved systems of similar masses.\\

\begin{figure}
\begin{center}
\includegraphics*[width=8cm]{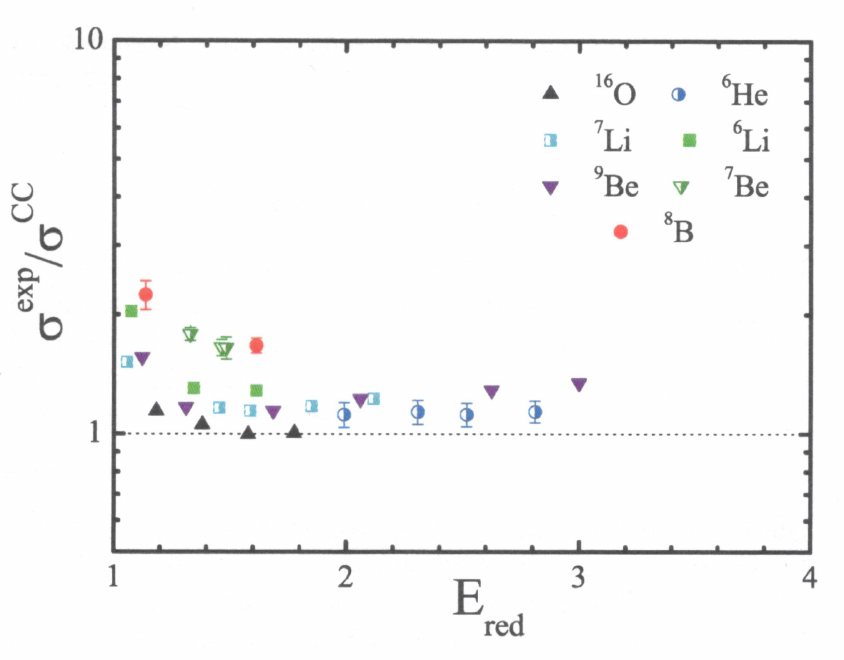}
\end{center}
\caption{(Color online) Reduced TR cross sections for collisions of several projectiles on $^{27}$Al, obtained by the procedure of Eq. (\ref{red-reac1}) 
(figure taken from Ref.~\cite{MLL17}).}
\label{Lub2}
\end{figure}
Recently, Morcelle {\it et al.}~\cite{MLL17} proposed a new procedure to reduce TR data, which emphasises the influence of the breakup channel. This method was applied 
to collisions of several weakly and tightly bound projectiles on $^{27}$Al. The method consists of the transformations,
\begin{equation}
E\ \rightarrow\ E_{\rm{red}} = \frac{E}{V_{\scr B}} ;\ \ \  \sigma ^{\scr{exp}}_{\scr{R}} \ \rightarrow\ 
\sigma _{\rm {red}} = \frac{\sigma ^{\scr{exp}}_{\scr{R}}}{\sigma ^{\scr{CC}}_{\scr{F}}},
\label{red-reac1}
\end{equation}
where $\sigma ^{\scr{CC}}_{\scr{F}}$ is the fusion cross section obtained by a CC calculation including couplings with the low-lying collective states of the $^{27}$Al target,
and $\sigma ^{\scr{exp}}_{\scr R}$ is the total reaction cross section obtained from fits to elastic scattering data. They considered data at above barrier energies for the 
following systems: $^{8}$B + $^{27}$Al~\cite{MLL17}, $^6$He + $^{27}$Al~\cite{BLJ07}, $^6$Li + $^{27}$Al~\cite{FFA07},  $^7$Li + $^{27}$Al~\cite{FAF06},  $^9$Be + 
$^{27}$Al \cite{GAM04,MGR05}, $^7$Be + $^{27}$Al~\cite{MLL14}, and $^{16}$O + $^{27}$Al~\cite{Cre79}. Sub-barrier energies were left out because, in this energy
region, the angular distribution corresponds, basically, to Rutherford scattering. In this way, the determination of the nuclear S-matrix tends to be inaccurate, leading to large error bars 
in the TR cross section. Breakup couplings, which hinder complete fusion, has the opposite effect on the total reaction cross section. It gives an additional contribution, 
which enhances TR. Therefore, $\sigma _{\rm {red}} $ must be larger than one. 

\smallskip

The reduced cross sections obtained in Ref.~\cite{MLL17} are shown in Fig. \ref{Lub2}. First, one notices that the reduced cross sections systematically increase as 
$E_{\rm{red}}$ decreases. This is a trivial consequence of the fact that the contribution from direct processes is dominant at low energies. Second, one observes that
comparing data of different systems in the same energy range, the largest reduced cross sections are for the projectiles with the lowest breakup thresholds.

\smallskip

\begin{figure}
\begin{center}
\includegraphics*[width=8cm]{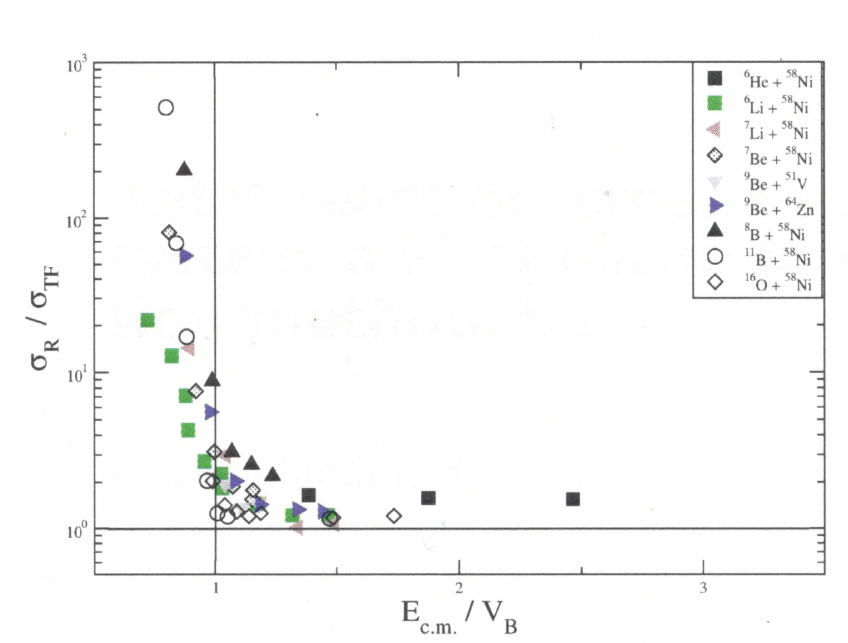}
\end{center}
\caption{(Color online) Comparison of the reduced total reaction cross sections for several projectiles on medium mass range targets., obtained through coupled-channel calculations. The experimental data were taken from Ref.~\cite{MPR14} for  $^6$He + $^{58}$Ni, Refs.~\cite{AML09,BRS08,PSB73} for $^6$Li + $^{58}$Ni, Ref.~\cite{PSB73} for $^7$Li + $^{58}$Ni, Refs.~\cite{AML09,MTP15} for $^7$Be + $^{58}$Ni, Ref.~\cite{MMB16} for $^9$Be + $^{51}$V, Ref.~\cite{GRM05} for $^9$Be + $^{64}$Zn, Ref.~\cite{AML09} for $^8$B + $^{58}$Ni, Ref.~\cite{DGC15} for $^{11}$B + $^{58}$Ni, and Ref.~\cite{KCC95} for the $^{16}$O + $^{64}$Zn system (Figure taken from Ref.~\cite{DeL18}).}
\label{Lub3}
\end{figure}
Deshmukh, Lubian, and Mukherjee~\cite{DLM19,DeL18} used the same reduction procedure to investigate the TR cross sections in collisions of stable and radioactive weakly bound projectiles
with intermediate mass targets, also considering data below the Coulomb barrier. Their results are shown in Fig. \ref{Lub3}. As could be expected, the reduced cross section below $V_{\scr B}$
grows abruptly, reaching  very high values. However, the correlation between the reduced cross section and the breakup threshold of the projectile emerges clearly, in the whole energy interval. 
The data points for the stable weakly bound nuclei ($^6$Li, $^7$Li, and $^9$Be) are clearly above those for the tightly bound $^{16}$O, and the ones for the radioactive projectiles ($^6$He, $^8$B 
and $^{11}$Be), which have breakup thresholds below 1 MeV, are still higher.\\

 
\section{Summary}


We have presented an account of recent theoretical and experimental developments in the study of nuclear reactions at near-barrier energies, with emphasis on 
weakly bound systems and in evaluations of the total reaction cross sections. Nuclear reactions with weakly bound nuclei have large breakup cross sections, and
other reaction channels are strongly influenced by the breakup process.\\

We began reviewing the main features of the quantum mechanical treatment of potential scattering, and its semiclassical approximations. Then, we have discussed
the coupled channel approach, with emphasis in the discretization of the continuum, which is essential to describe collisions of weakly bound nuclei. The influence of 
core and target excitations on the elastic and reaction cross sections were then reviewed, and a few applications were considered.\\

We addressed also the surrogate method to describe inclusive cross sections in collisions of weakly bound nuclei. Different formulations of the method were
discussed and a few examples were given.\\

A detailed discussion of the available methods to measure the total reaction cross section has been presented, with emphasis on the most recent experimental
techniques to measure cross sections in experiments with low intensity secondary beams of unstable nuclei. We then discussed reduction method to allow meaningful
comparisons of data for different systems. We pointed out that the fusion function reduction method works very well for fusion data, even for very different systems,
but satisfactory reductions of total reaction data can only be achieved for very similar systems.







\end{document}